\documentstyle[12pt,aasms,tighten]{article} 

\begin{document}
                                                                                
\title{The Northern ROSAT All-Sky (NORAS) Galaxy Cluster Survey I: X-ray        
  Properties of Clusters Detected as Extended X-ray Sources
\footnote{Results reported here are based on observations made with the 
Multiple Mirror Telescope, a joint facility of the Smithonian Institution
and the University of Arizona} 
}                
                                                                                
\author{H. B\"ohringer$^1$, W. Voges$^1$, J.P. Huchra$^2$,B. McLean$^3$, 
R. Giacconi$^4$, P. Rosati$^4$, R. Burg$^3$, J. Mader$^2$, 
P. Schuecker$^1$, D. Simi{\c c}$^1$, S. Komossa$^1$, 
T.H. Reiprich$^1$, J. Retzlaff$^1$, J. Tr\"umper$^1$}

\altaffiltext{1}{Max-Planck-Institut f\"ur extraterrestrische Physik, 
D-85740 Garching,Germany}
\altaffiltext{2}{Harvard-Smithonian Center for Astrophysics, 60 Garden Street,
Cambridge, MA 0213}                 
\altaffiltext{4}{Space Telescope Science Institute, San Martin Drive, 
Baltimore, MA 02138}
\altaffiltext{3}{ESO, D-85748 Garching,Germany}
\begin{abstract}

In the construction of an X-ray selected sample of galaxy clusters
for cosmological studies, we have assembled a sample            
of 495 X-ray sources found to show extended X-ray emission in the           
first processing of the ROSAT All-Sky Survey.                                   
The sample covers the celestial region with declination
$\delta \ge 0\deg $ and galactic latitude                  
$|b_{II}| \ge 20\deg $ and comprises sources with a count rate $\ge 0.06$       
counts s$^{-1}$ and a source extent likelihood of 7. In an optical follow-up        
identification program we find 378 (76\%) of these sources                  
to be clusters of galaxies.
                                                                                
It was necessary to reanalyse the sources in this sample with a new X-ray        
source characterization technique to provide more precise values
for the X-ray flux and source extent
than obtained from the standard processing. This new method, termed
growth curve analysis (GCA), has the advantage over previous methods
to be robust, easy to model and to integrate into simulations, 
to provide diagnostic
plots for visual inspection, and to make extensive use of the X-ray
data. The source parameters obtained assist
the source identification and provide more precise X-ray fluxes.
This reanalysis is based on data from the more recent second             
processing of the ROSAT Survey. We present a catalogue of the
cluster sources
with the X-ray properties obtained as well as a list of the
previously flagged extended sources which are found to have a non-cluster
counterpart. We discuss the process of source identification from
the combination of optical and X-ray data.
                                                                                
To investigate the overall completeness of the cluster sample as a              
function of the X-ray flux limit, we extent the search                          
for X-ray cluster sources to the data of the second processing of the           
ROSAT Survey for the northern sky region between $9^h$ and $14^h$ in            
right ascension. We include the search for X-ray emission of known              
clusters as well as a new investigation of extended X-ray sources.           
In the course of this search we find X-ray emission from additional          
85 Abell clusters and 56 very probable cluster candidates among 
the newly found extended sources. A comparison of the 
X-ray cluster number counts of the NORAS sample 
with the REFLEX Cluster Survey results leads to an estimate
of the completeness of the NORAS sample of RASS I extended 
clusters of about 50\% at an X-ray flux of  
$F_x(0.1-2.4 {\rm keV}) = 3 \times 10^{-12}$ erg s$^{-1}$ cm$^{-2}$. The 
estimated completeness achieved by adding the supplementary sample in 
the study area amounts to about 82\% in comparison to REFLEX.
The low completeness introduces an uncertainty in the
use of the sample for cosmological statistical studies which will be
cured with the completion of the  
continuing Northern ROSAT All-Sky (NORAS) cluster survey project.               
                                                                                
\end{abstract}                                                                  
                                                                                
                                                                                
\section{Introduction}                                                          
                                                                                
Galaxy clusters are important tracers of the large-scale structure              
of the matter distribution in the Universe. As the evolution of clusters is     
closely linked to the overall evolution of the cosmic large-scale structure,    
important tests of cosmological models can be performed with statistical        
data on the cluster population. The mass distribution and the spatial           
clustering of clusters are particularly interesting measures in such studies    
(e.g. Henry et al. 1992, Bahcall \& Cen 1992, Eke et al. 1998, 
Thomas et al. 1998, Borgani et al. 1999).
The construction of well defined cluster catalogues and the compilation        
of their properties is therefore an important task for observational            
cosmology.                                                                      
                                                                                
Galaxy clusters were first detected and are continued 
to be cataloged from              
optical observations of galaxy density                                          
enhancements in the sky (e.g. Abell 1958, Abell et al. 1989 (ACO), 
Zwicky et al. 1961 - 68, Shectman 1985, Dalton et al.
1992, Lumbsden 1992, Collins et al. 1995, Couch et al. 1991, Bower et al. 1994,
Postman et al. 1996, Olsen et al. 1999, Scodeggio et al. 1999). 
X-rays have also successfully been used           
to detect galaxy clusters and to                             
conduct clusters surveys (e.g. Picinotti et al. 1982,                        
Kowalski 1984, Lahav et al. 1989,                                               
Gioia et al. 1984, 1990, Edge et al. 1990, Henry et al. 1992,                   
Romer et al. 1994, Pierre et al. 1994, Ebeling et al. 1996, 1998,               
Castander et al. 1995, Rosati et al. 1995, 1998, Burns et al. 1996
Collins et al. 1997, Burke et al. 1997, Vihklinin et al. 1998,                  
Scharf et al. 1997, Jones et al. 1998, B\"ohringer et al. 1998, 
De Grandi et al. 1999, Ledlow et al. 1999, Romer et al. 1999). 
The use of samples of clusters detected and characterized by their X-ray        
emission for cosmological studies has two major advantages over samples based   
on optical observations. 
First, the optical observations (without very extensive redshift           
measurements) provide only the projected galaxy distribution and not            
all galaxy density enhancements in the sky are bound, three-dimensional        
entities. In fact, in   
the course of the ESO Nearby Abell Cluster Survey 
(Katgert et al. 1996, Mazure et     
al. 1996) it was found that of the order of 10\% of the rich                    
clusters from the catalogue                                                     
of Abell, Corwin, and Olowin (1989) in the nearby redshift range $z \le 0.1$    
were spurious clusters without obvious clustering peaks in redshift space.      
For the optical surveys the reliability 
has improved, however, with the advent of multi-color surveys and machine based
matched-filter selection techniques (e.g. Postman et al. 1996, Olsen et al. 1999).
Extended X-ray emission from the hot intra-cluster plasma       
of galaxy clusters is a more clear indication of the presence of a large   
gravitationally bound mass aggregate since otherwise the hot plasma would        
have been dispersed immediately. And secondly,
the X-ray luminosity is a parameter much more    
tightly correlated with the mass of clusters than the usual richness           
parameter measured in the optical (e.g. Reiprich \& B\"ohringer 1999).          
Thus X-ray emission gives evidence for            
the presence of galaxy clusters within a certain mass interval.
(The correlation of the X-ray luminosity and cluster mass actually shows a 
dispersion of about a factor of 1.6 if one wishes to determine the mass
for a given luminosity - Reiprich \& B\"ohringer, in preparation). The one
exception is the case where the X-ray emission is not clearly extended and     
where the cluster emission could be confused with the emission of an AGN        
within the cluster or with a possible foreground or background source.          
This confusion is a problem for a very small fraction                           
of the cluster sources, but in general 
X-rays are a very useful indicator of a true cluster.     
In addition projection effects are minimized in X-ray surveys
since the X-ray surface brightness is more centrally concentrated than the
galaxy distribution.        
                                                                        
The ROSAT All-Sky Survey (RASS), the only large scale X-ray survey              
conducted with an X-ray telescope (Tr\"umper 1993, Voges et al. 1999),          
provides an ideal data base to detect large numbers of clusters and             
to compile an all-sky cluster catalogue with homogeneously applied
selection criteria.                                                             
To exploit this unique data base we are conducting an optical                   
follow-up identification program and redshift                                   
survey of RASS X-ray clusters in the northern hemisphere,                       
the Northern ROSAT All-Sky (NORAS) Cluster Survey project.
A complementary survey, the REFLEX (ROSAT-ESO Flux Limited X-ray)
Cluster Survey, is conducted
for the southern part of the RASS (B\"ohringer et al. 1998,
Guzzo et al. 1999). The NORAS identification program 
was started in 1992 in a first step
with a list of extended X-ray sources extracted from the                        
general source list of the first RASS processing                                
(RASS I; Voges et al. 1992).  
Apart from the selection for extent the following extraction                    
criteria were used: northern declination, a minimum distance of                 
20 degrees to the galactic plane, and                                            
a minimum count rate of 0.06 cts s$^{-1}$ in the ROSAT broad                    
band (0.1 to 2.4 keV). The criterion of X-ray source extent                     
was chosen for the selection of promising cluster candidates,                   
because early tests have shown that such a sample would be highly                
enriched (by about 70 - 80\%) in galaxy clusters. Contrary to the
cluster selection scheme used for the REFLEX Survey which is based
on the correlation of X-ray sources with galaxy overdensities,
the present sample selection is purely based on X-ray information.
With this different bias the NORAS Survey has also the potential 
to find more distant and possibly ``opitcally dark'' clusters.                               
                                                                                
The identification of these sources is now complete. In this paper              
we present a catalogue of the X-ray properties of the 378                       
cluster sources and 117 non-cluster sources of the primary                      
candidate list. An accompanying paper by Huchra et al. (1999)           
provides a detailed catalogue of the optical identifications                    
and redshift measurements of this sample, and scientific aspects                
of this survey are discussed in a paper by Giacconi                             
et al. (1999). While this survey was ongoing some of the brighter sources of    
this sample as well as some X-ray emitting Abell and Zwicky clusters were       
spectroscopically observed for the ''BCS program'' (Ebeling et al. 1998)       
by Allen et al. (1992) and Crawford et al. (1995, 1999). The region
with the deepest exposure in the northern RASS, the north excliptic 
pole with exposure times ranging from 2000 to over 40000 sec, is also
the subject of a dedicated survey which has identified all X-ray sources 
(c.f. Henry et al. 1995;  Gioia et al. 1995; Bower et al. 1996; and 
Henry et al. 1997).
                                                                                
Further studies on the X-ray properties of a sample of                       
these extended sources in $2\deg \times 2\deg$ sky fields extracted 
from the RASS revealed, that in the first standard processing 
of the RASS the count rate and the extent of the cluster 
sources are severely underestimated (see also                                   
Ebeling et al. 1996, DeGrandi et al. 1997). Therefore                                     
a detailed reanalysis of the sources in the present sample was necessary.    
                                                                                
Here we also report the results of the detailed reanalysis of the sources 
using a new X-ray source characterization technique. 
In 1996, a second revised processing of the RASS (RASS II; Voges et al. 1999)
with greatly improved attitude quality and with                                 
a fully merged photon data base became available. Our reanalysis 
is based on these new data.

Since the incomplete assessment of the X-ray count rate and extent in RASS I        
not only leads to an underestimate of the X-ray fluxes 
for extended sources but
also to an incompleteness of the sample extracted from the data base            
with certain limiting parameters, we have also used the new RASS data base      
to explore the incompleteness of the present cluster sample in terms of         
a flux-limited X-ray selected sample of galaxy clusters. In this study          
we selected a subregion covering the right ascension range from 9$^h$ to 14$^h$.      
In a first step we use the Abell cluster catalogue to study the           
completeness provided by the RASS I extent criterion.                           
We further study the prospects                                                   
of finding more clusters with a more comprehensive extent criterion based on the
new X-ray source analysis technique. This study also points the way to          
a more complete selection of galaxy clusters from the RASS X-ray sources.        
We now apply this algorithm in the ongoing NORAS cluster redshift survey.      

Since we will probably not be able tp rapidly complete the identifications
of the newly found sources, we decided to publish the first part of the
survey for which identifications are now complete and redshifts are nearly
(all but 9) complete. The main emphasis here is 
not to publish a catalog of a complete, flux-limited sample, but
to compile a cluster catalogue with reliable identifications
based on a wealth of X-ray and optical data which are included 
in the identification process in a comprehensive way.
The present sample contains many newly
found objects, some of which are interesting targets for further
astrophysical studies. 
                                                                                
The paper is organized as follows. In Section 2 we summarize the                 
properties of the                                                               
primary RASS I source list of extended sources, and in Section 3                 
we describe the techniques used       
to reanalyze the X-ray properties of the sample sources. In Section 4 we        
present the X-ray source catalogue with detailed X-ray properties of the        
495 sources. Major X-ray properties of the sources which help                  
in the identification of the objects are discussed in Section 5.                
The completeness of the sample is                                               
addressed in Sections 6 and 7 where we report the results of a               
rigorous search for X-ray emission                                              
from all ACO clusters and search for more extended X-ray sources 
with an improved analysis algorithm in a test region ranging 
in right ascensions from 9$^h$ to 14$^h$. 
In Section 8 we compare the present results to the previous ROSAT
Bright Cluster Survey by Ebeling et al. (1998). Section 9
provides a summary and conclusions. Throughout the paper we are using
a Hubble parameter of $H_0 =  50$ km s$^{-1}$ Mpc$^{-1}$ and
$h_{50} = H_0 / (50$ km s$^{-1}$ Mpc$^{-1})$ and further
$\Omega _0 = 1$ for the density parameter and $\Lambda _0 = 0$
for the cosmological parameter.

\section {The RASS I list of extended sources}                                  
                                      
During the ROSAT mission the first All-Sky Survey was conducted with
an X-ray telescope (Tr\"umper 1992, 1993).
The RASS was performed over a period of six months from August     
1990 to January 1991 with two follow-up auxiliary survey missions carried out  
to fill the gaps in the survey in February and August 1991.
The first processing of the survey (RASS I) provided a source list of
49441 sources (Voges et al. 1992, 1996). For this first analysis 
the survey data received were sorted into one of 90 2-degree wide strips
(oriented in the direction of constant ecliptic longitude)
while the satellite was still scanning the sky. As a consequence of this,
strips are overlapping in regions outside the equator. Photons
are exclusively sorted into only one of the overlapping regions.
Therefore the exposure time across the strips is quite homogenous, but
no advantage can be taken of the high total exposure in the ecliptic
pole regions. The survey product resulting from this first processing
of the RASS will be referred to as RASS I data base.
(The situation is different in RASS II, the second processing, where the 
$6.4\deg \times 6.4\deg$  sky regions contain the full exposure
data from the RASS). The RASS was conducted with the X-ray telescope
(Aschenbach et al. 1988) and Position Sensitive Proportional Counter (PSPC;
Pfeffermann et al. 1986) providing a high sensitivity
and a very low internal background. A histogram of the exposure 
time distribution of the NORAS survey area in RASS I is shown in Fig. 1
(where it is also compared to the exposure distribution in RASS II).
For this statistic we use the maximum exposure in any of the strips
for those cases where a sky pixel is covered by several survey strips.
The mean and median exposure times are 397 and 402 sec, respectively.

\begin{figure}                                                                  
\plotone{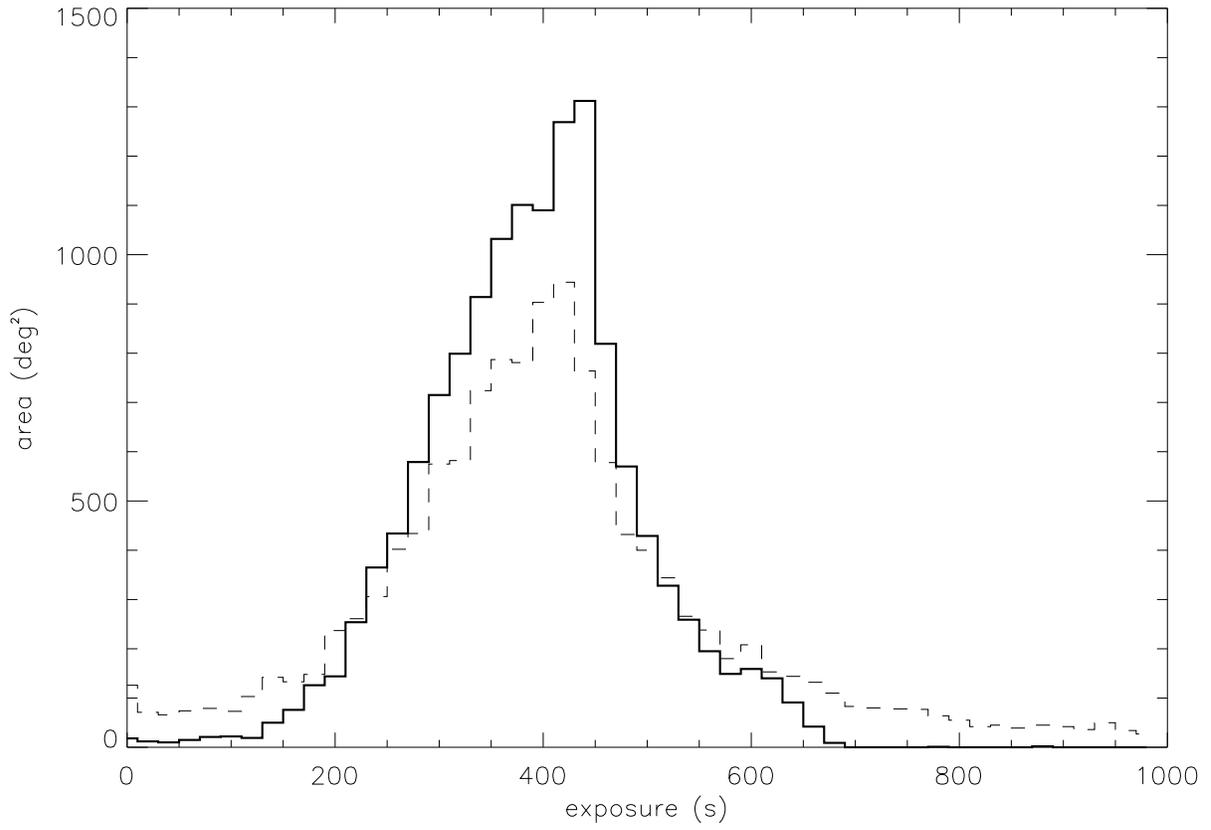}
\caption{Exposure time distribution of the RASS I for the NORAS survey area.
For comparison the distribution of the exposure time in the RASS II
involving the merged photon data from all survey strips is also shown
as broken line.}
\end{figure}   

The source detection procedure was based on detections using two          
alternative sliding window techniques plus a subsequent evaluation               
of the source detection significance and quality based on a maximum likelihood  
method (Voges et al. 1992, for aspects of the maximum likelihood method see     
also Cruddace et al. 1991). Only sources with a likelihood of 
detection larger than    
$L = 10$ were accepted into the RASS I source list. 
(The likelihood value here and throughout the paper is defined as
$L = -\ln P$, where P is the probability for a spurious source detection --
or a spurious extent detection in the case of the extent likelihood.)   
This threshold was
chosen such that an estimated fraction of less than 1\% spurious
sources enter the source catalogue. The detections       
are based on the source counts in the broad ROSAT PSPC energy      
band covering the detector channels 11 -- 240 which roughly corresponds to       
an energy range of 0.1 to 2.4 keV. Further qualities of the sources             
evaluated during the maximum likelihood assessment in three                     
energy bands comprise the source count rate, an estimated source extent         
in excess of the broadening of the sources due to the telescope-detector        
point spread function (PSF), and two hardness ratios based on the counts        
measured in the soft (channel 11 -- 40) and hard energy band                     
(channel 52 -- 201) or in the hard band 1 (channel 52 -- 91) and hard band 2      
(channel 92 -- 201), respectively. The source extent which is of special
importance here was determined within the maximum likelihood analysis
by assuming for the source image shape a convolution of two 
two-dimensional Gaussian functions for the PSF and the source shape
(Voges et al. 1999),
respectively. The result of this analysis is then a value for the excess
extent in terms of a $\sigma$-radius of the second Guassian.
In this approximation the Gaussian wings are less extended
than both the wings of the PSF and the wings of a King-type surface
brightness model (e.g. Cavaliere \& Fusco-Femiano 1976, Jones \& Forman
1984). This is one, probably minor, reason for the effect that
some of the X-ray flux in the outer X-ray halos is underestimated in 
the RASS standard analysis.

\begin{figure}                                                                  
\plotone{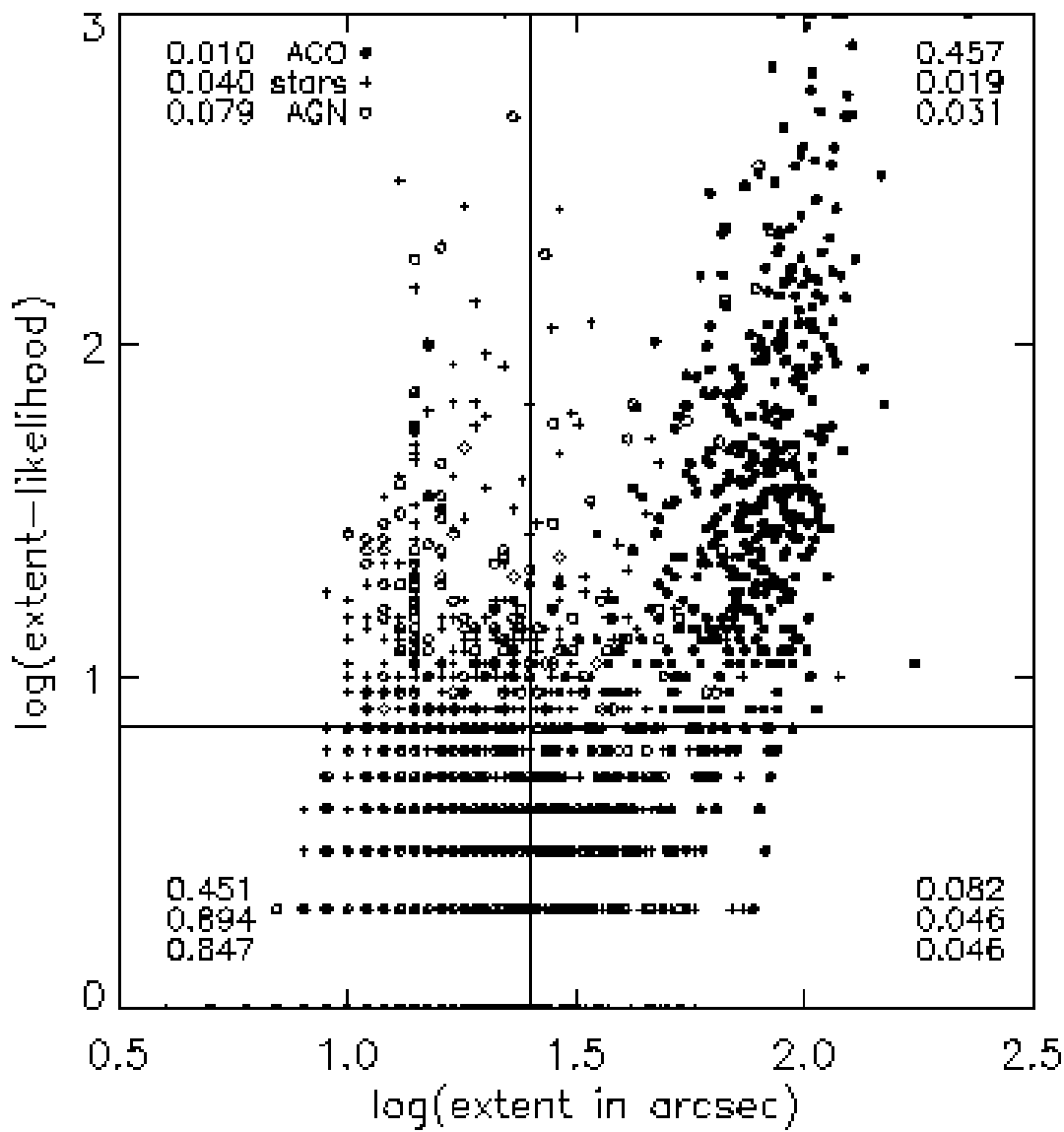}                                                              
\caption{A comparison of the distribution of the values for the extent and      
extent likelihood for a set of RASS sources identified with ACO clusters,
stars, and AGN. The data are taken from the ROSAT Bright Source catalogue
compiled from RASS II and the cross identifications with optical catalogues
as described by Voges et al. (1999). The coincidence radius used for
the cross-identification is 5 arcmin. The vertical and horizontal lines
give the extent limit of 25 arcsec and the extent likelihood limit of 7. The
fraction of sources for each category located in the different quadrants
is given in each quadrant. The upper number in each quadrant gives 
the fraction of all ACO clusters falling into this quadrant, the middle 
number gives the corresponding fraction for stars and the lower number
the fractions for AGN. The fractions for each class summed for all
quadrant adds up to unity, thus, for example, less than 2-3 \% of the 
non-cluster sources are found in the upper right quadrant.  
The selection criteria for the present sample
correspond to the upper right quadrant which is highly enriched in
ACO clusters.  
}                                                                               
\end{figure}

First tests of the source quality parameters in RASS I 
and ready identifications          
with existing source catalogues showed that many clusters of galaxies          
featured significantly extended X-ray emission in the RASS. It             
was known from previous X-ray surveys (e.g. the Einstein Medium Sensitivity      
Survey, e.g. Gioia et al. 1994, Stocke et al. 1994) that slightly more than     
about 10\% of the X-ray sources at the depth of the RASS should be              
galaxy clusters emitting in X-rays (see also B\"ohringer et al. 1991).          
Therefore it was clear that an X-ray sample highly enriched in             
galaxy clusters could be obtained by selecting those RASS sources               
featuring a significant source extent. For example, this is demonstrated         
by a comparison of the extent properties of the X-ray sources for               
ACO clusters, stars, and AGN taken from the ROSAT Bright Source catalogue
and the correlation with optical catalogues from Voges et al. (1999)
shown in Fig. 2 for RASS II data
(a comparable figure is also shown in Ebeling et al. 1996 for RASS I results).
Even though the present study is concerned with RASS I results we are
showing a statistic for RASS II in Fig. 2 because there is no principle 
difference and there is a larger data base of correlations with 
catalogued objects available for RASS II.  
About half              
of the galaxy clusters occupy an almost exclusive parameter space               
characterized by an excess source extent larger than 25 arcsec               
with a reasonably high extent likelihood
(with a value of $L=7$). Only about 2-3 \% of the 
non-cluster sources are found in this region. Since galaxy clusters account
for about $10 - 15\%$ of all X-ray sources we can expect a contamination 
of the order of $20 - 30\%$ by non-cluster sources if the sample is selected
from the upper right quadrant. This is approximately what is found
below.
That this small value 
of 25 arcsec for the excess
extent radius threshold shows a significant effect is somewhat surprising,
since the mean half power radius of the survey point spread function
is 70 arcsec, much larger than this threshold. This can
be explained by the fact that in the RASS analysis likelihoods
are calculated separately for each photon before they are summed and 
therefore each photon can be weighted by its own PSF according to the 
place in the detector where it was registered. In this way photons
registered in the central part of the detector with a half power radius
of the PSF of 15 - 20 arcsec give a high weight to the maximum
likelihood analysis. This makes the RASS source analysis
very sensitive to the recognition
of small deviations from the expected shape of point sources.
In the following analysis we will not make use of the information
on the detector positions of individual photons. This is a disadvantage
when compared to the standard RASS maximum likelihood analysis,
but other advantages more than compensate for this. 
                                                                                
A first search for X-ray selected clusters
was made with RASS I sources flagged as extended. The selection criteria 
were as follows: For the extent parameters a minimum       
threshold of 25  arcsec for the extent radius and a minimum value        
of  7 for the extent likelihood was chosen, as indicated by the dividing
lines in Fig. 2.       
Further a lower count rate limit of 0.06 cts s$^{-1}$ in the ROSAT broad        
band was set and the sky area was restricted to the        
region $\delta \ge 0\deg$ and $|b_{II}| \ge 20\deg$. The sources
at the count rate threshold are thus typically characterized by about
25 source photons. This leads
to a source fraction of 76\% galaxy clusters among the sources selected.
The advantage of this approach is that 
it yields a low fraction of contaminating sources, provides an effective
way to detect galaxy clusters, and involves
relatively simple selection criteria. The disadvantage
is that the selection by source extent is much more difficult to
quantify and to model than for example a purely flux-limited selection
technique. 
                                                                                
In total 537 sources matching the selection criteria were extracted from the    
RASS I data base. 40 of these sources have been found to be detections of       
secondary maxima or fragments of clusters which are already in the list due to  
a detection at the main maximum. The largest fraction of these fragment         
sources is located in the very extended, diffuse emission region of the Virgo   
cluster (B\"ohringer et al. 1994). The fragment sources were removed from    
the list after a careful check that they are not associated with another        
distinct X-ray source in the line-of-sight. We have also excluded from
the present catalogue the two detections in the Virgo cluster at the 
position of M87 and M86, because a useful flux measurement in the Virgo
region requires a more detailed approach.
Thus we report results for these parts of Virgo separately. 
In the following we will therefore discuss the analysis 
and identification of the remaining 495 sources.                
                                                                                
Early 1996 a new ROSAT Survey product, RASS II, became available at MPE.         
This version which is based on a greatly revised attitude solution              
for the pointing of the satellite during the survey and also uses a much        
more stringent quality threshold for the times with acceptable attitudes         
was used to create a new RASS II source list from which the RASS Bright          
Source Catalogue (RASS BSC) was created (Voges et al. 1999). 
In this survey product the data are sorted in 1378 sky fields with 
sufficient overlap ($\sim 0.23$ degrees) to guarantee an  
undiscriminating source assessment in the boundary regions. Each field now      
contains all the photons registered for this part of the sky during the entire  
survey. The resulting exposure distribution in the NORAS survey area is also    
shown in Fig. 1. As expected this distribution features a tail of high          
exposures up to about 40000 s. The reanalysis of the X-ray sources
of the present sample makes almost exclusively use of the RASS II data
base.                                                 
                                                                                
\section {Reanalysis of the X-ray source properties}                            
                         
To reanalyse the sources we apply a novel technique that is essentially
based on measuring background-corrected source counts as a function 
of a growing circular aperture
and checking for saturation to determine the observed source counts.
The growth curve of the counts as a function of aperture radius is 
also used subsequently to analyse further source properties. We therefore
term this method the growth curve analysis, GCA. We preferred to apply
this method over techniques applied in earlier studies. The Steepness
Ratio Technique used in De Grandi et al. (1997) has some similarity to
the present analysis, but makes only restricted use of the available
photon data in only extracting the source counts in aperture radii of
3 and 5 arcmin. A comparison shows that the uncertainties 
in the determined count
rates are usually higher for that technique than for GCA. 
Our preference of the GCA method 
over Voronoi-Tesselation
and Percolation (Ebeling et al. 1996, 1998) is due to
the fact that the GCA technique is simple to reproduce in models 
and simulations, 
the resulting count rates are quoted for a known
aperture radius for each source allowing a better assessment of the results
in subsequent modeling, and the GCA technique provides a set of 
very essential diagnostic plots which make the interactive evaluation 
of the reliability of the GCA results easy and transparent. In addition
the VTP technique needs two counteracting steps to correct for the unobserved
flux. The first step relies on an extrapolation based on the assumption
of spherically symmetric sources and leads to a significant overcorrection
which is then compensated in a second step with an {\it a posteriori}
recalibration based on a comparison with pointed data.
The present method achieves a good agreement with pointed data in 
one relatively minor {\it ab initio} correction in a first step 
as shown below.
The presentation of more details and tests of the GCA method is 
planned for a future
publication (B\"ohringer et al. in preparation), while the essential 
features of this method are described in the following.    
                                                       
The reanalysis of the X-ray properties was conducted for all 495 X-ray sources  
in the sample using RASS II data in fields of $1.5\deg \times 1.5\deg$ 
centered on    
each X-ray source. For 17 nearby clusters featuring a large extent the
analysis is performed in larger fields of $4\deg \times 4\deg$ 
or $8\deg \times 8\deg$. For 7 sources, where the exposure in RASS II is less
than 70 sec, data fields were extracted from RASS I which
features a higher exposure for these cases. The reason for the reduced
exposure in RASS II is the tight quality constraint which leads to the
rejection of some RASS photon data in RASS II as compared to RASS I.
The 7 fields of RASS I used here were carefully checked and did not show
any peculiarities as e.g. double or otherwise distorted images of
bright sources which would indicate a problem with the attitude control
during the observation.
                                                                              
The primary data set used for each field consists of a photon event file        
containing all data of the photons                                              
registered in the field area and the corresponding exposure map, providing      
the exposure time as a function of sky position with a resolution of            
45 arcsec pixels. The exposure maps include a broad band               
correction for vignetting and the effect of the shadowing of the support        
structure of the PSPC window. (The possible difference between the broad
band vignetting correction and the proper correction for the specific
source spectrum introduces an error no larger than 2\% and no further
correction is applied).
The photon event files provide information        
on the sky position and energy channel as well as the time and the detector     
position for each registered photon. In the following analysis only the         
first two parameters are used.                                                  
                                                                                
The three energy bands defined for our data reduction procedure are identical 
to those used in the RASS analysis (Voges et al. 1999):    
broad band, soft band, and hard band. 
For the source count rate determination and the shape characterization we use   
exclusively the photon counts in the hard energy band. In this band the         
soft X-ray background is reduced to about one fourth, while -- depending 
on the value of the interstellar column density -- 
60 to 100\% of the cluster emission   
is detected. Therefore the analysis in this energy band provides the highest    
signal-to-noise ratio and the most reliable count rates. Another quite          
important aspect of the choice of this energy band is that it    
minimizes the contribution of contaminating sources to the count rate.          
Since the majority of all sources in the RASS are softer than the cluster       
X-ray sources, their contribution to the hard band is usually                   
less significant than the contribution to the broad band counts.                
                          
\subsection{Source position and count rate}                                     
                                                                                
Prior to the evaluation of the source count rate the                 
source center position and the sky background brightness is determined.         
The input field centers chosen are the X-ray positions               
provided by the maximum likelihood technique of RASS II. 
These positions are not optimal            
for extended sources and in particular extended sources are sometimes 
multiply detected in the RASS. Therefore    
a redetermination of the source position is performed based            
on a moment method which determines the two-dimensional
``center of mass'' of the photon distribution within an aperture of 3 arcmin
around the input value for the center. This procedure is iterated
with the newly found center position until the process converges
to a stable center position. The small aperture of 3 arcmin 
gives a high weight to local maxima. We check all the center positions
interactively and correct those cases in which this method 
has settled on a secondary maximum or where the local maximum
is obviously offset from the large scale symmetry of the cluster.
Those 5 sources are marked in the catalogue. We have also applied the
moment method for the determination of the center position using
larger apertures (5 and 7.5 arcmin). In 7 cases we preferred
to quote these centers in the catalogue. Also these cases are marked.

\begin{figure}[h]                                                                  
\plottwo{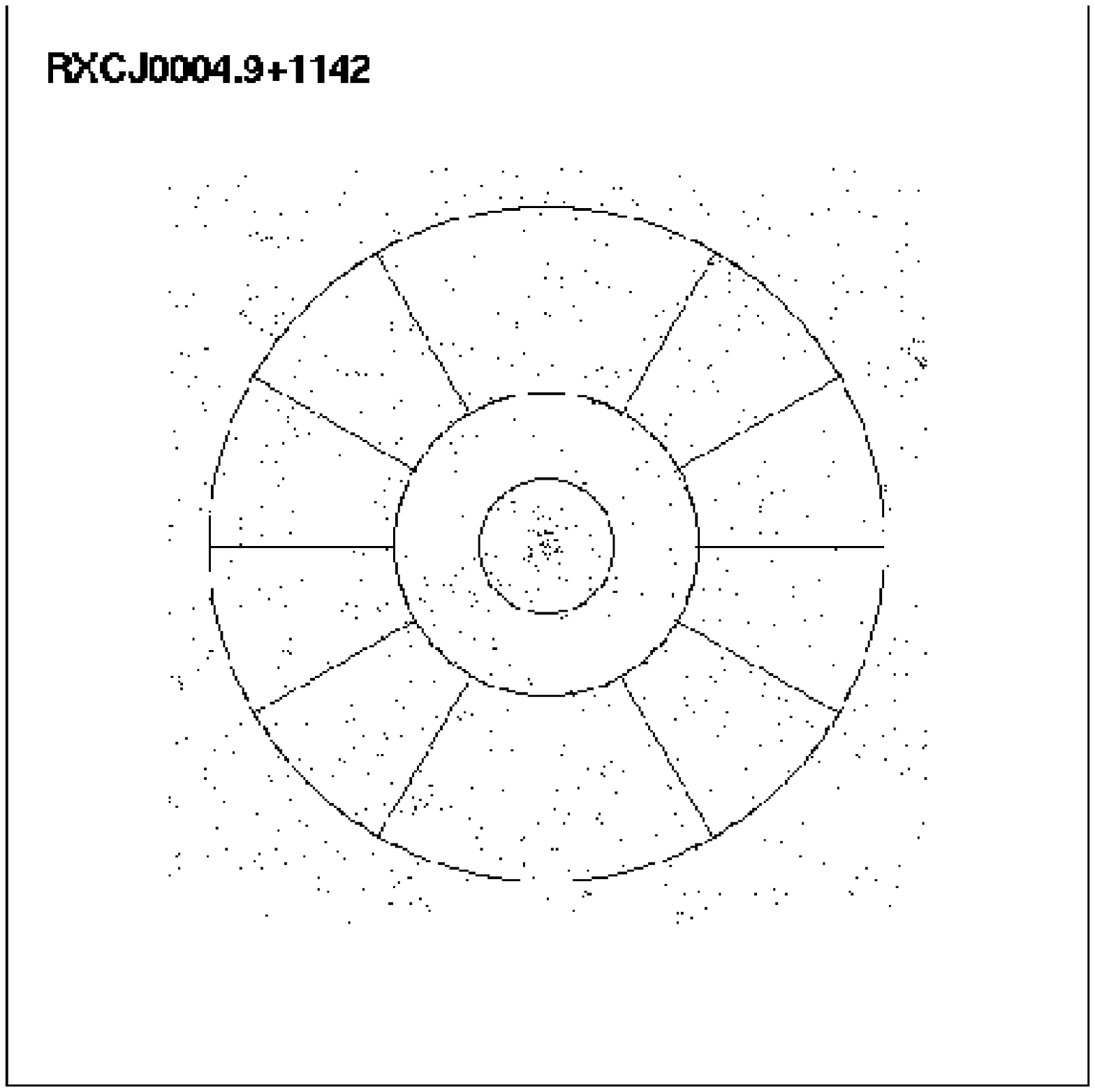}{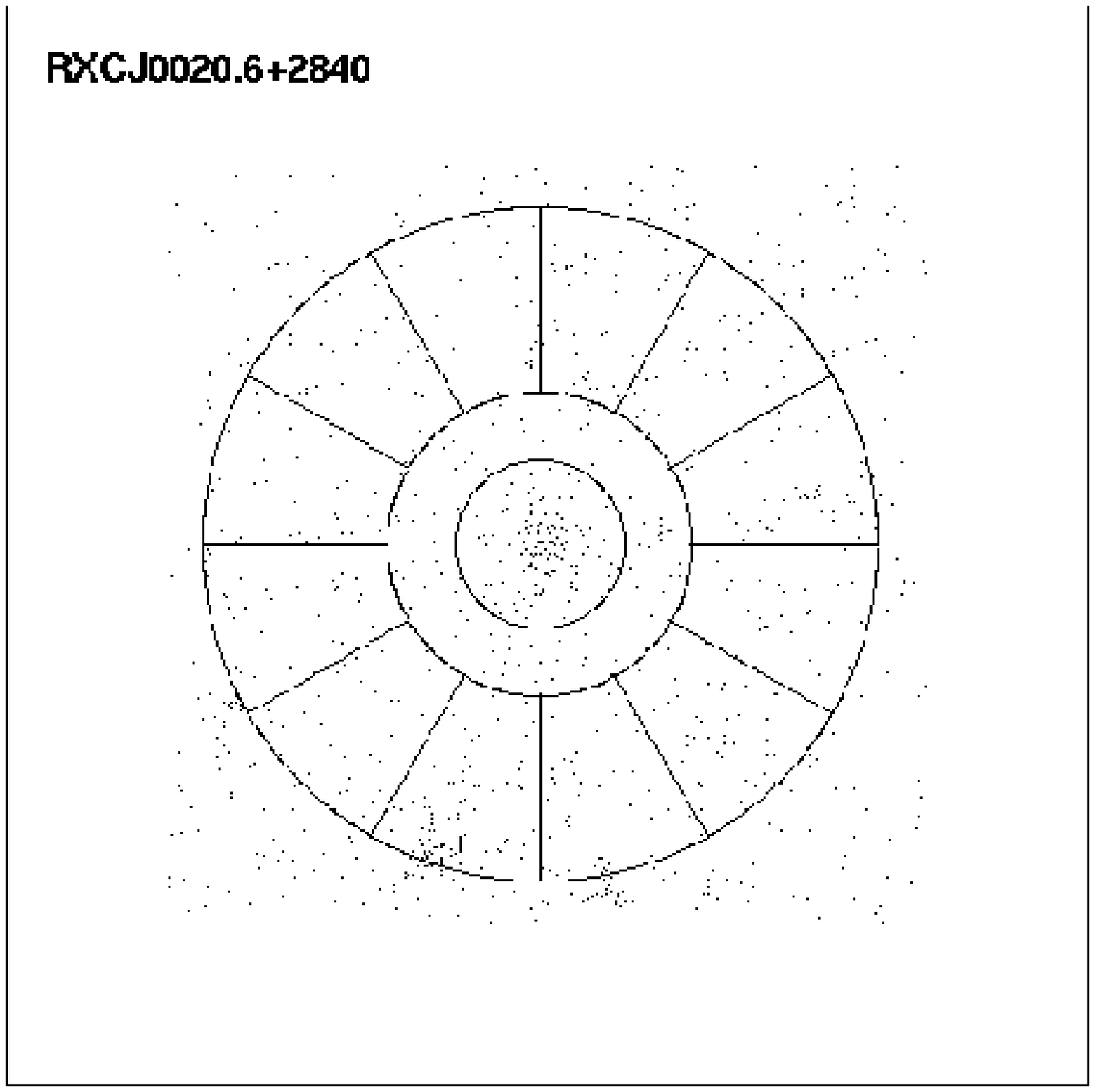}                                    
\caption{Distribution of the photons detected by the ROSAT PSPC                 
in the hard band ($\sim $0.5-2.0 keV) in the regions of the X-ray                
cluster sources RXCJ0004.9+1142 and RXCJ0020.6+2840.                            
The ring segments of the area                                                   
used for the background estimate and the source region as determined             
from the analysis are marked. The photon field extracted from the ROSAT           
All-Sky Survey as shown here covers an area of 90 by 90 square arcmin.
The ring segment of the source field of RXCJ0020.6+2840 marked by a 
cross is excluded from the background determination due to
contamination.}               
\end{figure}

\begin{figure}[h]                                                   
\plottwo{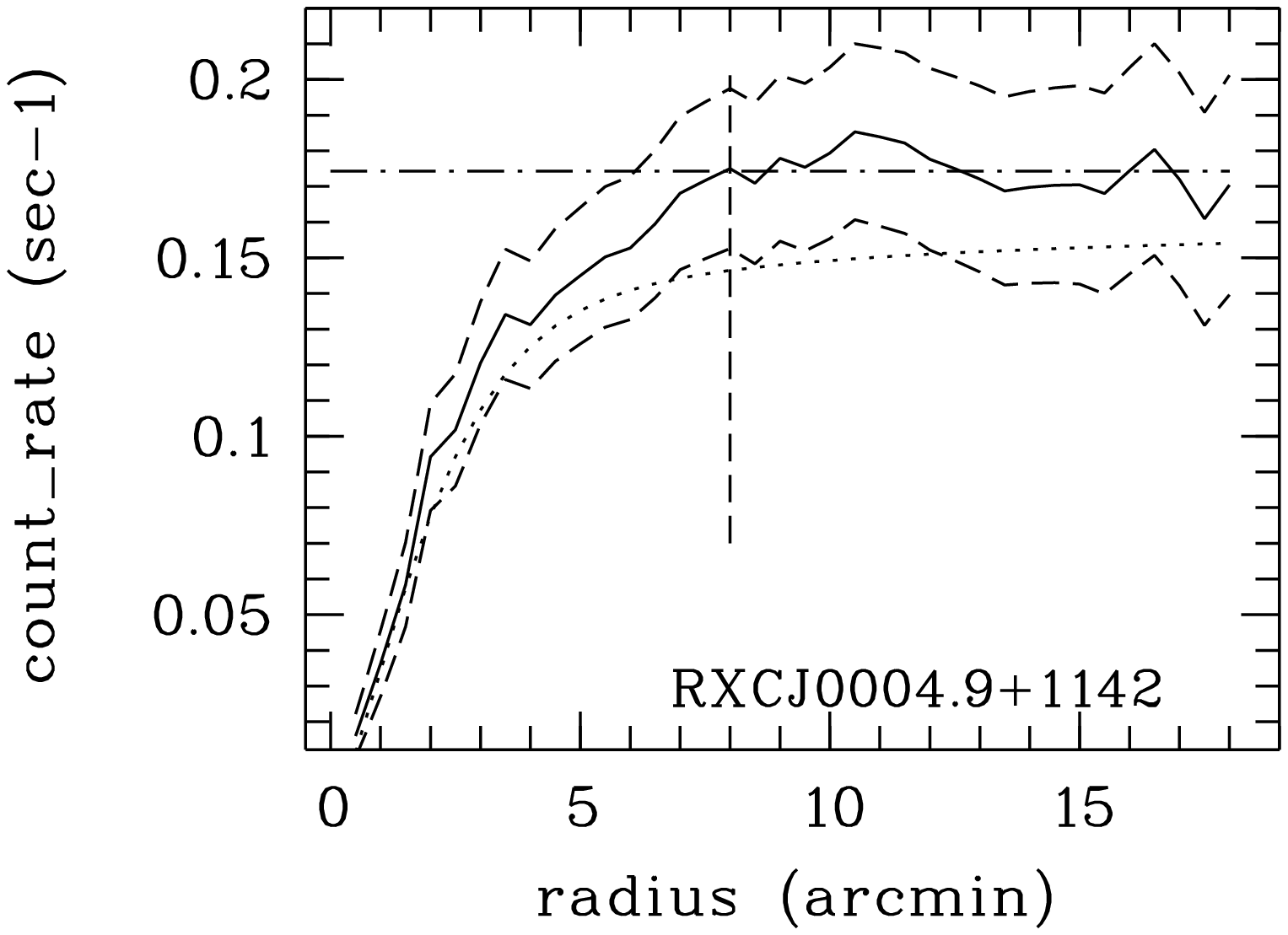}{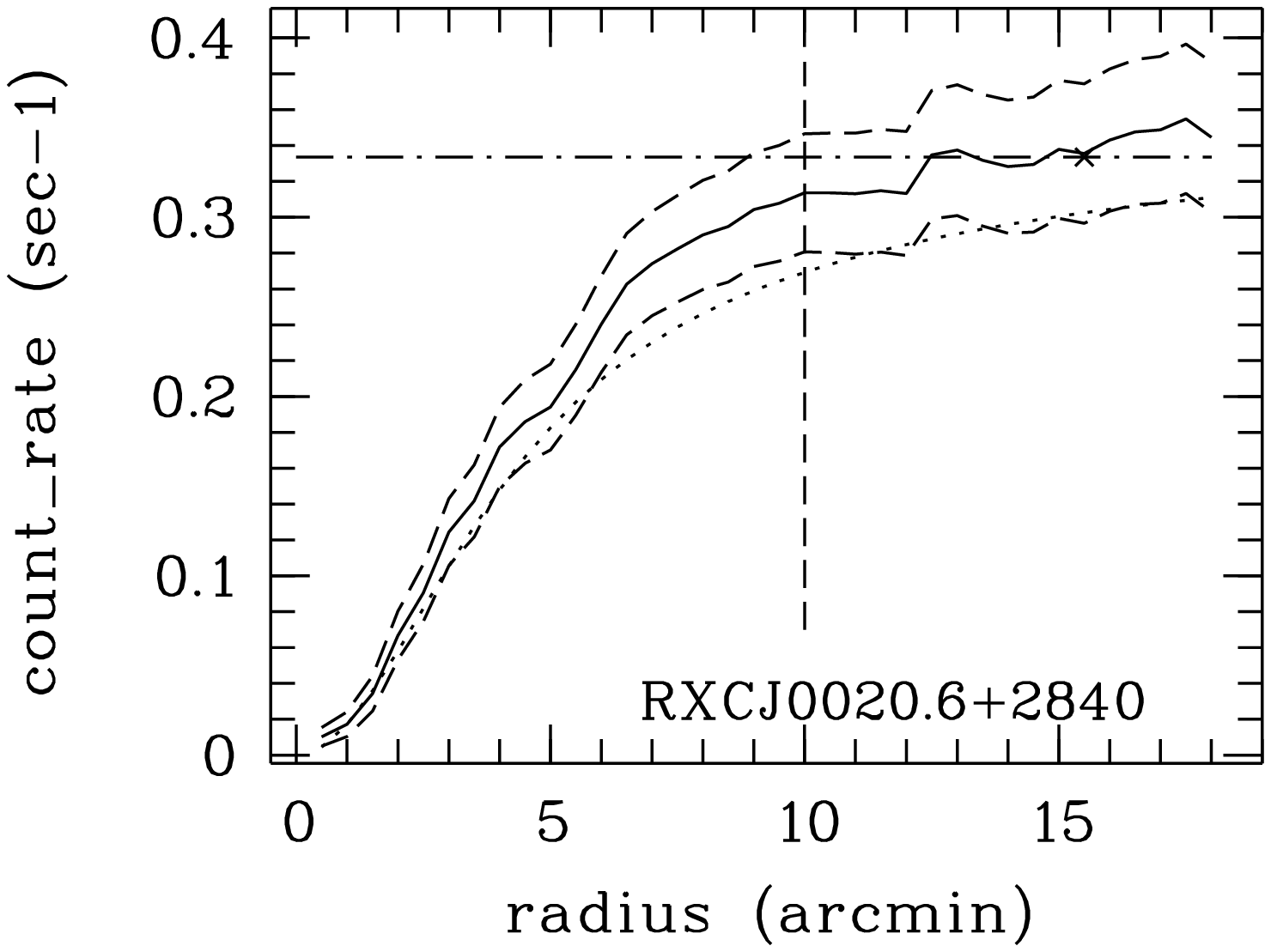}                                    
\caption{Radially integrated count rate for the cluster sources                  
RXCJ0004.9+1142 and RXCJ0020.6+2840 in the hard band                            
($\sim $0.5-2.0 keV).                                                           
The two techniques used to determine the total observed source    
count rate are shown: the vertical dashed line shows the                         
limiting radius outside which the $1\sigma$ uncertainty of                  
the signal increases faster than the signal itself; the                         
horizontal line shows the fit to the saturation plateau
(as explained in the text below).                        
Also shown as dotted line is the best fit of a $\beta$-model                    
profile convolved with the PSF with a core radius of 
0.5 arcmin and a normalization of                 
0.158 cts s$^{-1}$ for RXCJ0004.9+1142 and 2.5 arcmin and                       
0.360 cts s$^{-1}$ for RXCJ0020.6+2840, respectively. Note
that the actual fits were performed to the differential count
rate profiles and not to the cumulative ones as shown
here. Thus the deviations of the fits at large radii
have a low significance. They are often found and may
indicate irregular structure at the outskirts of many 
clusters.}                        
\end{figure}

The background of the field is then determined from a ring area centered on     
the source with an inner and outer ring radius of 20 arcmin and 41.3 arcmin,    
respectively, as shown in Fig. 3. 
The inner ring size has been chosen such that it is
outside the outer radius of the X-ray emission for the majority of the
clusters. The outer radius is chosen to make almost full use of the
$1.5 \times 1.5$ degree fields. This large background area ensures
that the number of photons used for the determination of the background
surface brightness is large and introduces an almost negligible
Poissonian error into the source flux determination.
There are 17 nearby clusters in the sample which exceed the
inner background ring radius in size. For these clusters larger
fields have been extracted from RASS II and larger background rings
have been used in the analysis. 
   
The ring is subdivided in 12 sectors. The photons in each         
of the three energy bands in all sectors are counted and the exposure time for  
each photon position is obtained from the exposure map. The count rate in each  
sector and the surface brightness are then calculated by averaging in
count rate:     
                                                                                
\begin{equation}                                                                
C = \sum _i {1 \over t_i} ,                                                        
\end{equation}                                                                  
                           
\noindent                                                     
where $C$ is the count rate, $t_i$ is the exposure time at each               
photon position, and the summation is over all photon events in the sector.
The surface brightness is obtained by division with the sector area.      
An uncertainty for the surface brightness in each sector is calculated          
from Poisson statistics.                                                        
                                                                                
To avoid that discrete sources located in the background ring are included     
in the background estimate, the median of the sector count rates is determined  
and sectors featuring a larger than $2.3\sigma$ deviation from the              
median are discarded from the further calculations. Even though sources         
are only expected to cause large enhancements, we also exclude                  
sectors that have count rates which are too low by more than $2.3\sigma $,      
since some of the positive deviations are due to fluctuations,                   
and to avoid a negative bias in the background the negative fluctuations        
have to be discarded for reasons of symmetry.                                   
The clipping threshold of  2.3 $\sigma$ guarantees a successful      
removal of sources with typical count rates above about 0.04 s$^{-1}$
which would otherwise introduce a typical error $\ge 1\%$ in the
background determination. The chosen clipping threshold
leads in general to the clipping of not more than 1 - 3 sectors     
which preserves most of the background area for averaging resulting in          
a small photon statistical error in the background of typically about
5\%. This also shows that the variations in the background on that scale
are generally small and hardly larger than what is expected from 
Poisson statistics.           
                                                                                
The procedure of the background determination is illustrated in Fig. 3.
For the data in Fig. 3a there is no interference of background sources and      
the background is smooth enough that all sectors were included in the background  
measurement, while for the data in Fig. 3b one of the sectors had to be         
discarded due to the presence of a significantly disturbing source.            
The two figures show the photon distribution of the hard band counts for each   
source field. 
The two sources, which have net source counts of 75.2 and 127.9,  
respectively, will be used in the following to illustrate the further analysis. 
(These two sources selected as 
the first sources in the list illustrating the features we
like to show: RXCJ0004.9+1142 is the first source in the list with
less than 100 photons which features a small but significant extent
and RXCJ0020.6+2840 is the first source showing the search for the
plateau in the count rate in the presence of steps in the plateau
region, see below).              
                                                                  
The cumulative source count rate as a function of radius starting               
from the earlier determined                                                     
central position is then found by integrating the source counts                 
in concentric rings outwards while subtracting the background contribution.
The integration is         
performed using a ring-width of 0.5 arcmin (a reasonable resolution for the      
given PSF of the instrument). The source count rate is determined for each      
ring by weighting each photon with the local exposure time according to eq.(1). 
The integration is performed for the three selected energy  
bands. The results are count rate profiles as shown in Figs. 4a and 4b.  
The uncertainty corridors in
these count rate profiles resulting from Poisson photon statistics       
are indicated as dashed lines in the figures. They also include the Poisson  
error of the background determination.                                          
                                                                                
In most of these cumulative profiles the count rate levels               
off to a plateau value which gives the total observed source                    
count rate. 
The total observed count rate is determined in the automated             
source characterization program in two alternative ways. 
In the first approach  
we determine the radius outside of which the source signal increases less       
than the $1\sigma$ uncertainty in the count rate. This radius, which we call    
the outer radius of significant X-ray emission, $R_x$, is indicated             
by the vertical dashed line in Figs. 4a and 4b. The count rate with its 
statistical uncertainty at this radius provides the value of the  
significantly detected count rate of the source. The radius $R_x$    
is shown for the two sources as the inner circle in Figs. 3a and 3b . 

Alternatively the                                                               
total source count rate is measured by getting an estimate of the plateau       
level. In the simplest case it is the average of the flat plateau outside       
$R_x$. In practice we determine the mean value of the plateau as well as the    
slope by means of a linear regression method for the profile part              
outside $R_x$. If the slope of the plateau is less than $0.8$\% of the total    
count rate per arcmin radius, the plateau value is accepted. If the plateau is  
decreasing the count rate is determined from the mean of three bins around      
$R_x$. If the plateau is increasing, another effort is made to find the best    
flat part of the plateau by iteratively excluding the outermost and in a       
second step also some of the innermost bins. This procedure helps in excluding  
an outer rise of the count rate profile due to a neighboring source or by      
skipping a few bins if the count rate curve has not completely saturated to a   
plateau at $R_x$. Both effects can be seen in Fig. 4b, where the outer       
radius of the considered plateau region before a secondary rise of the 
profile is indicated.     
The source analysis is checked in each case on the basis of 
diagnostic plots as those shown in Figs. 3 - 6. 
About 35\% of the plateaus can be characterized by the first step
and about 83\% by the further iterative trials.
There is a residual fraction of about 17\% of the sources for which no 
satisfactory plateau can be established automatically often as a result 
of contaminating nearby sources. These cases are analysed individually. 
For all sources we determine the              
radius, $R_{out}$, out to which the plateau count rate was measured by 
simply following   
the profile until the plateau value is reached.                                
                                                                                
For the count rates and their uncertainties quoted in the present catalogue     
we have adopted the following approach. For the count rates we take the         
results from the fitted plateau values, which are in general higher by a few
percent than the count rates determined at $R_x$. This is mainly due to an 
insignificant further rise of the cumulative count rate 
profile beyond $R_x$. This rise is in most cases much
smaller than the statistical error in the count rate.  
The error in the count rate is then   
calculated from the root mean square of the shot noise error and the deviation  
of the plateau value from the value at $R_x$. Taking the root mean square       
would be justified for errors which are statistically independent and Gaussian  
distributed. Since the second error is a systematic deviation, this does not    
apply in this case. Nevertheless this is for the present case 
a practical approach which integrates the two errors by putting 
a larger weight on the larger one of the two uncertainties.
The correction of the measured count rate of the source to a total 
count rate is discussed below.

\subsection{Spectral hardness ratio and source extent}                          
                                                                                
For the determination of the spectral 
hardness ratio the count rate in the soft band is also       
determined for the same radius as for the hard band value of $R_x$.     
The hardness ratio, HR, used here as well as                  
in the RASS data base is defined as                                             
                                                                                
\begin{equation}                                                                
HR = {H - S \over H + S} ,                                                        
\end{equation}                                                                  
                                                                                
where $H$ is the hard band and $S$ the soft band source count rate. The         
expected values for the hardness ratio for cluster sources is roughly in the    
range $0 - 1$ as shown in Fig. 8.                                              
                                                                                
The source extent is addressed in two ways: quantifying the source size
and testing the probability that the extent is real, respectively. 
In the first analysis a King profile   
with the surface brightness distribution,                                 
                                                                                
\begin{equation}                                                                
S_x(R) =  S_0  \left(1 + {R^2 \over r_c^2} \right)^{-1.5\beta +0.5} ,             
\end{equation}                                                                  
                                           
(where $R$ is the projected radial distance from the soure center and $r_c$
is the core radius of the X-ray surface brightness distribution)   
convolved with the averaged survey PSF (as calculated by G. Hasinger from the   
ROSAT XRT/PSPC PSF averaged over the detector area with a correction for the    
vignetting effect) and azimuthally integrated is fitted to the differential   
count rate profile. For the profile parameters a fixed value  
$\beta = 2/3$ is taken according to the most typical  value found
in X-ray cluster observations (e.g. Jones \& Forman 1984). In the $\chi^2$
fit the core radius is varied in steps of 0.5 arcmin, and the             
normalization is a free fitting parameter. Due to the very low count            
statistics we have thus limited the fitting parameters per step to one,
the normalization,  
by choosing the          
most common value for $\beta$. The results for the two example sources are      
shown in Figs. 5a and 5b, and the best fitting profiles are indicated by          
dotted lines in Figs. 4a and 4b. Note that, while the figures show fits to the    
cumulative profiles for a better diagnosis of the results, the actual            
calculations are conducted for the differential profiles to assure statistical  
independence of the count rates in rings as required by the $\chi^2$              
fitting method. Together with the best fitting value we also keep the           
minimum radius which is still consistent within the $2\sigma$ uncertainty       
limit. Comparing the $\chi^2$ values to the 1, 2, and 3$\sigma$ limits
shown in Fig. 5 we find that for the two examples the first source is only 
marginally extended ($2\sigma$ result), while the second source 
features a clear and large extent.

\begin{figure}                                                                  
\plottwo{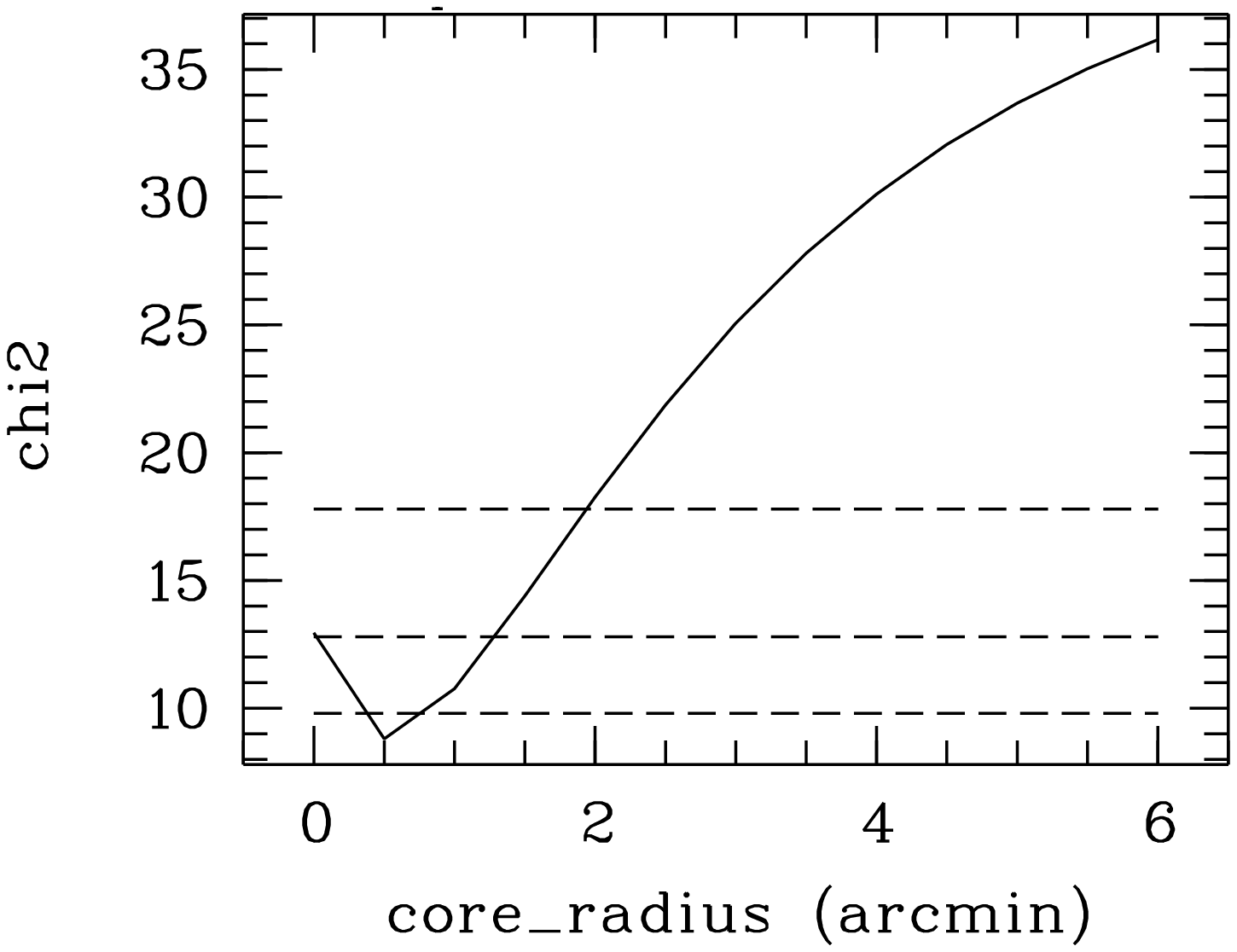}{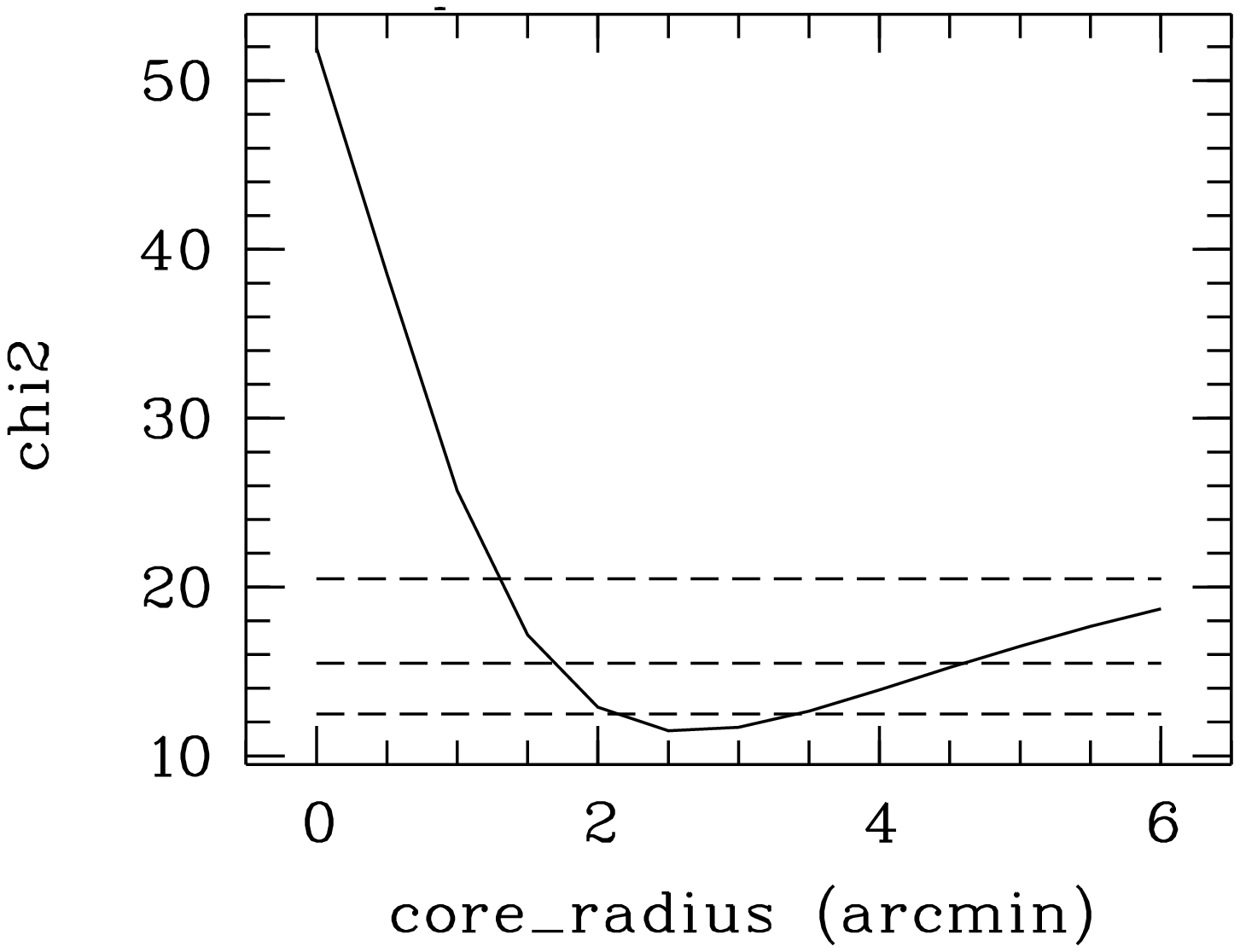}                                    
\caption{Evaluation of the source extent by means of a $\chi^2$ fit of a King   
profile convolved with the mean RASS PSF. The $\chi^2$ value is determined for  
varying core radii in steps of 0.5 arcmin. Here the best fitting core radii     
are found to be 0.5 and 2.5 arcmin. Only the second source is found by this     
test to be clearly extended with high significance. The horizontal dashed lines 
show the 1, 2, and $3\sigma$ uncertainty limits, respectively.}
\end{figure}                                                                    
                                                                                
The second, more sensitive method is used as a test for the            
probability that the source has an extent at all. For this we use a             
Kolmogorov-Smirnov test comparing the expected cumulative count rate 
profile of a point source including background with the given instrument PSF 
with the radially sorted, cumulative, and unbinned photon         
counts out to a radius of 6 arcmin. This radius is larger than the 90\%
power radius of the survey PSF and thus provides enough leverage to
display the deviations of extended sources, but is small enough to minimize
the influence of possible background errors. This test does not depend
on the assumption that the uncertainties are Gaussian distributed.
Since the pure photon counts also contain   
the background counts the previously determined background surface brightness   
has also to be added to the expected point source profile. Examples of the     
expected point source profile and measured curves are shown 
in Figs. 6a and 6b. The background      
contribution to the expected profile is also indicated and we can see that it   
is generally a minor contribution at these small source radii. Tests with       
known point sources have shown that the misclassification 
of point sources as extended sources is generally less     
than about 5\% if we take an upper limit for the KS probability 
of 0.01 to classify a source as extended (see B\"ohringer 1999, in preparation).
For the two examples we find  
probability values of $0.005$ and $1.3 \times 10^{-14}$, respectively,           
and therefore both sources will be classified as extended in the catalogue.
In the following we will use these results in the form of the extent parameter
defined as $P_{ext} = -\log _{10}$(KS probability).

\begin{figure}                                                                  
\plottwo{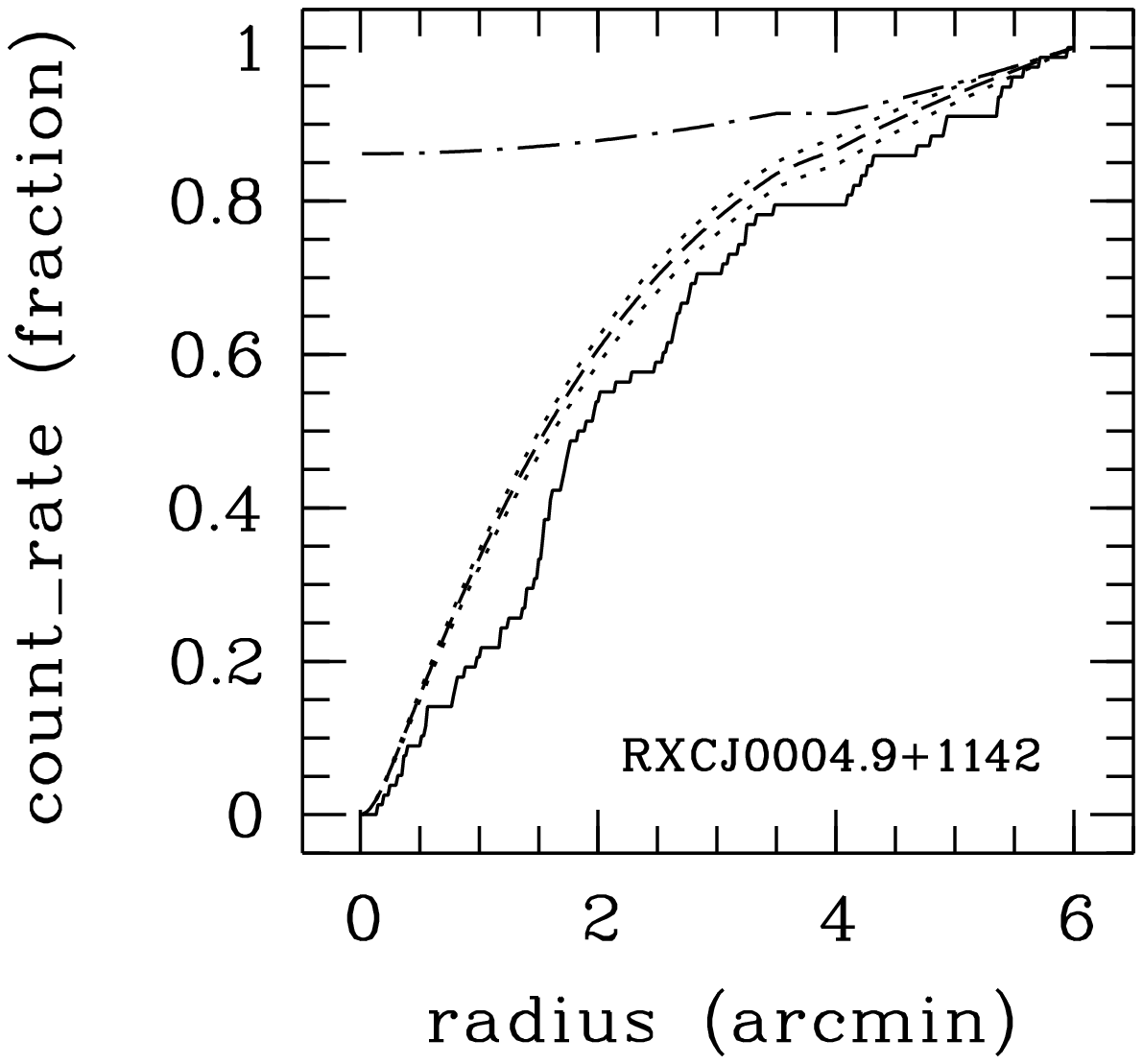}{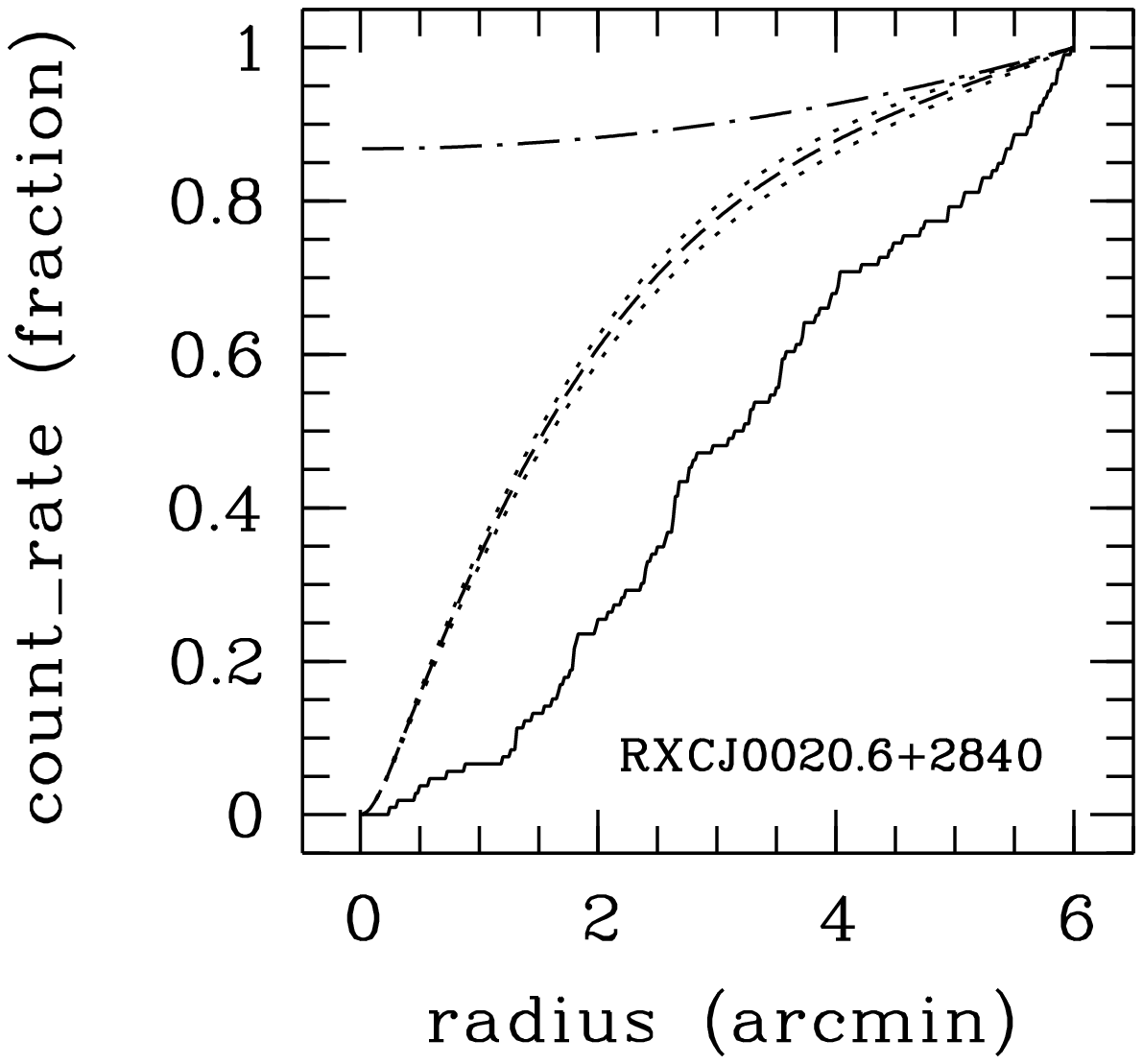}                                    
\caption{Test for the probability of a source extent by a KS test comparing  
the cumulative count rate profile out to a radius of 6 arcmin with 
the expected profile for a point source plus background. 
Both profiles are normalized to 1 at the outer radius. The long dashed curve
gives the expected point source profile plus background and the dotted curves
show the effect of a 30\% background variation. The dotted-dashed curve
indicates the background contribution.}                                    
\end{figure}                                                                    
                                                                                
\subsection{Deblending and analysis of very extended sources}                   
                                                                                
The visual inspection of the diagnostic plots of the GCA results for all    
the sources showed that in 19 cases the source analysis 
suffered from the       
blending of the cluster source with another nearby source probably not          
associated with the intracluster X-ray emission. These sources were scheduled   
for another analysis including a deblending technique. The correction
by deblending was performed for all the sources where the contamination
was clearly recognized as due to point sources. Tests show that a single
contaminating source is usually easily recognized if its contribution
to the total hard band count rate is larger than 5 - 10\% and if the
source is outside the central 3 arcmin radius. The deblending 
is not applied to irregular clusters or clusters with substructure
where the non-symmetric emission region is most probably
part of the diffuse intracluster X-ray emission (and not likely
to be due to a point source).  
In this second analysis   
the source region is divided in two sets of twelve sectors for the radial       
region 3 to 8 and 8 to 15 arcmin. The variation of the surface brightness in    
the different ring sectors is analyzed in a similar way as done for the         
background ring sectors. Contaminating sources are best flagged 
by selecting those sectors with a more than $3.5\sigma$ deviation 
from the median. For weaker sources this detection threshold corresponds
roughly to sources with fluxes larger than 
$3 \times 10^{-13}$ erg s$^{-1}$ cm$^{-2}$ and in general guarantees
the deblending of sources with contributions larger than 10\%. 
The clipping technique is illustrated in 
Figs. 7a and 7b. In the further analysis the        
marked sectors are interpolated, that is, they are assigned a value for 
the surface brightness equal to the mean of the remaining sectors. 
The automated clipping works well for most sources but in about 30\%
of the cases the clipping either did not remove the contaminating source
completely or removed in addition other parts of the cluster. For these     
cases we preferred to determine the count rate of the contaminating    
source directly (with the analysis centered directly on the contaminating
source) with the same deblending algorithm and subtracted it 
from the blended source count rate. The 19   
sources that required a deblending are marked in the catalogue.            

Since the outer radius of the region in which the source profiles are analyzed  
is fixed in the automated source analysis routine, and since the cut out      
regions per source are restricted to a size of $1.5\deg \times 1.5\deg$ , 
some nearby clusters are too extended to be covered 
completely by the analysis. For these clusters, 17 in total, 
larger files of photon data from RASS II were requested with fields covering 
$4\deg \times 4\deg $ or even $8\deg \times 8\deg$ around the source. To these  
data the same source analysis was applied as described above with an extended   
radial range. The clusters that required a reanalysis in a larger sky field     
are also marked in the catalogue and can also be recognized by their large      
values of $R_x$ and $R_{out}$.                           
                                                                                
\begin{figure}                                                                  
\plottwo{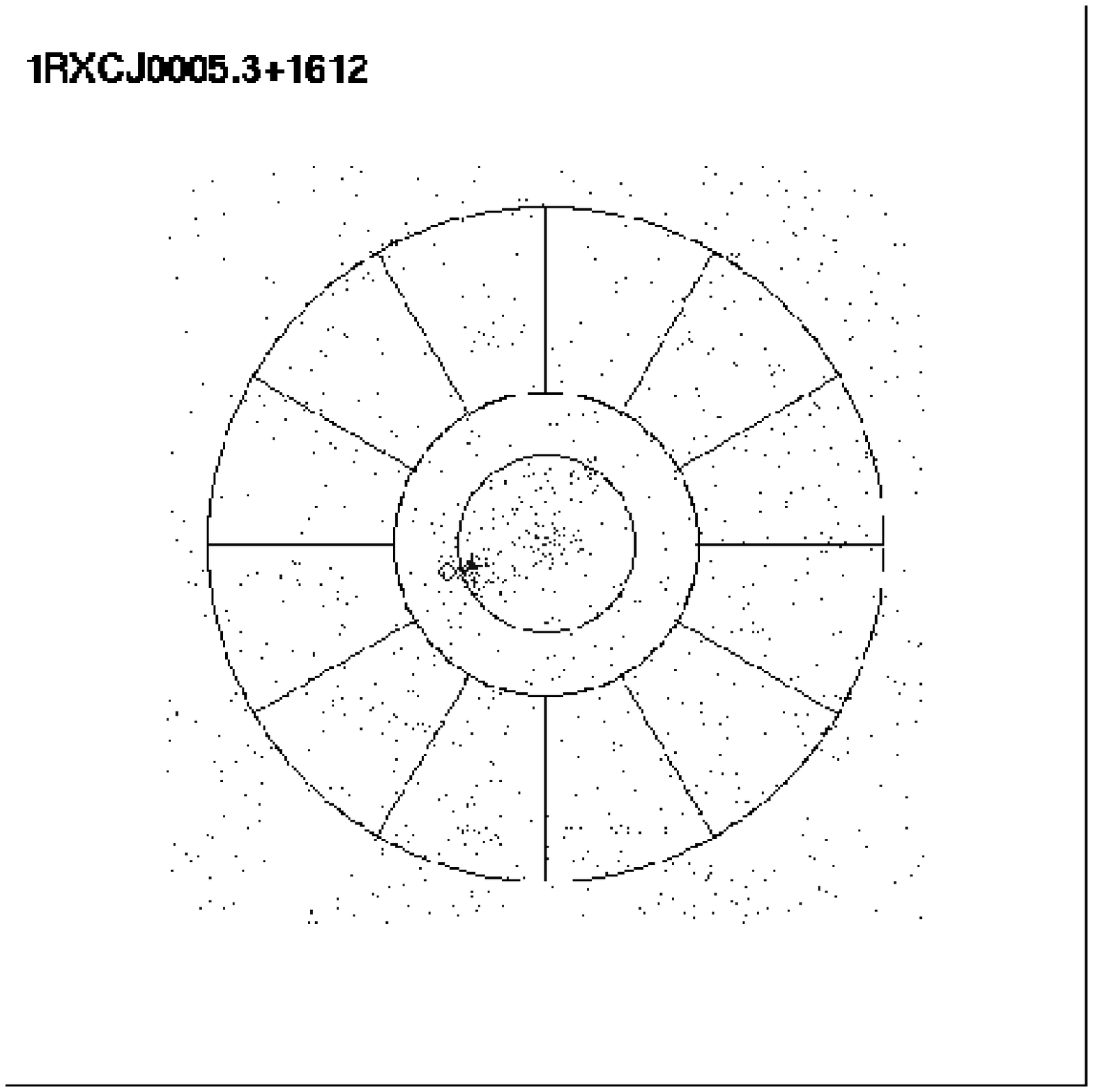}{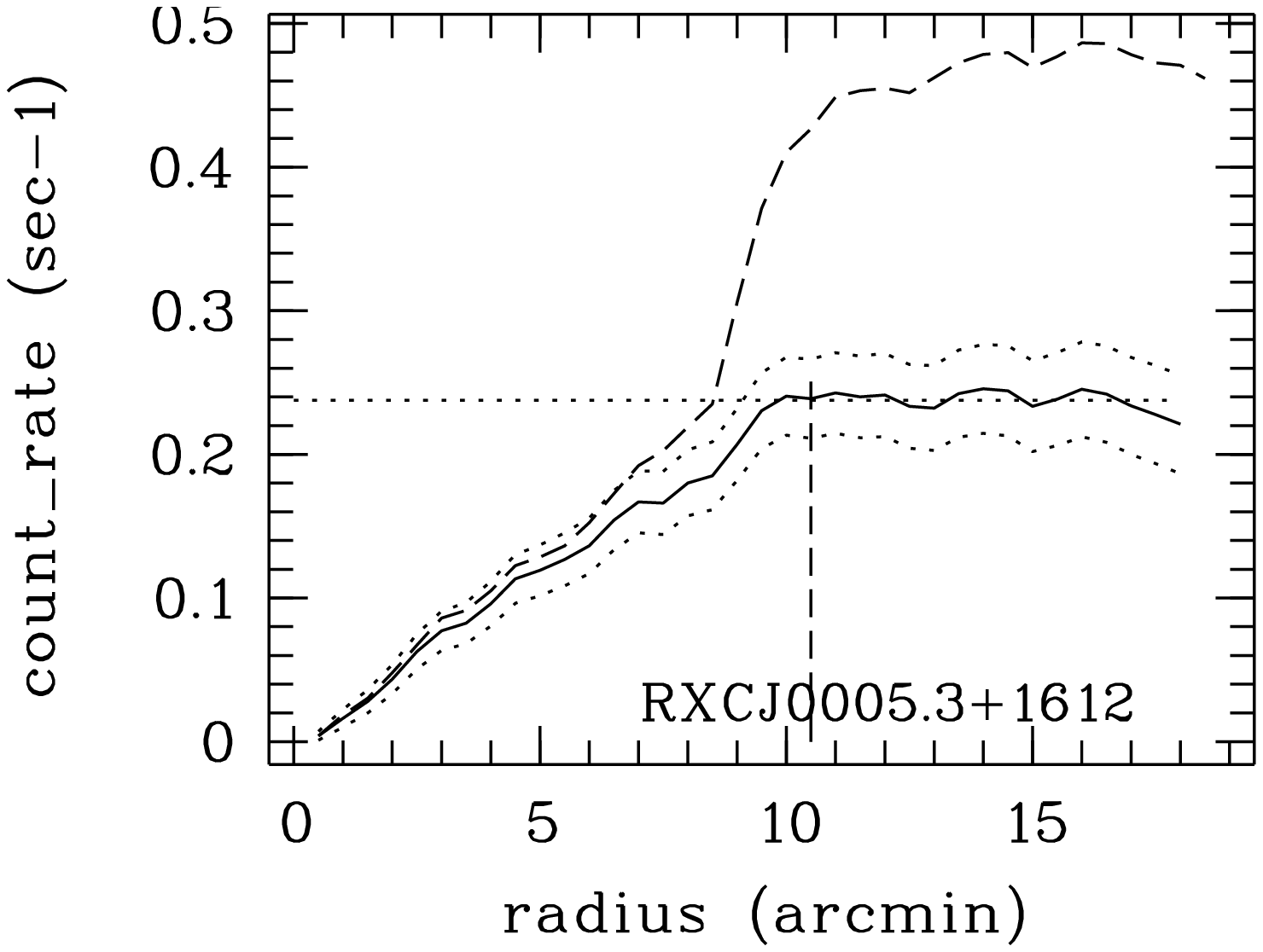}
\caption{The left figure shows the photon distribution of a cluster
X-ray source which is clearly contaminated by a point source. The small circle
near the point source marks the sector which was excised from the
source analysis to deblend the contaminating source. The right figure 
shows the result for the count rate analysis. The upper dashed curve gives
the cumulative source count rate without deblending. The solid curve shows
the result after the sector with the source was cut out and interpolated.
The two dotted curves show again the $1\sigma$ Poisson error for the count rate
including the effect of the interpolation. The horizontal 
short dashed curve indicates the plateau value determined and the
vertical dashed curve shows the outer radius of significant X-ray
emission from the source.}                              
\end{figure}

\subsection{Flux and luminosity determination}                                  
                                                                                
To determine the flux and the luminosity of a source we first obtain the value  
of the interstellar hydrogen column density as measured at 21cm (Dickey \&      
Lockman 1990, Stark et al. 1992) for the direction of the source                
by means of the EXSAS software                                                  
package routine (Zimmermann et al. 1994). The flux is then determined in a      
first step by calculating the conversion factor from count rate to flux for a   
source with a thermal spectrum and a temperature of 5 keV (based on a 
modern version of the radiation code by Raymond \& Smith 1977), a metal abundance  
of 0.3 of the solar value (Anders \& Grevesse 1989), a redshift of zero,
and an interstellar absorption according to the measured 21 cm value 
(Dickey \& Lockman 1990) and the absorption tables of Morrison \& McCammon 
(1983). We calculate the flux in the nominal ROSAT       
energy band: 0.1 - 2.4 keV. All fluxes and luminosities quoted in this
paper and in the catalogue refer to this energy band. 
Fig. 8a shows the conversion factors as a function   
of absorbing column density for the three plasma temperatures 2.5, 5 and 8 keV.    
We note that the main dependence is on the column density. The variation with   
temperature makes a difference of less than $7\%$ in the temperature       
range (2 - 10 keV). Only for temperatures below 1.5 keV larger  
corrections occur,       
which applies for the smallest groups of galaxies in the sample. The            
dependence on the metal abundances is even less, about 1\%. Prior to any   
further knowledge about the redshift and the nature of the cluster source the   
estimated flux value represents a good first approximation.               
Note, however, that for non-cluster sources a different spectral shape and      
thus a different conversion factor is expected.                          
                                                                                
Once the redshift of the cluster source is known we can determine its           
luminosity. 
The luminosity is also calculated for the ROSAT energy band.                   
The calculation is performed iteratively. In a first step we calculate   
a trial luminosity from the estimated flux value and the redshift 
and use its value to estimate a cluster plasma temperature           
using the X-ray luminosity-temperature relation of Markevitch (1998)    
                                                                                
\begin{equation}                                                                
T_x = 2.34~ L_{44}^{1/2} \times h_{50}~~~~~ ,            
\end{equation}                                                                  
                                                                                
where $T_x$ is in keV, $L_{44}$ is the X-ray luminosity in 
units of $10^{44}$ erg s$^{-1}$ (in the 0.1 - 2.4 keV band). We make use
of the relation which was derived by Markevitch without any correction
for cooling flows in $L_{44}$ and $T_x$. This applies to our case
since we are only dealing with integral count rates and average spectral
properties. (Note that we have approximated the exponent of 1/2.02 found
by Markevitch by 1/2).
The temperature estimate allows the calculation of a new count rate-flux        
conversion factor for which we now also take the redshift of the source         
spectrum into account and          
calculate the source rest frame value for the X-ray luminosity
(which involves the equivalent to the cosmic ``K-correction'').
The K-correction term for example increases up to about 6\%
out to a redshift of $z = 0.3$ for clusters with a temperature of 
about 2 keV and up to about 15\% for clusters with 10 keV. The iteration   
is repeated twice but found to actually converge to the final solution in the   
first step. The final rest frame X-ray luminosity is the value quoted in the     
catalogue.                                                                      
                                                                                
\begin{figure}                                                                  
\plottwo{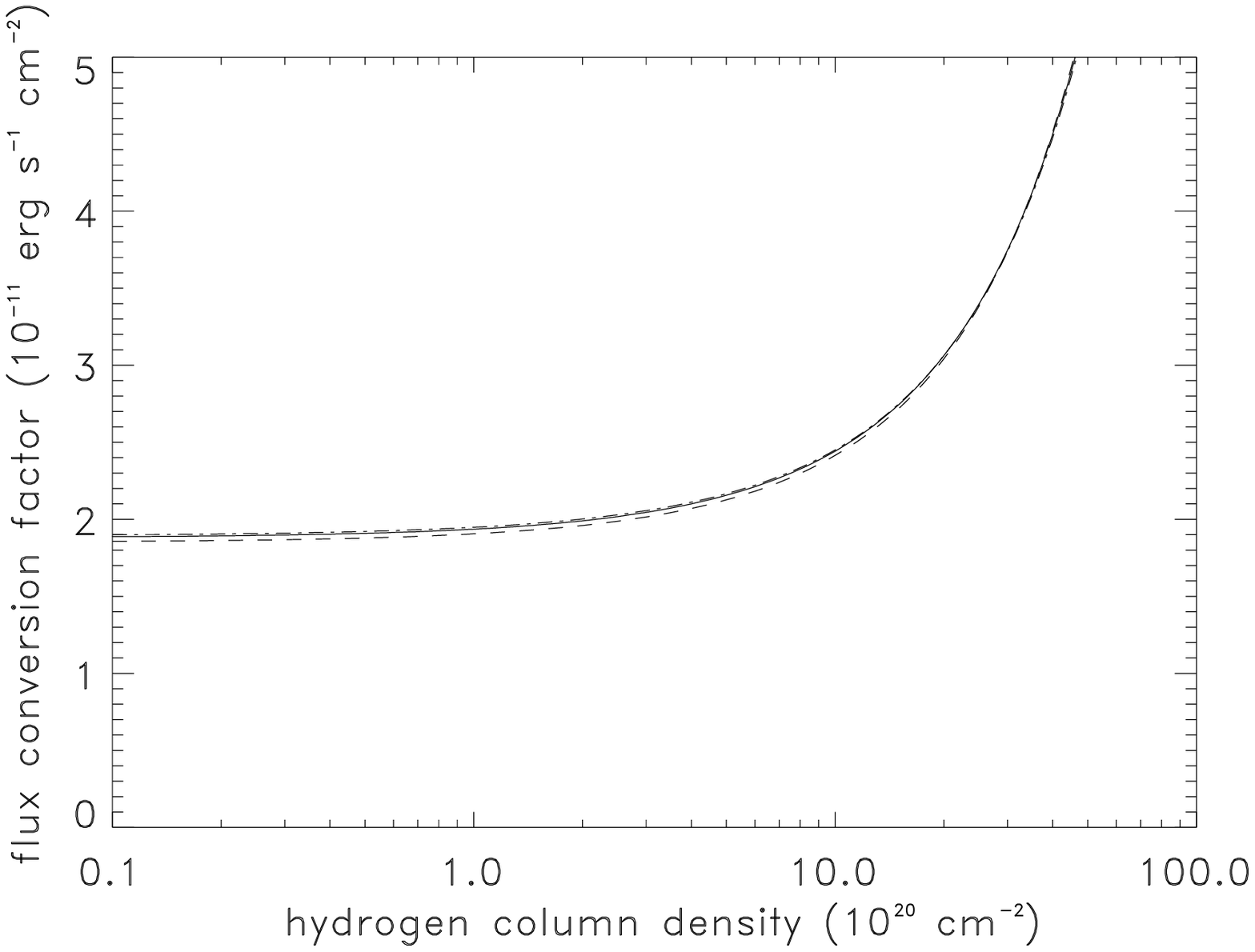}{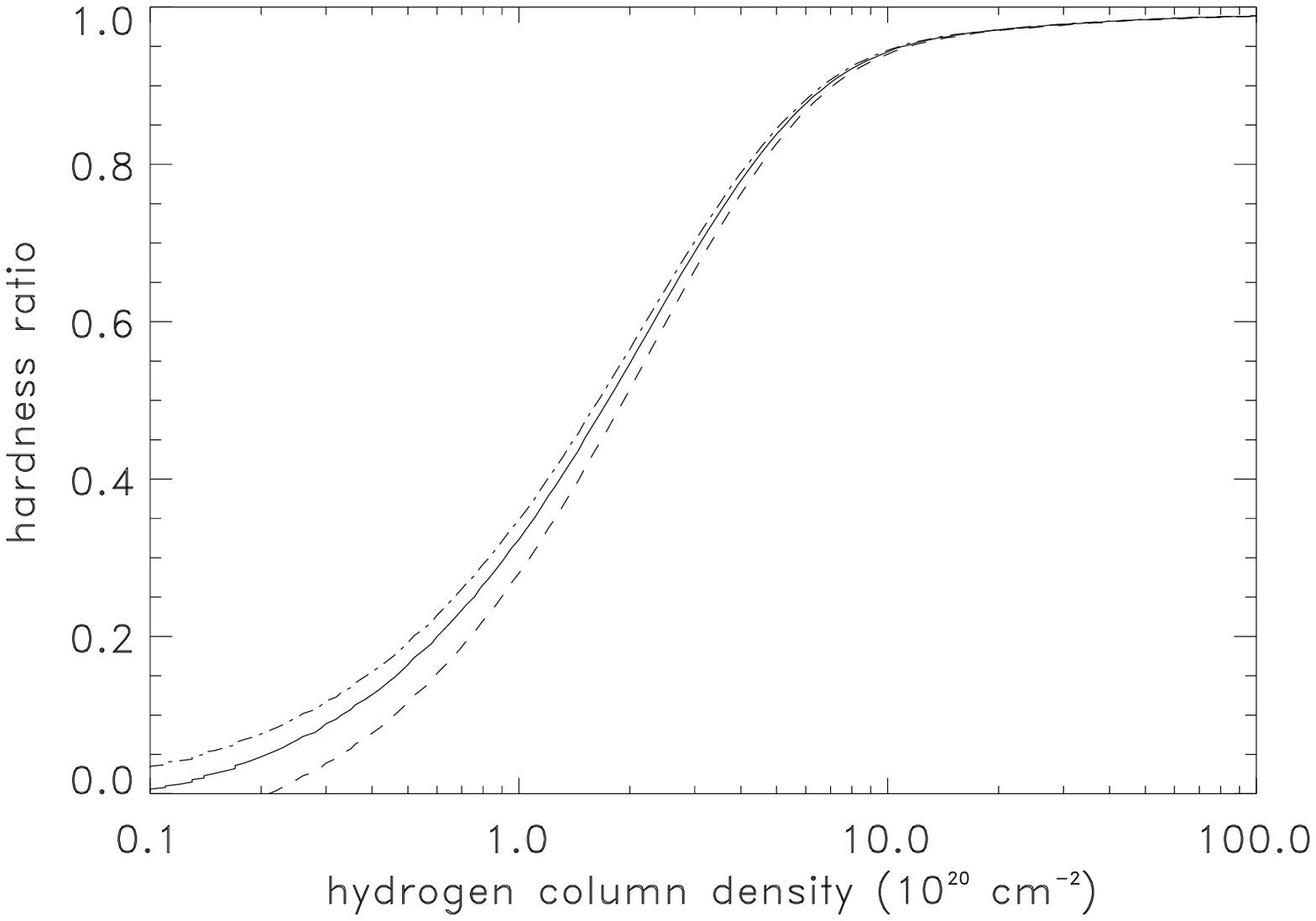}                                              
\caption{Flux conversion factor and hardness ratio as a function of              
the galactic hydrogen column density, $N_H$, for three different                
plasma temperatures: 2.5, 5, and 8 keV. }                                       
\end{figure}                                                                    

We have compared our results for the count rate to flux conversion
based on EXSAS software and special programs using a similar data base 
and radiation code as EXSAS with results obtained from
XSPEC, the online PIMMS software, and the conversion factors used
in Ebeling et al. (1998). The differences are always less than $3\%$.
Thus the use of different flux evaluation software does not constitute
a significant source of potential differences between different
RASS cluster surveys.

\subsection{Estimates of the total flux}

The X-ray fluxes determined from the observations may still be biased 
low compared to the total flux coming from the cluster, since part of the
flux in the faint outer regions is lost in the background. We can use
the fact that we obtained the flux within a well defined angular
aperture to make an estimate of the flux that may lie outside this
aperture. Note that we use this approach here as a tentative estimate of
the flux lost and we will therefore make  
no further effort in this paper to use the
results for a correction, since the underlying assumption that all
clusters have the same self-similar shape is not realized precisely
enough to make a case by case correction useful without further 
tests and justifications. The aperture radius that corresponds
to the generally used plateau value of the count rate is $R_{out}$.

To obtain a rough estimate of the flux possibly missed 
outside the aperture we adopt
the following generic cluster model characterized by a $\beta$-model
surface brightness distribution as given in eq.(3) with $\beta = 2/3$
to extrapolate the surface brightness profile outside $R_{out}$.
For the core radius we are not using the results of the $\chi^2$ fit,
since they have too large uncertainties. We rather prefer to
make a rough estimate of the cluster size from its X-ray luminosity.
From the studies by Reiprich \& B\"ohringer (1999) we find that
the cluster mass is well correlated with the X-ray luminosity 
according to the relation

\begin{equation}
L_x \propto M_{grav}^{1.2} ~~~~~.
\end{equation} 
     
We further assume that the self-similar relation of the core radius
and mass of the form $r_c \propto M_{grav}^{1/3} \propto L_x^{1/3.6}$
holds (see e.g. Kaiser 1986). Taking a Coma-type cluster 
with $L_x \sim 7\times 10^{44}$ erg s$^{-1}$ (0.1 - 2.4 keV) 
and a core radius of 300 kpc to normalize the relation we find

\begin{equation}
r_c = 0.3~ {\rm Mpc} \left( {L_x \over  7\times 10^{44}{\rm  erg
s}^{-1}} \right)^{1/3.6}~~~ .
\end{equation} 

Rather than integrating the X-ray flux of the $\beta$-model to
infinite radius, we stop the integration at 12 core radii which 
is about as large as the virial radius of a Coma-type model
cluster. The difference between an integration to infinity
(as for example performed in Ebeling et al. 1998 and 
De Grandi et al. 1999) to the cut-off radius at 12 $r_c$ is about
8\%. This overestimate of the flux for integration to infinity
is in general smaller than the individual uncertainties
but not negligible if one is concerned with the global bias of the
sample.

Applying this model to our cluster sample we can calculate the 
fraction of the X-ray flux missed for each of the sample sources.
Fig. 9 shows these missing fractions as a function of the 
source luminosity and of the detected number of source photons.
The mean missing flux is about 8.3\%. (The most discrepant point
with a missing flux of about 50\% and $L_x \sim 10^{44}$ erg s$^{-1}$
in Fig. 9a is for example a distant cluster observed at low flux 
($0.5 \cdot 10^{-12}$ erg s$^{-1}$ cm$^{-2}$) at low exposure
which would be excluded in a proper flux limited sample).

\begin{figure}                                                                  
\plottwo{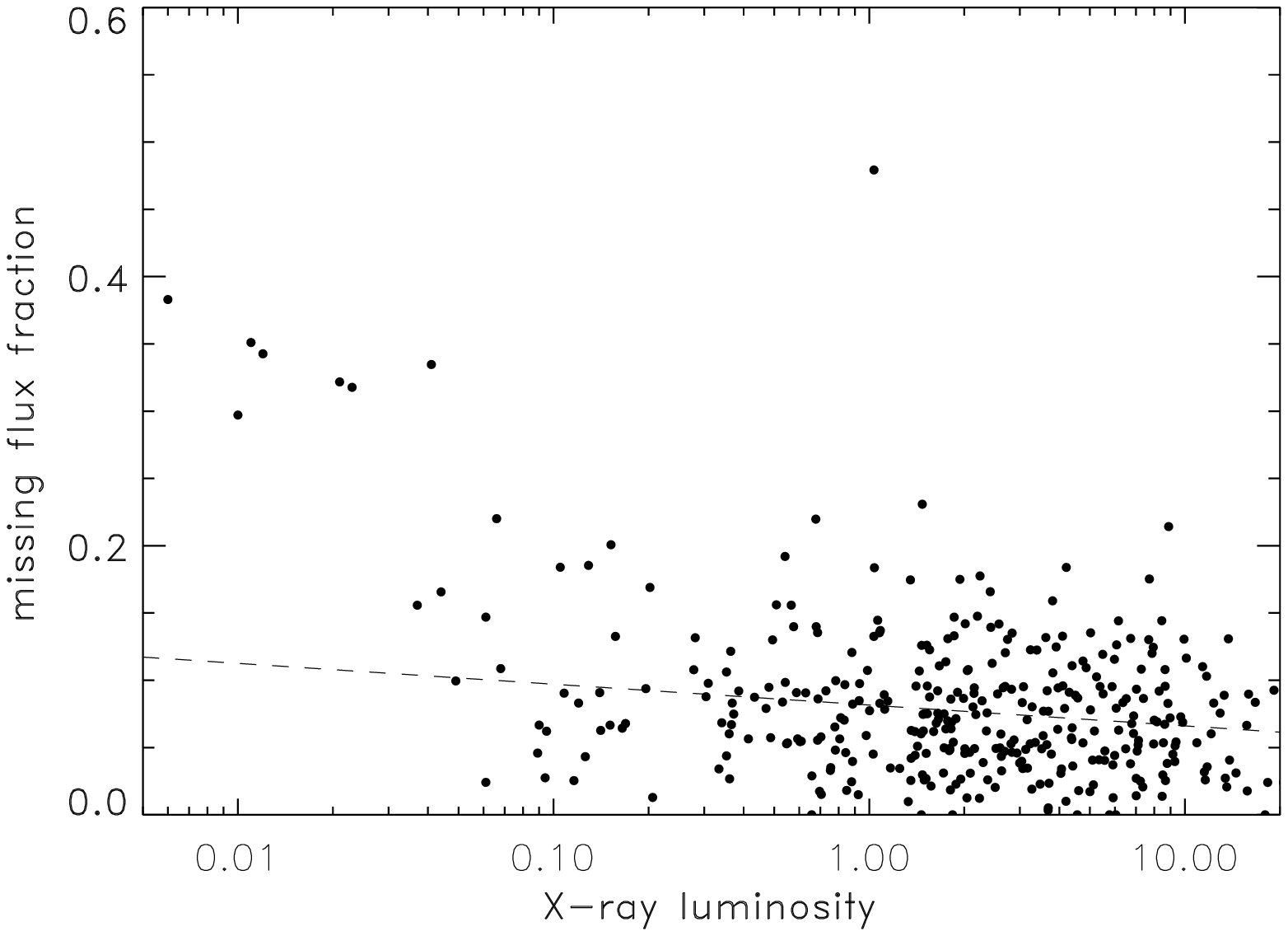}{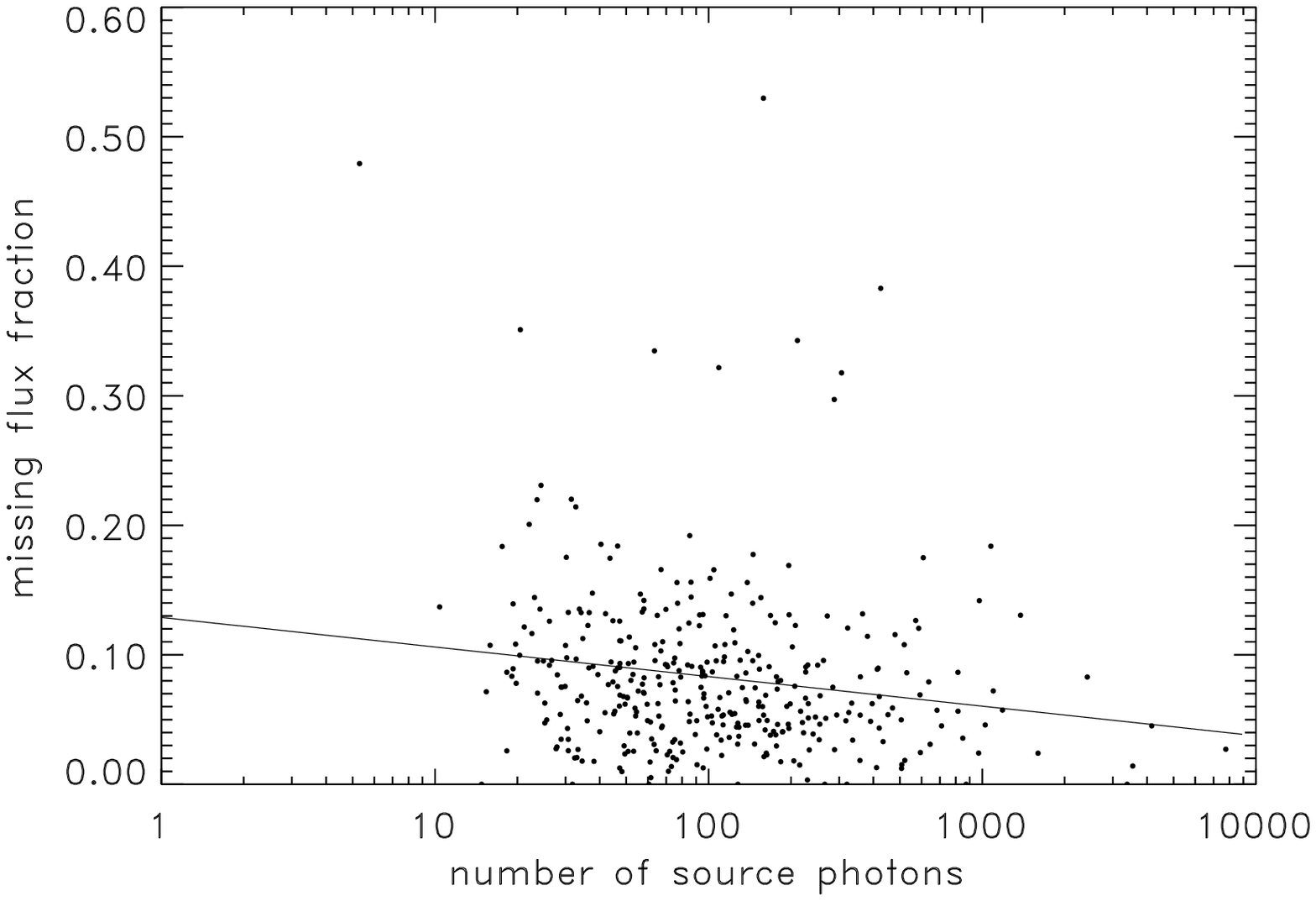}
\caption{Estimated missing flux of the NORAS cluster sources
as a function of a) the source luminosity (in units of $10^{44}$ erg s$^{-1}$) 
and b) the number of detected photons per source.} 
\end{figure}

The missing flux fraction, $fr$, features only a very weak dependence
on X-ray luminosity. The linear regression fit to the function
$fr = f(\log(L_x)$ shown in Fig. 9a decreases from 8.3\% for
$L_x = 10^{43}$ erg s$^{-1}$ to 7.9\% for $L_x = 10^{45}$ erg s$^{-1}$.
A more significant dependence is found for the number of source photons,
as could be expected since this is the main parameter determining the 
significance of the source detection and how far out the count rate
integration can be performed. Here the linear regression fit of Fig. 9b,
$fr = f(\log(N_{ph})$ shows a decrease of $fr$ from 9.5\% for 30
source photons to 6.7\% for 500 source photons. Also this dependence is 
weak.

An exception to the relatively small values for the missing fraction
constitute some of the low luminosity sources as displayed
in Fig. 9a. These are elliptical galaxies
or very small groups dominated by elliptical galaxies, which have
much smaller core radii - as indicated by the GCA King-model fits -
compared to the values assumed after eq.(6). Thus for these small
objects the model assumption seems to break down and the actual
values for the missing flux is much smaller than estimated here.
The validity of this generic model for the extrapolation of the 
total flux will be pursued in more detail in a future publication. 
A comparison with pointed observations performed below gives 
already an encouraging confirmation of these estimates.
 
\begin{figure}                                                                  
\plottwo{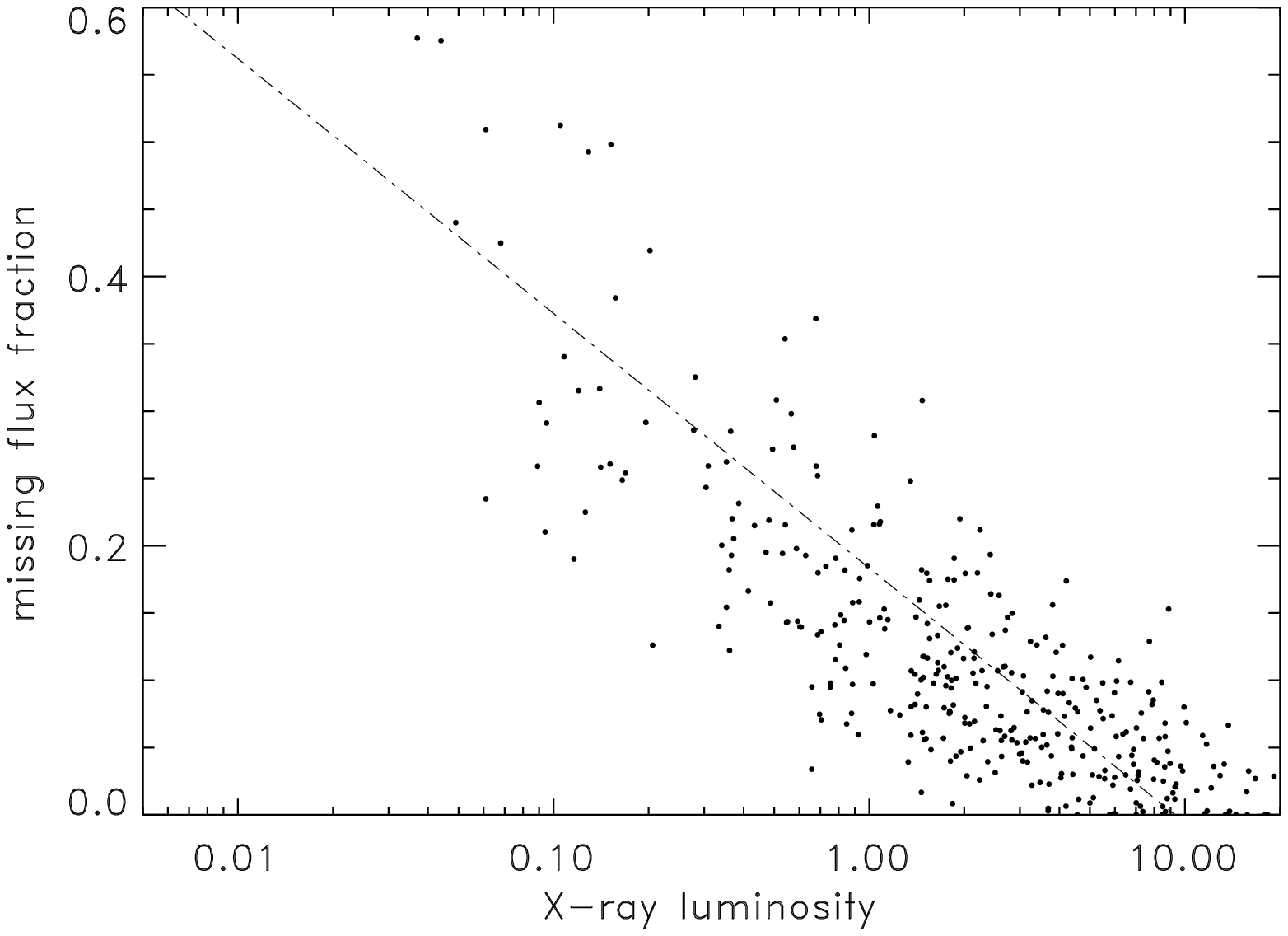}{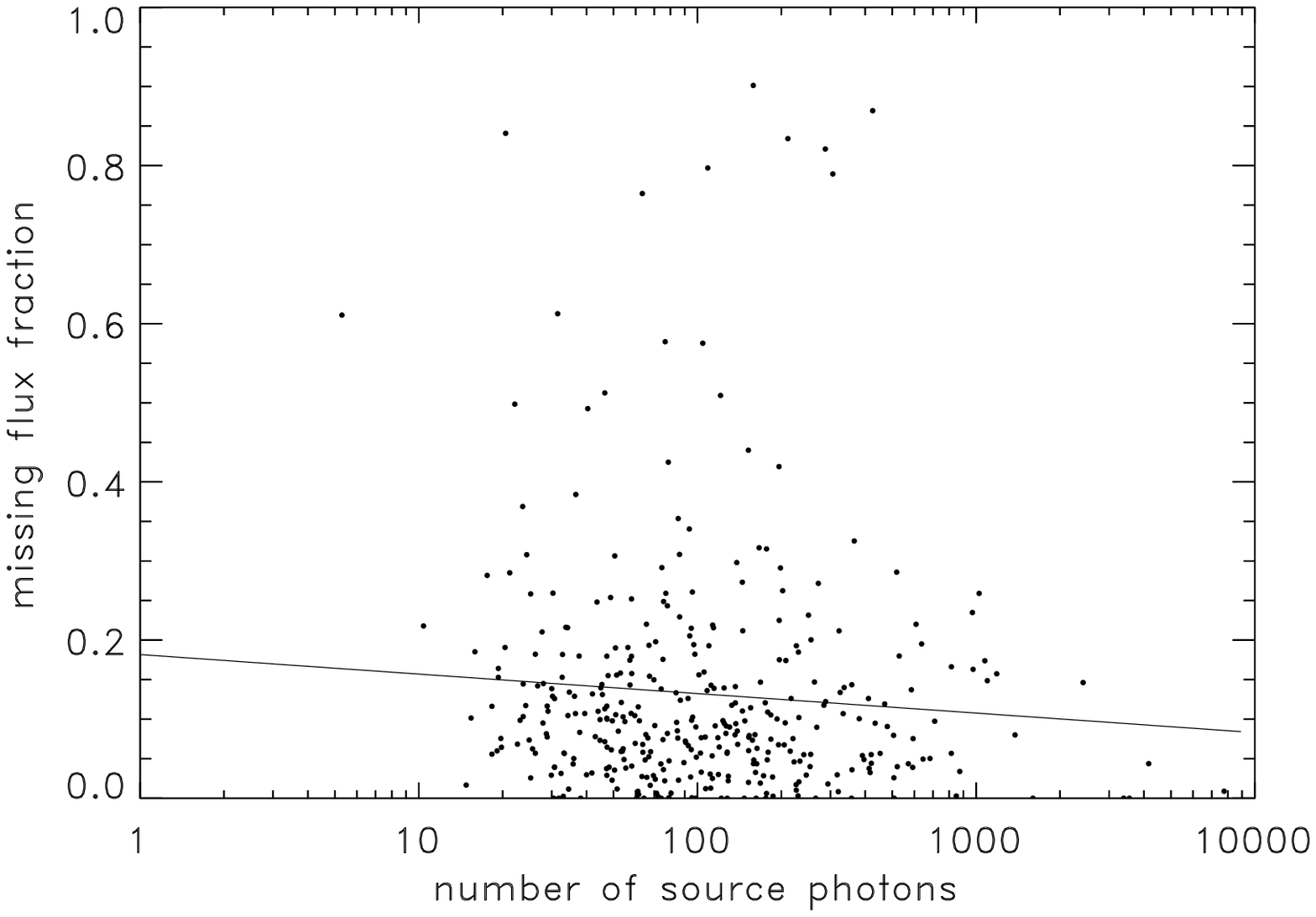}
\caption{Estimated missing flux of the NORAS cluster sources
as a function luminosity (left) and source photon number (right) 
if a constant core radius of 250 kpc is assumed.
The straight lines show the results of linear regression fits to the
data.}
\end{figure}                            

Alternatively we explore a second approach to estimate the missing 
X-ray flux by fixing the core radius in the $\beta $ model to 250 kpc.
This approach was used in earlier studies of X-ray cluster samples
(e.g. Henry et al. 1992). Figs. 10a and 10b show the results corresponding
to the results of Fig. 9. One clearly notes a very steep increase in the
missing flux for decreasing X-ray luminosity in Fig. 10a. The dependence
on the photon number for which we would expect the strongest dependence
is much less pronounced and not essentially different from the results
in Fig. 9. One notes, however, that the scatter in Fig. 10b has 
approximately doubled. The obvious interpretation of these results is
that the strong dependence on X-ray luminosity seen in Fig. 10a 
is artificial and results from an overestimate of the core radius for the 
less luminous objects. This inappropriate choice of the core radius
also increases the scatter in Fig. 10b. Thus we conclude that this 
approach is clearly inappropriate for our study and the above used scaling
of the core radius is a reasonable choice. 

\subsection{Comparison with the results of RASS I}

\begin{figure}                                                                  
\plottwo{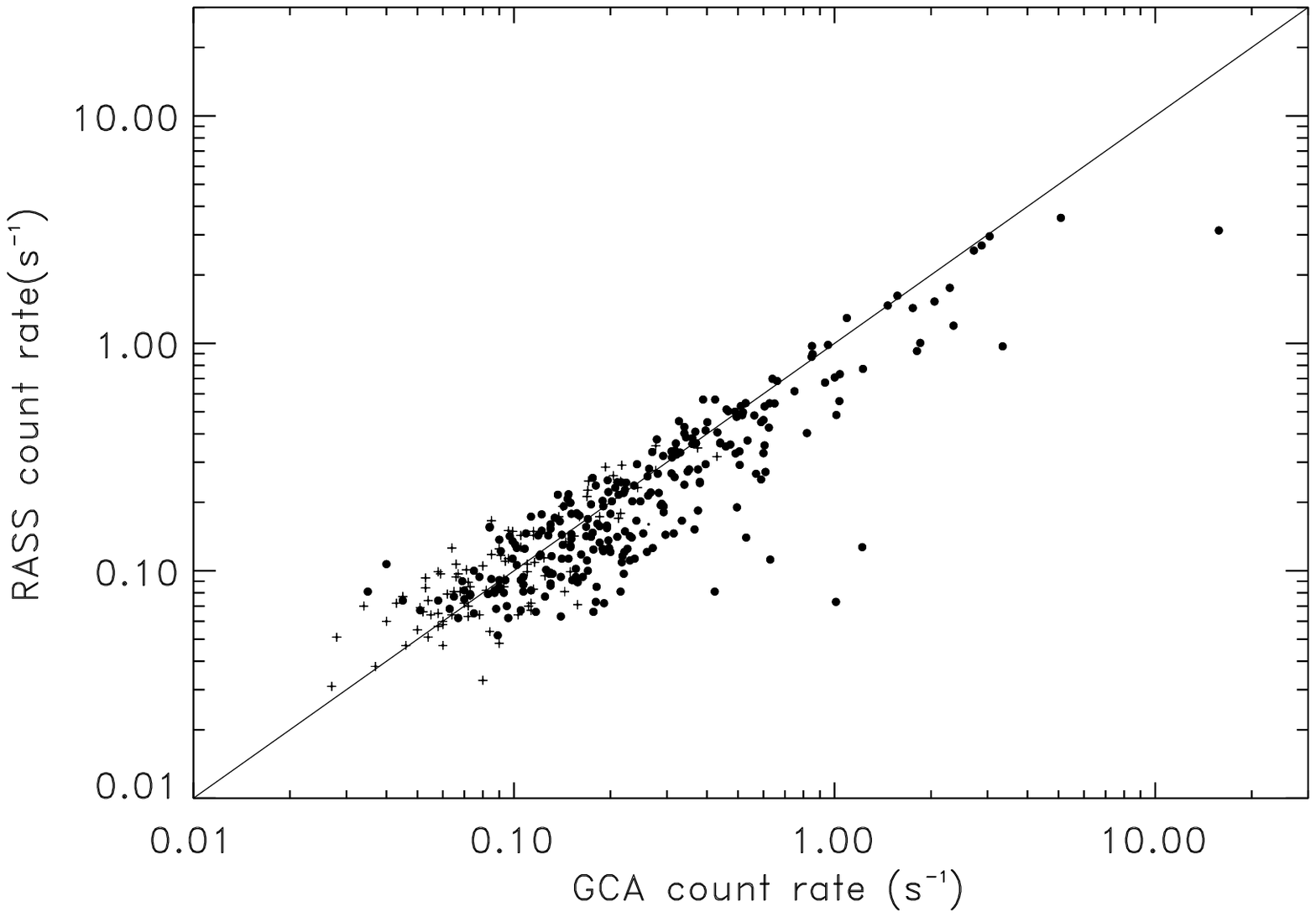}{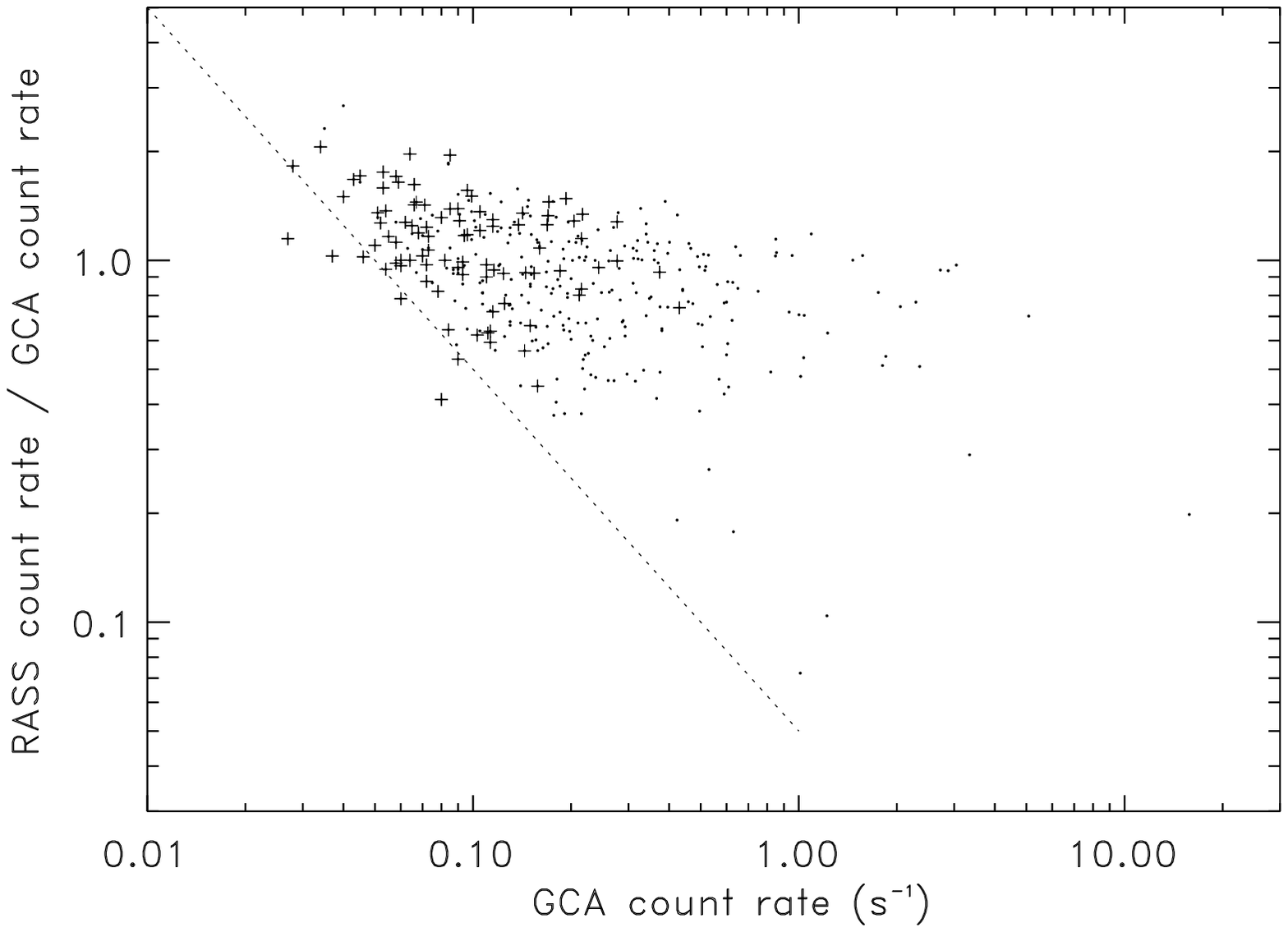}            
\caption{Comparison of the count rates determined by the                    
maximum likelihood method applied in RASS I and the 
results of the GCA. The straight line in the left pannel 
indicates equality. The sources found     
to be extended in the new analysis are shown as filled circles and the          
point-like sources as crosses. In the right pannel the extended sources are
shown as small dots. The dotted line in the right pannel indicates the
approximate count rate limit for the RASS I source results (the count
rate limit in the plot has a small uncertainty because the count rate 
conversion from broad band to hard band also includes a dependence on the 
interstellar column density).}                                                  
\end{figure}                                                                

A comparison of the count rates measured in RASS I for the 378 cluster
sources listed in Table 1      
with the results of the GCA reanalysis is shown in Figs. 11a and 11b.   
For the comparison the RASS I count rates measured in the broad band have been    
converted to hard band counts by means of the measured hardness ratio.          
Sources which feature a significant extent according to our new analysis are    
marked in the plot. 
There is a large fraction of sources for which the
count rates measured in RASS I or RASS II are underestimated by
up to an oder of magnitude, which are essentially the sources
marked as extended by the GCA method.
The pointlike sources scatter around the line of equal
count rate with an increasing scatter with decreasing count rate. The
increase of the RASS I to GCA count rate ratio for low count rates
seen in Fig. 11b is most probably an artefact produced by the previously
set count rate limit in the selection of the RASS I sources for this sample
(as indicated in the figure by the dotted line).               
                                              
The source of disagreement for the extended sources       
results from the design of the source analysis technique used for the RASS      
which is tuned to work optimally for point sources. Two effects discriminate    
against the proper accounting of the count rate of extended sources: 
i) the Gaussian kernel of the  
source shape fitting of the maximum likelihood analysis is bound to miss the    
outer wings of a typical cluster surface brightness distribution and ii) part   
of the outer X-ray halos of extended clusters may be treated as background by   
the background spline fitting process as used in the RASS standard analysis.
A source analysis technique tuned to process the extended 
sources properly is therefore required to avoid these 
problems and to obtain correct X-ray parameters for the
objects in our sample. The reanalysis of all the X-ray sources in the sample
was therefore a necessary prerequisite for the compilation of an
X-ray cluster catalogue to be used for astronomical and cosmological studies.

\subsection{Comparison to pointed observations}

To test the count rate determination of our new analysis 
technique against a more reliable standard we 
have analyzed 80 clusters of our sample in pointed observations
in which the better photon statistics allows a more detailed 
and precise analysis.
The results for the brighter
sources were taken  from the compilation of Reiprich \& B\"ohringer
(1999) of the ROSAT clusters with the highest flux. 
13 of the clusters from this sample were also analyzed in very 
large RASS survey fields 
($4\deg \times 4\deg $ or $8\deg \times 8\deg $) because of the large 
cluster sizes. In the analysis by Reiprich \& B\"ohringer (1999)
a more refined source analysis was performed (where contaminating
sources are excised in a wide region in and around the cluster and
a possibly badly measured background is iteratively 
corrected by parabolic fits to
the azimuthally integrated background surface brightness profile 
outside the cluster). Therefore it is also interesting
to keep these objects in the list for comparison.
The other data were retrieved from the ROSAT archive. 
The analysis technique used is similar to the one described above
with a main difference that contaminating sources are excised 
interactively. We are using the count rate 
at $R_x$. The values for $R_x$ found in the pointed observations are generally
larger than $R_{out}$ found in the RASS data
as the flux can usually be traced further out into the background
in the deeper observations. 
Fig. 12 shows a comparison of the count rates found in the two data sets.
The objects analyzed in large RASS fields are marked with open symbols.
The mean deviation of the count rates determined for the pointed data
and the present results is 8.6\%.  
Thus we conclude that the missing flux is on average about $7 - 10\%$
without a significant bias as a function of X-ray flux.
This result is in excellent agreement with the estimates for
the missing flux in section 3.5. The validity of the GCA approach is
thus confirmed in two ways: the missing flux fraction is relatively small
compared for example to the measurement errors and the 
{\it ab initio} estimates for the missing fraction are approximately
correct.

\begin{figure}                                                                  
\plotone{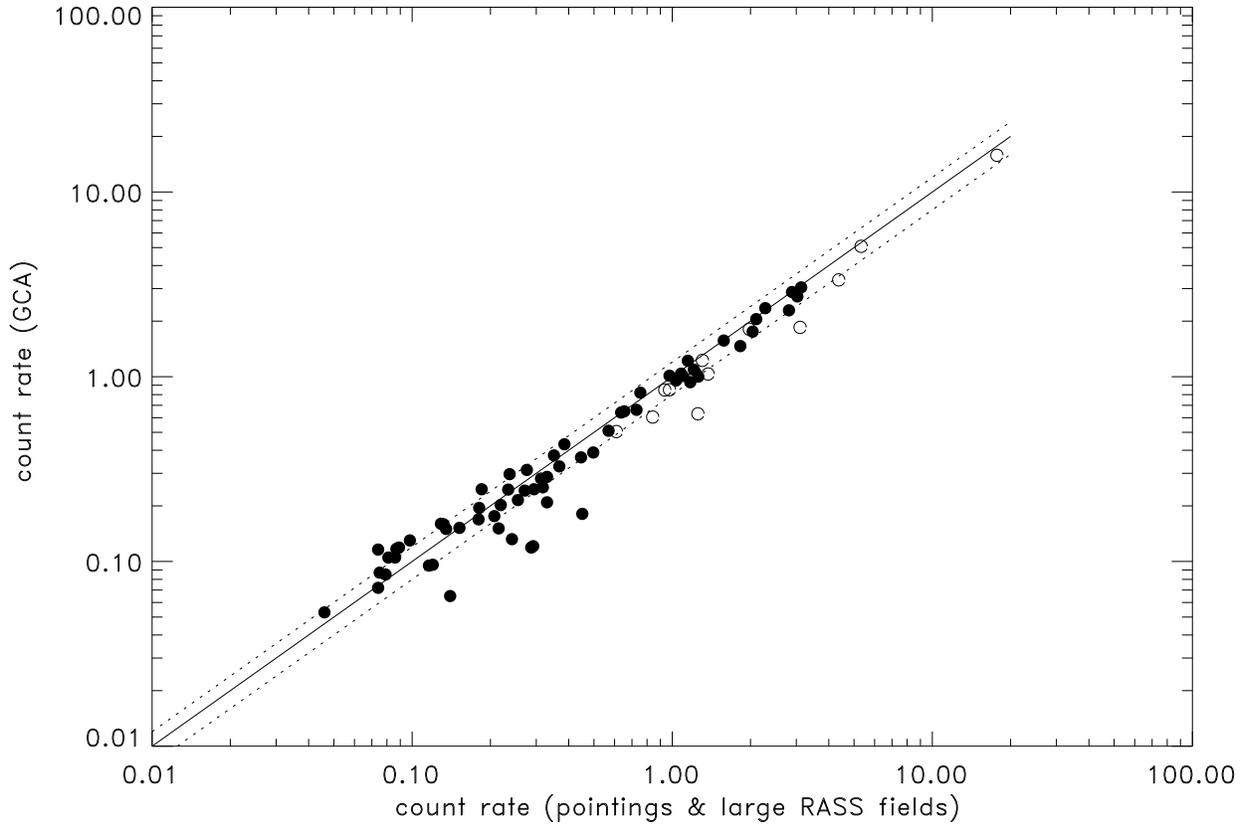}                                                           
\caption{Comparison of the count rates for 67 clusters as determined in the
present analysis and in deeper pointed observations (full circles). 
In addition we also show the comparison between the present results
and the interactive analysis by Reiprich an B\"ohringer (1999)
for 13 clusters in the same RASS fields (open symbols). The solid line 
indicates equality
and the two dashed lines indicate deviations of 20\%.} 
\end{figure}

\section{The catalogue of RASS I extended sources: 
clusters and non-cluster sources}
                                                                                
In the following we are presenting the catalogue of the sample of 495 RASS I
extended X-ray sources identified as galaxy clusters or as non-cluster
X-ray sources. We discuss further characteristics of 
the source properties      
and the sample in the subsequent sections. We split the catalogue in three        
parts, the list of 378 sources identified as clusters,          
the list of 99 non-cluster sources, and a list of 17 X-ray AGN and one star
located in clusters or in the line-of-sight of clusters, where the cluster
is not the main source of the X-ray emission.
The first tables, Table 1 - 4 list the major properties of the sources.
Further X-ray parameters for the galaxy clusters are given in Table 5.

The parameters of the table columns of Table 1 
are described as follows. Column (1) 
lists the source name given by the following scheme:  
we use the prefix RXCJ for the reanalyzed RASS sources that have                
been identified with a galaxy cluster, where the ''C'' stands for cluster. 
This prefix will be exclusively used for all RASS clusters 
analyzed by the above GCA technique. In particular this prefix has 
also been assigned to the RASS clusters identified within the REFLEX Survey 
(B\"ohringer et al. 1998, Guzzo et al. 1999). 
The remaining part of the name refers to the source coordinates for the
epoch J2000 in hours (RA) or degrees (DEC), minutes, and fractions of a 
minute. Since the cluster sources are 
usually extended by more than one arcmin we use an arcmin precision in 
converting the cluster coordinates into the remaining part of the source 
name, thus the total name has 15 digits. Note that the coordinates in the 
cluster source name can deviate from the official RASS source catalogue 
coordinates reflecting the difference in the source analysis technique used. 
Column (2) gives alternative source names for previously catalogued optical
counterparts to the X-ray sources, mainly Abell and Zwicky cluster names,
names of NGC and UGC galaxies forming the central dominant galaxies of groups,
and previously identified RASS or other X-ray sources. 
The Zwicky cluster names given in the 
table conform with the convention of the NED data base (note that the
coordinate reference in the name refers here to epoch B1950).
Columns (3) and (4) give the source position
in decimal degrees for the epoch J2000. Column (5)
gives the redshift of the cluster. Column (6) lists the measured count rate 
in units of counts s$^{-1}$. Columns (7) and (8) give the X-ray 
flux in units of $10^{-12}$ erg s$^{-1}$ cm$^{-2}$ for the flux estimated
for a temperature of 5 keV in the first step and the finally calculated
corrected flux, respectively. This correction includes the recalculation
of the count rate flux conversion for the best temperature estimate 
and the redshift of the spectrum, but not the addition of the estimated
missing flux. 
The fractional uncertainty in percent for
the count rate, the fluxes and the luminosity are given in Column (9). 
Column (10) gives the rest frame X-ray luminosity of the clusters in the 
0.1 - 2.4 keV energy band. Column (11) lists the value assumed for the
absorbing column density in the line-of-sight to the source
in  units of $10^{20}$ cm$^{-2}$, and 
column (12) indicates with the sign ``?'' if the identification as a
cluster leaves some residual doubts. 
In some cases a second flag provides information on the way the source position
was determined as explained below. Column (12) also lists the flags
for the sources which have been deblended, flag $B$, analyzed in extra 
large fields, flag $L$, and sources which may 
be contaminated, flag $C$. The last column, (13), gives the reference
number for the redshift as listed in the table caption. The objects
marked by ``?'' or $C$ are commented below.

For the position we have used the coordinates determined by the        
moment method with a 3 arcmin aperture radius. 
As an exception in the case of 12 cluster sources we have chosen  
to either redetermine the center position by a moment method with a larger      
aperture of 5 or 7.5 arcmin or we have determined a center position by         
hand. In these sources the automatic detection has selected a maximum which is    
significantly offset from the global center of symmetry of the clusters. These       
clusters are marked by the flags b, c, o in column (13) for the 5 arcmin
and 7.5 arcmin moment method and the determination by eye, respectively.

Table 2 lists the equivalent parameters for the RASS I extended sources 
identified as non-cluster objects. Here column (2) gives the source type of 
the identification except for cases with popular object names. A more
detailed identification will be given in the second paper by 
Huchra et al. (1999). Columns (3) to (8) have the same definition as
these columns in Table 1. 
Columns (9) to (14) give the extent parameter $P_{ext}$ from the KS test,
the best fitting core radius, its minimal value consistent with the $2\sigma$
uncertainty limit, the hardness ratio, its error, and the deviation
from the expected value of $HR$ in units of $\sigma$. The hardness ratio
does occationally exceed the value of one in the tables, which occurs
when the soft photon counts in the source region fall short of the
background expectation.  
Note that for the parameters $P_{ext}$ and $\Delta HR$ an upper limit to
the numerical value of 30 and 10, respectively, was set, which was
also used in plotting the data. Column (15) gives the interstellar
HI column density. 
In the last column (16) objects with no certain optical 
identification are marked with ``?''. 
The identification strategy for the non-cluster sources
is explained in section 5. The last column also contains the flag
for the alternative center determination.

Table 3 lists X-ray sources where the X-ray emission is obviously originating
from an AGN and in one case from a star, but where a cluster or group of 
galaxies is also visible on the Palomar Sky Survey images or on the CCD
frames taken for this project. In several cases which are commented 
in Table 4
the redshift information indicates in addition the existence of a cluster
and that the AGN is at the same redshift. The meaning of the columns
of Table 3 is the same as that for Table 2. Table 4 provides 
the redshifts of the AGN as far as known and comments on the source
identification.

Figure 13 shows the distribution of the sources from Table 1 - 3 in the sky.     
Since we have not yet imposed a strict flux limit to the survey and 
also due to the incompleteness of the sample we cannot
necessarily expect a homogeneous coverage of the sources in the sky.
Near the coordinates $RA = 270\deg$ and  $DEC = 70\deg$ we note a significant
concentration of sources, which is due to the concentration of
overlapping survey strips at the north ecliptic pole (NEP). (Note that this is 
not due to a pile up of exposure time at the NEP, since each survey strip was
analysed independently, but due to the fact that the chance for detection of
low flux, extended sources is increased in the multiple strips. A homogeneous
coverage is achieved here only after imposing a proper flux cut.) 
There are also several low density regions in the source distribution as 
for example at the northern tip of the south galactic cap region.  

\begin{figure}                                                                  
\plotone{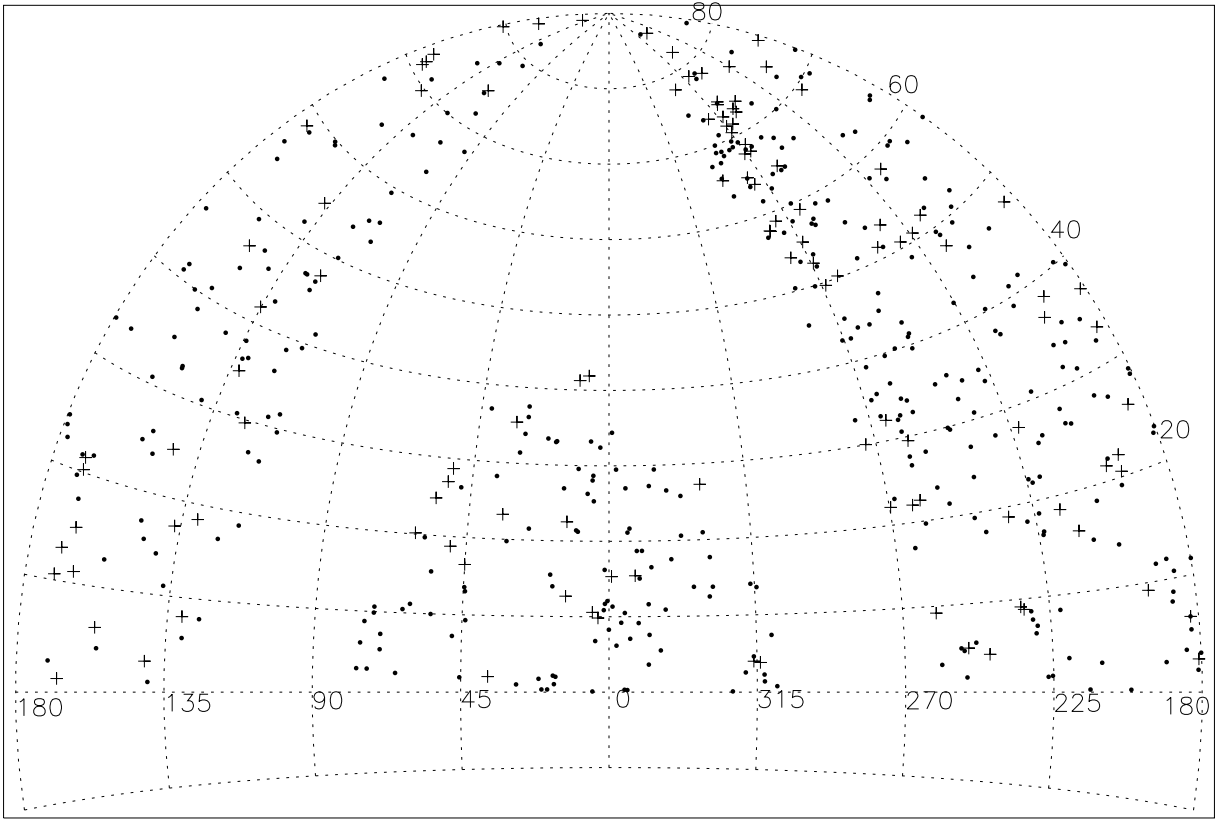}                                                             
\caption{Distribution of the RASS I extended sources selected 
for the NORAS sample.
The dots mark the NORAS cluster sources while the crosses mark the non-cluster  
sources identified in the course of the NORAS survey.}                           
\end{figure}

The distribution in X-ray luminosity and redshift of the sources in the
cluster sample is shown in Fig. 14. The parabolic curves indicate limiting
fluxes of $10^{-12}$ erg s$^{-1}$ cm$^{-2}$ and 
$3\times 10^{-12}$ erg s$^{-1}$ cm$^{-2}$, respectively.
There are 5 
clusters with redshifts larger than 0.4 and 18 clusters with redshifts
between 0.3 and 0.4. About half (11) of these clusters have luminosities 
in excess of $10^{45}$ erg s$^{-1}$ and belong to the most X-ray 
luminous clusters in the Universe.

\begin{figure}                                                                  
\plotone{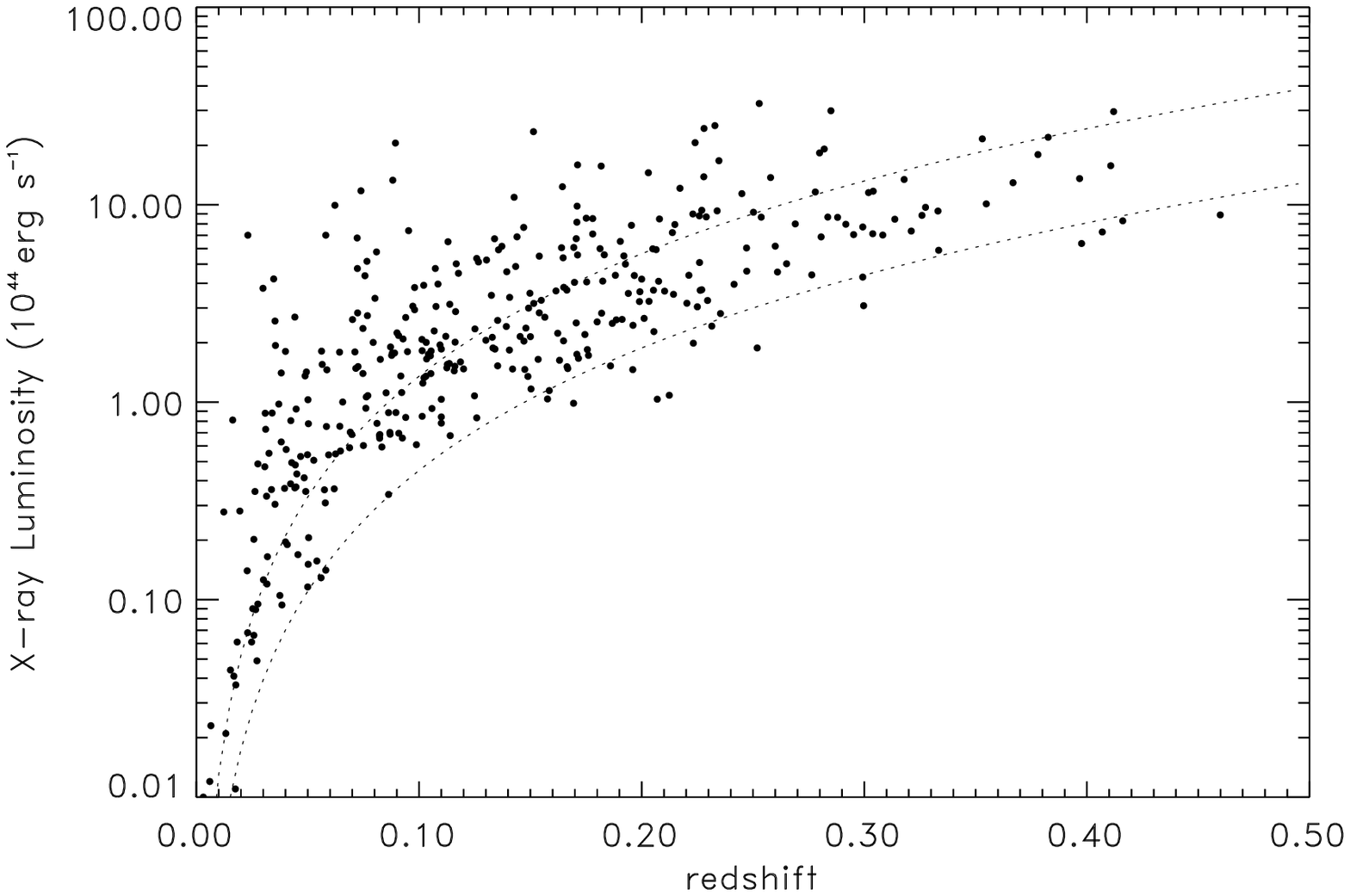}                                                                
\caption{X-ray luminosity and redshift distribution of the 
NORAS cluster sources. The two dotted lines indicate flux limits of 
$10^{-12}$ and $3 \times 10^{-12}$ erg s$^{-1}$ cm$^{-2}$.}
\end{figure}                                                                    

\subsection{Comments on individual objects}

{\bf RXCJ0106.8+0103} has an active galaxy in the cluster noted 
in the survey by Romer et al. (1994). The RASS image is consistent with a point
source and a cluster spectrum. The available ROSAT HRI image has a small
but significant extent. If the emission is mainly due to the cluster 
it would be very compact for the given high luminosity. A CCD exposure
shows a nice cluster image and therefore we expect that this object
is an X-ray cluster with a severe X-ray contamination by the AGN.

{\bf RXCJ0255.8+0918} is a pointlike X-ray source with spectral properties
consistent with intracluster medium emission centered on an elliptical
galaxy. The galaxy is also classified as Sy2/LINER (Pietsch et al. 1999). 
They also note that the galaxy is located in a group.
Without further information it is difficult to distingish between 
AGN or hot gas halo emission. The X-ray luminosity of 
$L_x = 0.7 \times 10^{43}$ erg s$^{-1}$ could well be that of a small group.

{\bf RXCJ0311.5+0714} is not certainly confirmed as an X-ray cluster and the
redshift is uncertain since only one galaxy redshift is available.

{\bf RXCJ0728.9+2935} is a cluster candidate in a crowded stellar
field featuring a pointlike X-ray source. Only one galaxy redshift
is available for the X-ray source region.

{\bf RXCJ0736+3925} shows extended X-ray emission, mostly due to
intracluster medium emission,  but obviously also some contribution
by a softer central point source.

{\bf RXCJ0921.1+4538} coincides with the radio galaxy 3C219. The X-ray 
source is point like but the hardness ratio is consistent with
hot gas emission. 3C219 is located in a galaxy group. Without
further information we cannot definitely decide if the X-ray emission
comes from the AGN or the group.

{\bf RXCJ1009.3+7110} is a case similar to RXCJ0106.8+0103 where a
Seyfert 2 galaxy is listed in the Veron catalogue (Veron-Cetti \&
Veron 1998) and the
ROSAT HRI image shows most probably a point source with a very
compact halo. Thus this object is also classified as cluster
contaminated by an X-ray AGN.

{\bf RXCJ1122.2+6712} shows in the available HRI image a small extended
halo around the galaxy VII Zw 392. The RASS source is contaminated by
a nearby point source. The X-ray halo is with a luminosity of less than 
$10^{43}$ erg s$^{-1}$ quite faint corresponding to a very small group.
It is one of the faintest sources in the sample.

{\bf RXCJ1157.3+3336}: For this X-ray source two cluster identifications are
possible. It is associated with Abell 1423 at a redshift of 0.0761
but there is also the possibility that a cluster is associated with the 
radio galaxy 7C 1154+3353 located at the center of the X-ray emission. 
The 7C galaxy is classified as cD galaxy. We adopt the identification 
of an X-ray cluster associated to the 7C galaxy because of the better positional
coincidence. Crawford et al. (1999) note the same identification.
             
{\bf RXCJ1229.7+0759} and {\bf RXCJ1243.6+1133} are the X-ray halos
of the two Virgo cluster galaxies M49 and M60, respectively. They are
included in this catalogue, even though these halos are embedded
within the low surface brightness structure of the Virgo cluster
emission. But since the two local halos stand out from the low surface
brightness emission environment and  can be reasonably characterized
by the present source characterization technique we have not excluded
them from our catalogue as we have excluded M86 and M87.            

{\bf RXCJ1510.1+3330, RXCJ1556.1+6621, RXCJ1700.7+6412} show a contamination 
by a point source in the available HRI image by no more than 10 - 15\%. 

{\bf RXCJ1518.7+0613} is contaminated by X-ray emission from an AGN in 
the cluster, which is also indicated by a softer hardness ratio than
expected for a galaxy cluster. 

{\bf RXCJ1554.2+3237} is not yet definitely confirmed as cluster
and the redshift is uncertain since there is only one galaxy redshift.

{\bf RXCJ1644.9+0140} and {\bf RXCJ1647.4+0441}
are cluster candidates in a crowded stellar field
where no conclusive redshift has been ontained so far. 
RXCJ1644.9+014 is possibly a distant cluster.

{\bf RXCJ1738.1+6006} is also a cluster candidate in a crowded stellar
field with extended X-ray emission at the significance threshold.
No redshift has yet been obtained. A Seyfert galaxy is known with
a distance of 1 arcmin from the X-ray maximum which is most probably
not the X-ray source because the offset would be unusually large.

{\bf RXCJ1800.5+6913} has a complex structure and is obviously contaminated
by emission from point sources. But these sources contribute less than 20\%
to the overall emission from the cluster.

{\bf RXCJ1832.5+6848} contains a BL Lac in the cluster center, which
most probably severely contaminates the cluster X-ray emission.

{\bf RXCJ1854.1+6858} is a cluster candidate featuring an extended
X-ray source. The redshift is uncertain since only one galaxy 
redshift is available.

{\bf RXCJ2035.7+0046} is an extended low surface brightness source
with a detection significance of 3 to 4 $\sigma$. It is a cluster
candidate in a very crowded stellar field for which no redshift is available
yet.

{\bf RXCJ2041.7+0721} is a cluster candidate in a crowded stellar field
with only one available redshift.

\section {Discussion of the major source properties}                            

Further properties of the X-ray cluster sources are
provided by Table 5.
The columns of the table are as follows. 
Column (1) and (2) repeat the name and rest frame ROSAT band X-ray luminosity 
of the sources from the previous tables. Columns (3) and (4) give the 
radius out to which the source count rate has been integrated, $R_{out}$, 
in units of arcmin and Mpc, respectively. Column (5) gives the
probability result of the Kolmogorov-Smirnov test for the source to 
be a point source in terms of the parameter $P_{ext}$. For very low 
probabilities for the consitency with a point source
the parameter  $P_{ext}$ was limited to a value of 30 for
the entry in the table. A source is considered very likely to be
extended if this parameter has at least a value of 2 (point source excluded
with 99\% probability). Column (6) and (7) give the best fitting core radius
for the King model fit and the minimal core radius still consistent within
2$\sigma $ error limits, respectively. Note that the core radii determined
here are a only qualitative measure for the source extent, since in general 
the errors are very
large and the fitting grid was coarsely spaced. Therefore we do not recommend to
use the results for $r_c$ as a measure of the statistics of the cluster shapes
at this point.
Column (8) and (9) give the spectral hardness ratio
defined by eq.(2) and its Poissonian error. Column (10) finally indicates the
deviation of the measured hardness ratio from the expectation value calculated
for the given $N_H$ and for a temperature of 5 keV. This deviation 
parameter is given in units of the $1\sigma$-error of the hardness ratio 
and the listed and plotted values were limited to a maximum numerical
value of 10.

One of the most interesting first questions about these parameters concerns the
discrimination power of the spectral and extent parameters in distinguishing
between cluster and non-cluster sources. This is analysed in detail for
statistical samples of sources with known identifications in the forthcoming
paper by B\"ohringer et al. (in preparation). The present source sample
should be highly biased since it was preselected from the RASS I data
with the criterion of showing a significant extent in the RASS I analysis. We 
have to expect for example that the subsample of point sources within the
present sample is already enriched in pathological cases including 
e.g. double and very bright sources. Therefore the present sample is 
used here only for a qualitative discussion while statistical numbers
can only be obtained from the analysis to be published in the following
paper.

Fig. 15 displays the distribution
of the spectral and spatial extent parameters for the cluster and
non-cluster sources for comparison. For the spectral discrimination
we plot the difference of the actually observed hardness ratio to
the theoretically expected hardness ratio as calculated for a 5 keV 
cluster as shown in Fig. 8. The difference is thereby quantified in terms
of the $\sigma$-deviation accounting for the uncertainty of the hardness
ratio measurement. To quantify the extent we use the probability
result of the KS-test for a point source in terms of the extent
parameter $P_{ext}$.
As we can see most of the clusters occupy the upper part of the
figure and cover a wide range of extent values. As expected, non-cluster
sources are concentrated in the region of small values of $P_{ext}$.
Clusters have harder spectra than the
average of the other X-ray sources being mostly stars or AGN, and therefore
clusters should be separable by the hardness ratio from the softer
part of the non-cluster population. On the other hand the extent parameter
should help in the discrimination against the non-cluster sources since
they are to the vast majority point-like sources at the angular resolution
of ROSAT.

This is clearly seen in Figs. 16 and 17 which show a blow-up of Fig. 15
and display the cluster and non-cluster sources
separately. Only 12 clusters out-off 378 show a more than $3\sigma$
deviation to the X-ray soft side. These are to about one half
very bright clusters (including Coma) where the relative deviation
is small in absolute sense and could reflect for example an incorrect 
value for the column density. For some of the sources we suspect
some contamination by AGN to the overall emission. One of these
cases is for example A1722 for which the HRI observation confirms
the contribution of a point source to the extended cluster emission
by an amount of about 30 - 40\%.
Anyway these pathological cases comprise only about 3\% of the
sample and the hardness ratio provides indeed a very powerful
diagnostics for the source identification.

A similarly interesting result is found for the extent parameter,
$P_{ext}$ for the case of the non-cluster sources shown in
Fig. 17. Only 22 out-off 117 sources show an extent if we
adopt the extent threshold criterion defined above, that is, 
a value of $P_{ext} \ge 2$. Some of these sources are
very bright and soft and are far from the expectation for
a cluster type spectrum. If we exclude these sources with
deviations larger than $-8\sigma$, there are only 13 sources
for which the nature of the extent should be investigated
to rule out a cluster nature. 
For more than half of these sources
we find a reason why they are featuring an extent: 2 are nearby
galaxies, M82, M106, which show extended X-ray emission, 7 of
the sources are double sources where the main, catalogued source
is consistent with a point source. Some of the remaining sources
are very bright in the RASS (more than 500 photons) and, therefore,
the relative deviation corresponds to a very small absolute deviation
which can be caused by small systematic effects. A pathological 
object in this subsample is the BL Lac RXJ1456.0+5048 which
has a value for $P_{ext}$ of 5.4. It has been analyzed in
a RASS-follow-up HRI observation by Nass (1998) and 
shows at most a very marginal extent in this observation with
a ten times higher angular resolution. Therefore the extent seen
in the present RASS analysis has to be spurious. We have also
checked the mean off-axis position of the source photons as a signature
of an imperfect scanning of the source which could produce a deviation
of the mean PSF for the source observation compared to the average
survey PSF, but found no obvious deviation. Therefore this case
demonstrates that systematic effects in the broadening of the PSF
for individual sources in the RASS exist, but they are obviously 
extremely rare. Subtracting the cases for which the extent is real,
we find that less than 10\% of the sources feature a spurious extent.
Applying a similar test to an unbiased point source sample shows
that the misclassification is usually less than about 5\%
(see B\"ohringer 1999 in preparation).

In Fig. 17 we have also marked the non-cluster X-ray sources which 
seem to be associated with an optical cluster either in projection
or located in the cluster. There is no indication that 
these sources show a different distribution in the source quality
parameters. Therefore the identification that in most of these 
sources the AGN or star is clearly the dominant X-ray source
is supported by this result.

\begin{figure}                                                                  
\plotone{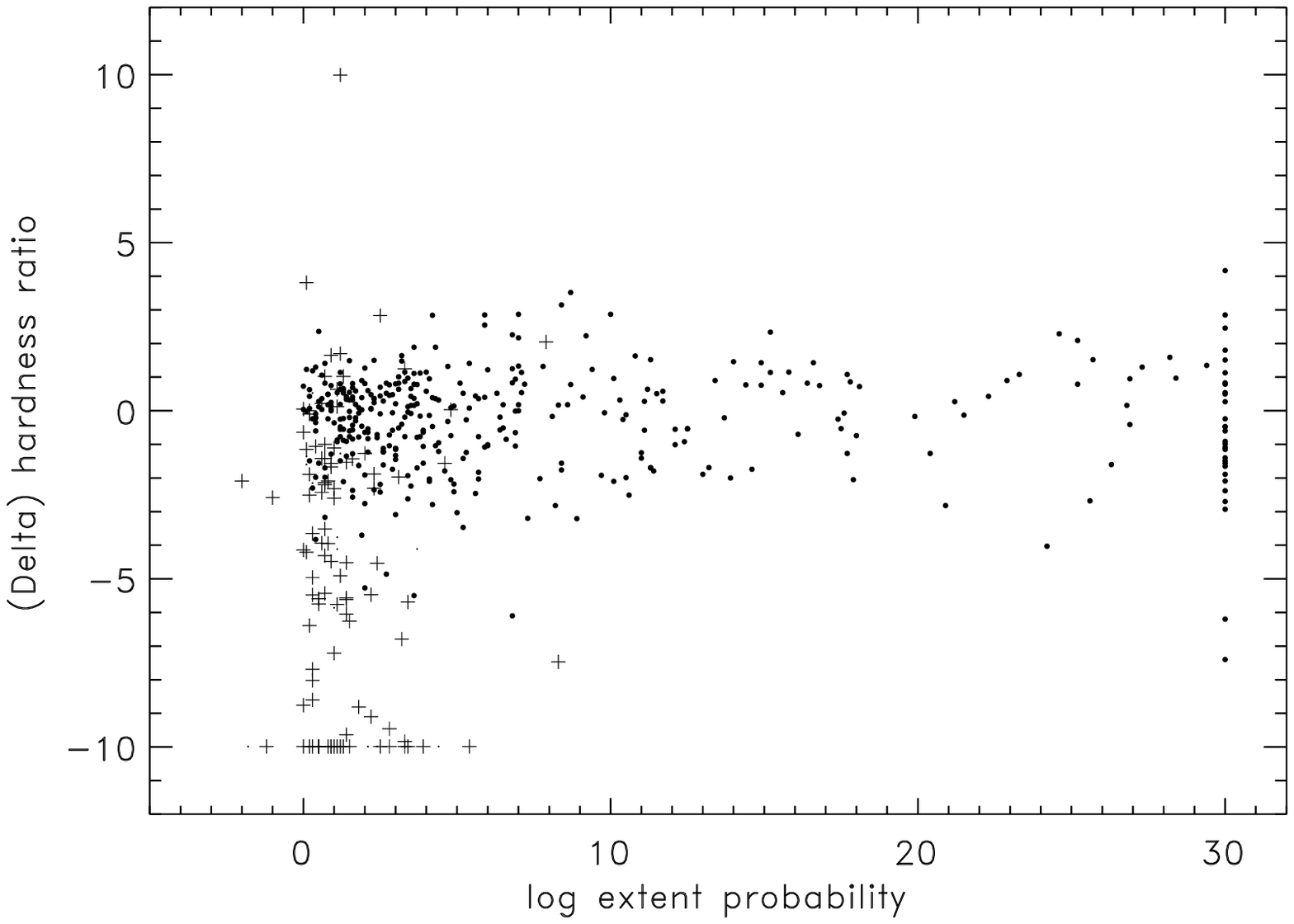}
\caption{Distribution of the values of the negative logarithmic 
probability for the KS-test for a 
point source (labeled log extent probability) and for the 
sigma deviation of the hardness ratio
from the expectation for given hydrogen column density and for 
a 5 keV cluster spectrum.
Filled circles mark the cluster sources while crosses label 
the non-cluster sources.} 
\end{figure}                                                                    
                                                                                
\begin{figure}
\plotone{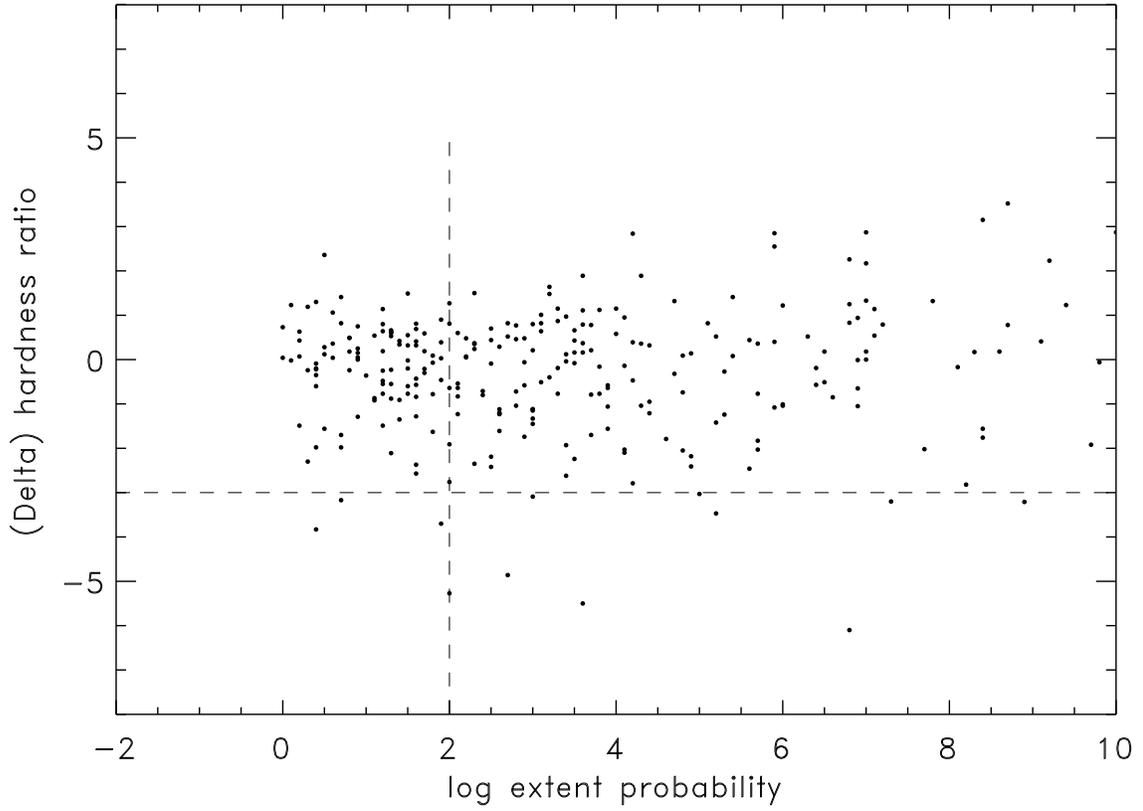}
\caption{Distribution of the log extent probability and the 
sigma deviation of the hardness ratio
from the expectation for the cluster sources only. 
Only about  3\% of the clusters show a more than $3\sigma$ 
deviation to the soft side. This hardness threshold and the threshold
for the extent parameter are marked by dashed lines.}
\end{figure}

\begin{figure}
\plotone{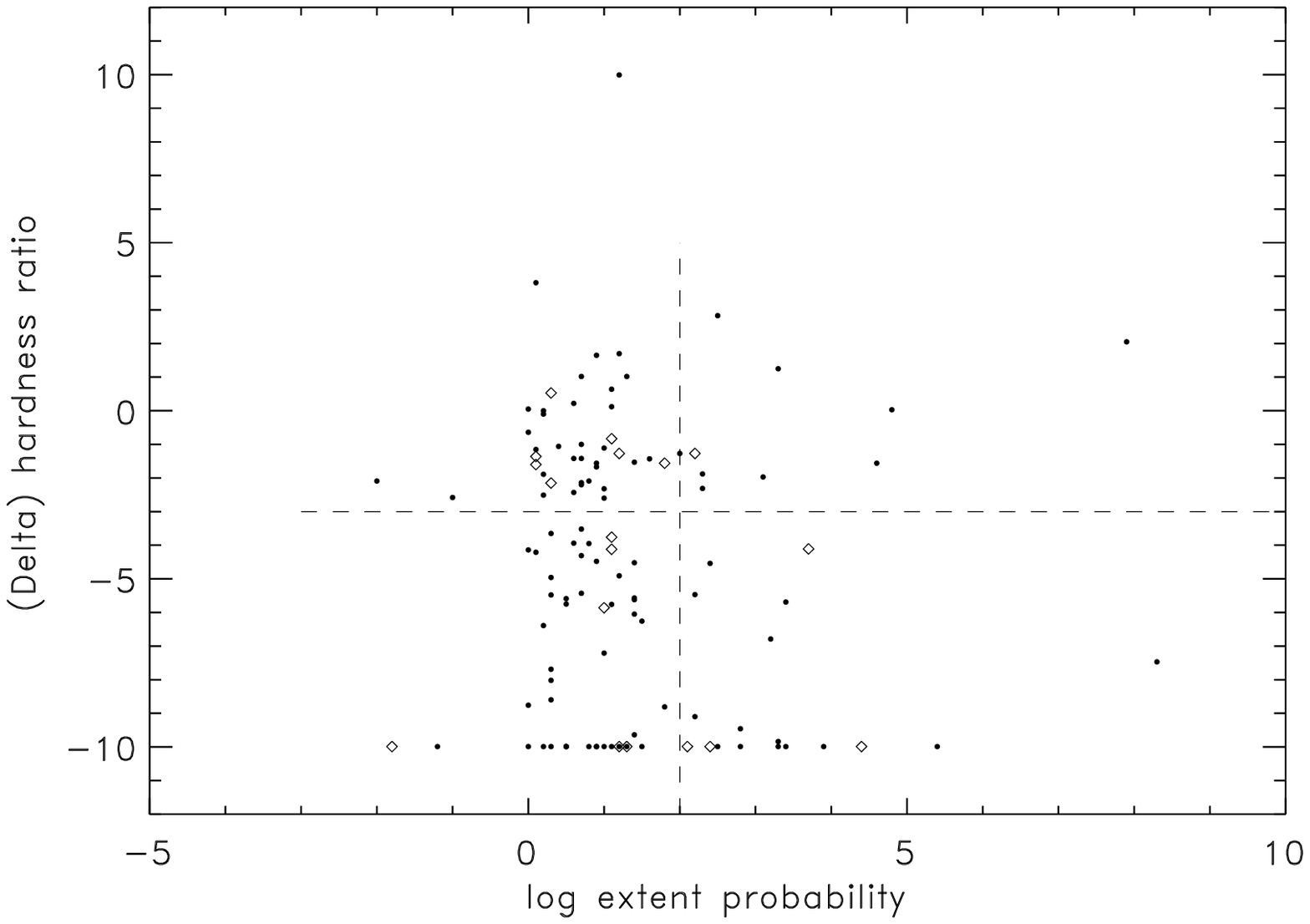}
\caption{Distribution of the log extent probability and the 
sigma deviation of the hardness ratio
from the expectation for the non-cluster sources only. 
Less than 19\% of the non-cluster sources show 
a log extent probability value larger than 2. }
\end{figure}

These results on the source quality can be compared with the prime
selection criteria of this sample as extracted from RASS I. We recall
that only sources which were characterized by 
a significantly large extent
parameter obtained in the maximum likelihood analysis were included
in the present sample. Using the new technique $\sim 25\%$  
of the sources do not
feature an extent. This could be partly due to the fact that the present 
method is sometimes less sensitive in recognizing the source extent for very
compact sources since it does not use the detector position information
and thus does not weight for photons imaged with different sharpness
as does the maximum likelihood method. Looking at the nature of the 
sources we conclude that about 23\% of the sources are genuine point
sources and thus the failure rate of the RASS I extent classification
is at least about 20\%. The results displayed in Figs. 15 - 17 show that
the present method provides a great improvement concerning the failure
rate of the method. This justifies the use of these
source quality parameters to assist the identification of the sources. 

We have made use of the above results in the identification process. The
criterion for classifying a source as a non-cluster source was
one of the following:

i) The optical counterpart is a bright star ($\ge 12.5$ mag) 
and the source is not extended. This is justified by the finding
of Voges et al. 1999 (in preparation) that there is a much less than
1\% chance coincidence of a RASS X-ray source with such a bright
star.

ii) There is an optically identified AGN at the source position
and the source is not extended. No cluster is seen on the CCD images
taken for this project.

iii) The source is a known, previously identified non-cluster
X-ray source.

iv) The source is clearly point-like and has a soft
hardness ratio deviation larger than $3\sigma $. There is no 
signature of an optical cluster in the digitized Palomar Sky 
Survey image or the CCD images taken for the project.

We noted 17 cases, where there is a signature of an optical cluster,
but the other criteria are inconsistent with a cluster identification.
These cases have been listed separately in Tables 3 and 4.

This identification scheme is only made possible by the
extensive spectroscopic follow-up and almost complete CCD imaging
of the targets in the sample of extended sources (no spectroscopy
was done for those objects for which this information is already
available from the literature or the CfA archives). A classification
only by identification of clusters on optical images or 
exclusion of sources with spectroscopically identified AGN without
further inspection would have failed in several cases.

\section{Further analysis of the 9 - 14$^h$ study region} 
                                          
While the above results show that the source identification 
has reached a high level of reliability, there are serious
concerns about the completeness of the present sample in
terms of a purely flux-limited X-ray cluster sample,
because the sample 
rests on the RASS I source extent selection criterion.
There are two principle sources of incompleteness: i)
medium distant and distant clusters may have too compact X-ray
emission regions to be resolved as extended X-ray objects 
in the RASS and ii) the RASS maximum likelihood analysis may
fail to recognize all the spatially resolved sources in the RASS
as extended X-ray sources. Both sources of incompleteness
affect the present sample. Figs. 2 and 16 show that there are 
galaxy clusters which appear as point-like X-ray sources for
both analysis techniques, the standard RASS analysis 
and the method used here.
In addition we will find below that there is a large fraction of
significantly extended sources which are missed by RASS I.

In an attempt to check for the incompleteness of the RASS I sample
we conduct two studies: we investigate
what fraction of ACO clusters with X-ray emission is missing in
the present sample and we search for extended sources in the RASS II
data base with the new analysis technique and inspect their nature
to find additional clusters or cluster candidates. 
Since this is meant as a statistical test we restrict the analysis
to a subregion of the study area: the region of the northern sky
between $9^h$ and $14^h$. The sky area of this region is 1.309 ster
as compared to the total NORAS survey area of 4.134. In this 
right ascension range the $|b_{II}| \le 20\deg$ band of the Milky Way
is completely located at negative declinations. We restrict the
cluster search by imposing an X-ray flux limit of 
$1.6 \times 10^{-12}$ erg s$^{-1}$ cm$^{-2}$.

Thus in the first step we have run our analysis on the 
sky positions in RASS II of the 901 clusters
listed by Abell, Corwin, \& Olowin (1989) in the sub-survey region.
Ten clusters of this sample were missed in the analysis since
RASS II has a too low exposure in the corresponding sky fields
for a significant source detection
(these clusters are: A917, A974, A986, A996, A999, A1011, A1042,
A1057, A1128, A1554).
93 additional ACO clusters are detected above the flux limit.
A closer inspection of the detections with the same procedures
as described above revealed that 8 of the detections have most
probably an AGN as the dominant X-ray source. These ACO clusters 
are: A763, A924, A1030, A1225, A1575, A1593, A1739, A1774. The 
results for the other 85 detections, where we identify the X-ray
emission to originate in the cluster, are listed in Tables 6 and
7. The columns of these tables have the same meaning as those of
Tables 1 and 5, respectively.   

In the second step we have selected all the sources from the
RASS II data base in the sub-survey region, reanalyzed them,
and extracted all the sources with a flux $\ge 1.6 \times 10^{-12}$
erg s$^{-1}$ cm$^{-2}$ and an extent parameter $P_{ext} \ge 2$.
In total there are 377 sources of which a large fraction is already 
contained in the NORAS I sample and the supplementary ACO sample.
Again we find another fraction which is identified as double
point sources or very bright point sources having a small absolute extent.

An inspection of all the remaining sources on the digitized
sky images and literature data from data bases leads to a sample of
52 very promising candidates of which 21 can readily be identified
as previously  listed galaxy clusters. The latter list of identified
cluster sources is given in Table 8 and 9. For the positive
identification of the remaining candidates further optical 
observations within the ongoing NORAS survey are in progress.

To understand which of the two sources of incompleteness discussed above
is more important for the loss of these additional cluster sources
in the primary candidate list of the NORAS Survey, we plot in Fig. 18
the distribution of the extent parameters of the X-ray clusters of the
two additional cluster lists. As we can see most of the additionally
found ACO clusters (74.2\%) feature an extent according to the new analysis
technique. The second supplementary list consists by definition
only of extended cluster sources. Therefore we have to conclude,
that the extent flag in the RASS I data base is not only not very
reliable but a large fraction of the well extended 
sources is also missed.  
                
\begin{figure}
\plotone{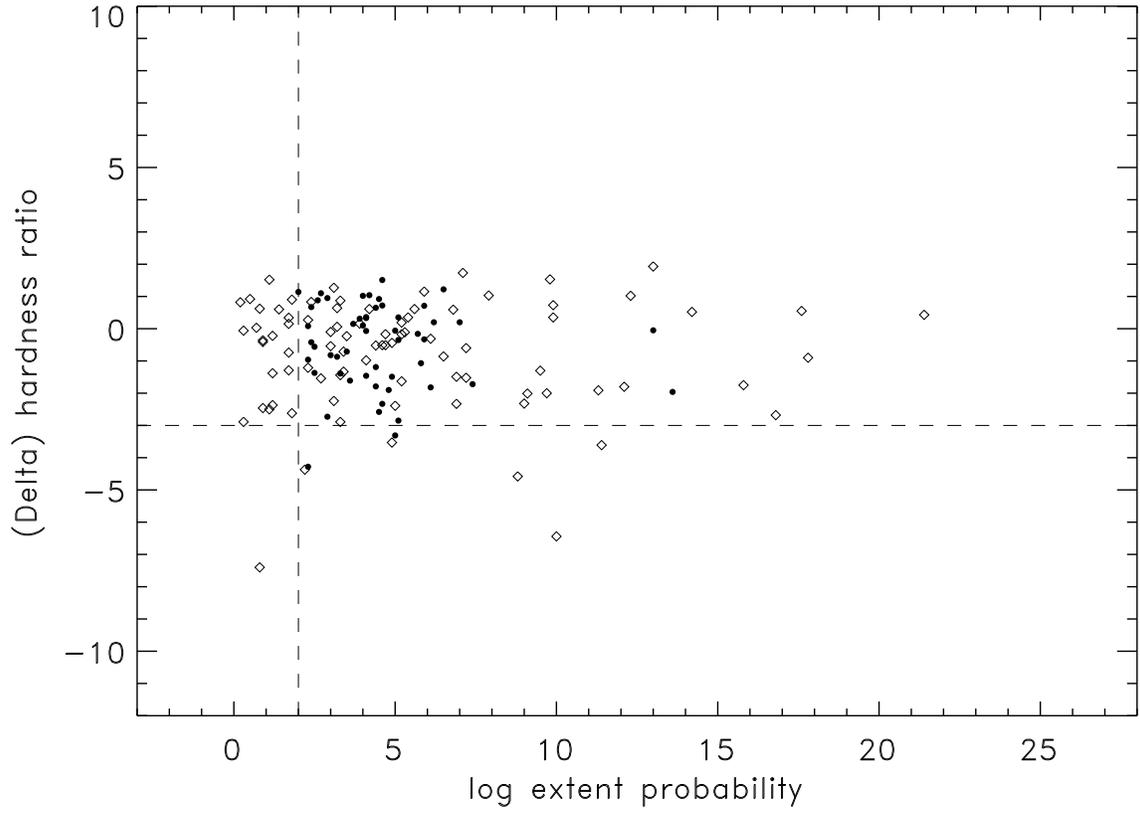}
\caption{Distribution of the log extent probability and the 
sigma deviation of the hardness ratio
from the expectation for the additionally found  ACO clusters
(diamonds) and extended sources (dots) from the 9$^h$ to 14$^h$ 
study region.}
\end{figure}

\section{Discussion of the sample completeness}

To test the completeness of the
catalogue, a comparison can be made to the southern RASS 
cluster survey project, the REFLEX Survey (B\"ohringer et al. 1998).
This sample has been constructed in a different way making extensive
use of the COSMOS data base to correlate X-ray sources with
the galaxy distribution. Therefore this sample does not rely 
on existing cluster catalogues nor on the selection of X-ray sources
featuring an extent. Internal statistical estimates for the REFLEX sample 
suggest a completeness larger than 90\% for the 
flux limit of  $F_X = 3 \times 10^{-12}$ erg s$^{-1}$ cm$^{-2}$. 
The number counts of the REFLEX survey and the NORAS survey are 
compared in Fig. 19. We note that the NORAS sample reaches a 
fraction of 50\% of the REFLEX number counts at the REFLEX flux limit.

A large fraction of the missing clusters has been recovered by the
supplementary sample in the study area. In Fig. 20 various
subsamples of the combined cluster catalog in the 9$^h$ to 14$^h$
region are compared to the REFLEX Survey. While the cluster
sample from Table 1 recovers with 40\% an even smaller fraction 
of the sky density compared to the REFLEX Survey, 70\% are
reached if the ACO clusters are added and 82\% are obtained for
the combined sample. Thus there is still a fraction of clusters
missing. This is easily understood, since the REFLEX sample contains
a fraction of 22\% of all objects which are not resolved as extended by
the GCA analysis. We can expect that about 10 - 20\% of the
clusters for the REFLEX flux limit are neither listed as ACO clusters
nor recognized as extended sources. Note that the combined sample in 
the study region completely recovers the previous RASS cluster 
sample by Ebeling et al. (1998) as discussed below. Most  
clusters not contained in Table 1 are easily found as ACO clusters and 
7 further clusters show a clear extent. 

Therefore to achieve a higher completeness in our continuing northern
cluster survey we are, in addition to including known catalogued
clusters with X-ray emission and newly classified extended X-ray
sources, conducting further imaging of promising non-identified
X-ray sources to recover the compact X-ray clusters missed in the
previous searches.

\begin{figure}
\plotone{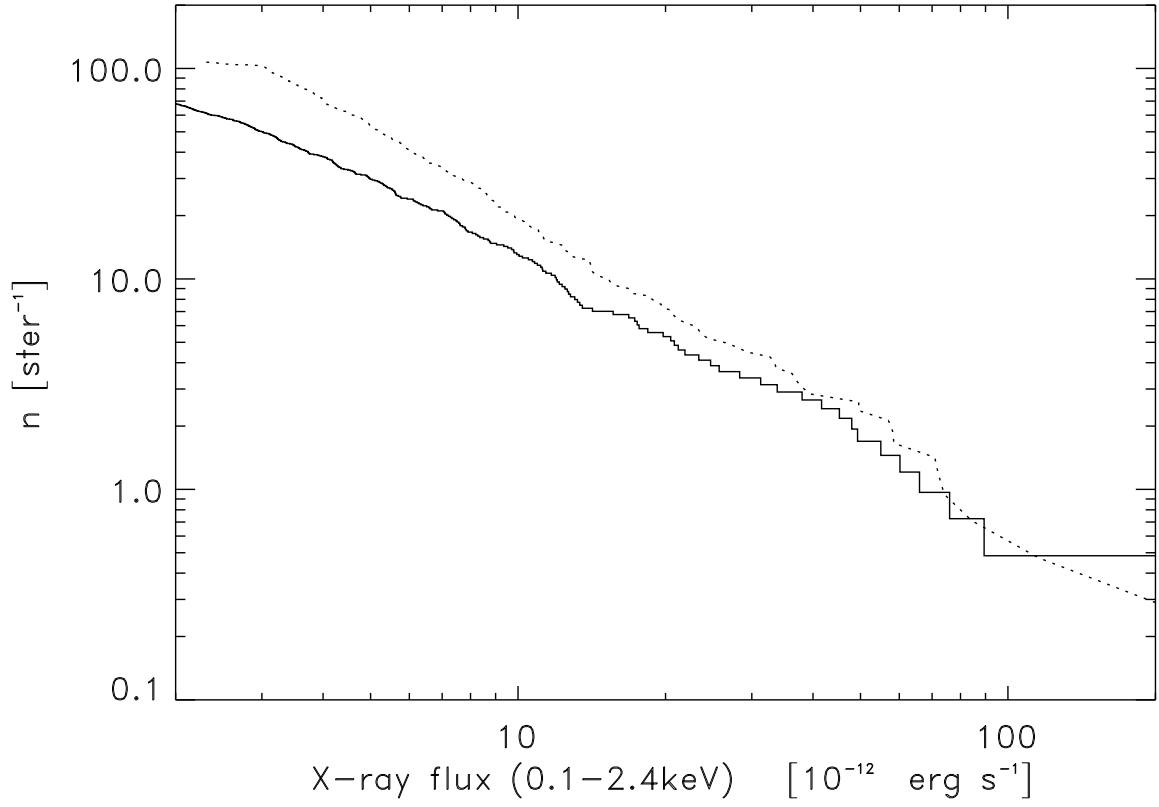}
\caption{
LogN-logS for the NORAS I sample (continous line) compared
to the REFLEX Cluster Survey (dotted line).}
\end{figure}

\begin{figure}
\plotone{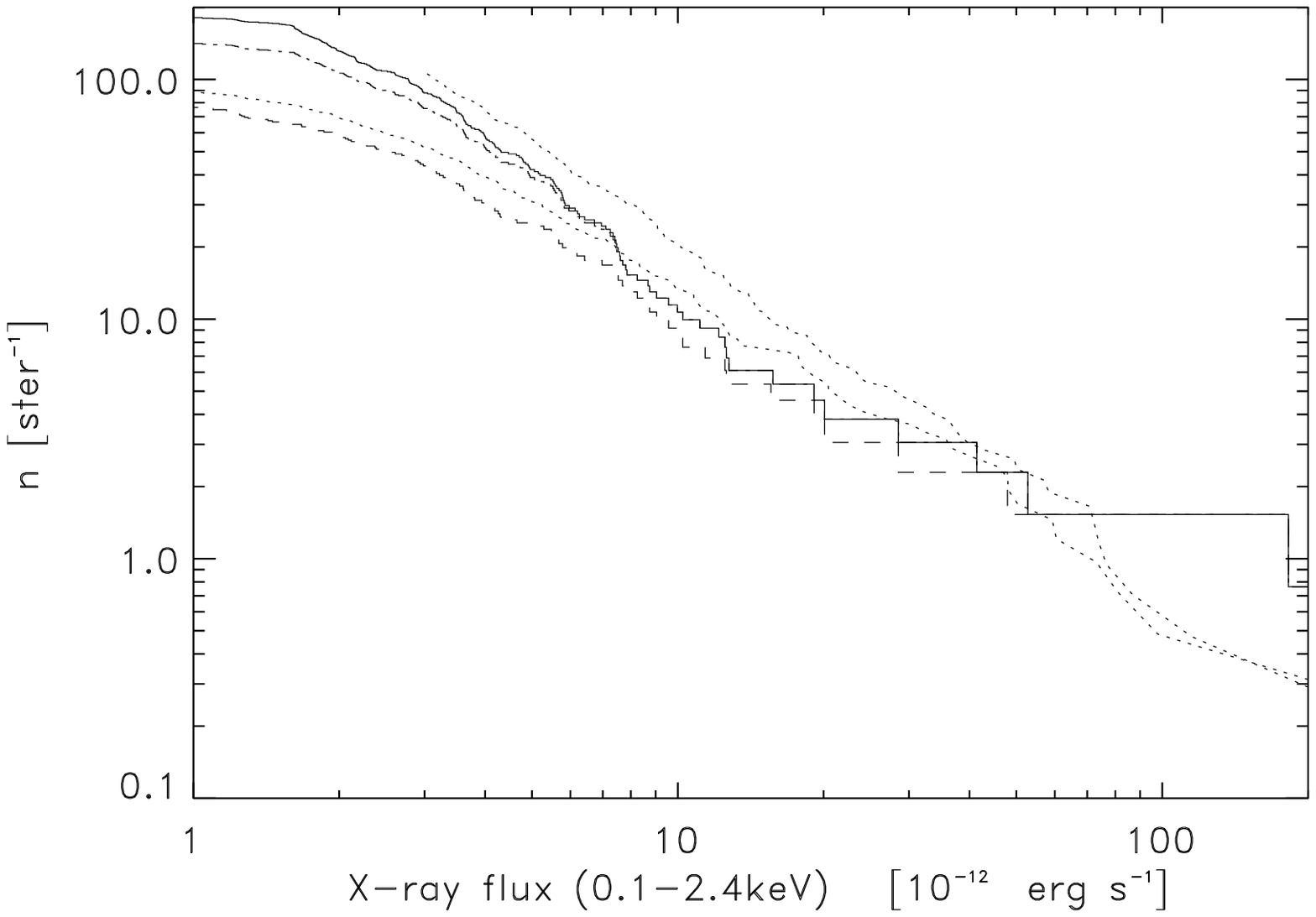}
\caption{
LogN-logS for the NORAS 9-14 hour region. NORAS I main sample (lower
dashed curve), NORAS I and ACO clusters (dashed-dotted curve),
NORAS I, ACO, and supplementary extended cluster candidates (continuous
curve), NORAS I sample for the whole northern sky (lower dotted curve),
REFLEX Survey (upper dotted curve).}
\end{figure}

\section{Comparison with the northern Brightest Cluster Sample}

In total 166 X-ray sources analyzed here overlap with the 
previous RASS galaxy cluster compilation by Ebeling et al. (1998).
142 sources coincide with the sources of Table 1, 22 sources with
sources in Tables 6 and 8 for the 9$^h$ to 14$^h$ region, and
two sources are identified with AGN in our analysis. Thus 40 sources
compiled by Ebeling et al. are not included here
(including the Virgo cluster). Since the
two compilations are made with the same intention we compare
the results of the two samples in some detail.

Fig. 21 shows a comparison of the X-ray fluxes determined 
by the two different methods used in the two surveys for the 166 cluster
in common. Note that in our compilation the
fluxes have not been corrected for the missing flux outside the
measurement aperture. Therefore we show both results from the
work of Ebeling et al. in the plots: the uncorrected measured fluxes
(with open circles) and the corrected final fluxes (full circles).
The agreement at high fluxes is very good and there is also quite
good agreement for most sources. The scatter is increasing, 
however, towards lower fluxes. Mostly at lower fluxes we find a 
few sources with fluxes larger by factors up to 2 
as determined  by the VTP method compared to the present results.
Note the bias introduced at low fluxes by the flux limit set in
the Ebeling et al. sample that suppresses the cases where the 
VTP to GCA flux ratio is lower than one. Thus, what appears like
a high bias of the VTP results at low flux in Fig. 21b is most 
probably explained by an increased scatter in the flux ratio
with decreasing flux. Thus in general the agreement is good.

\begin{figure}
\plottwo{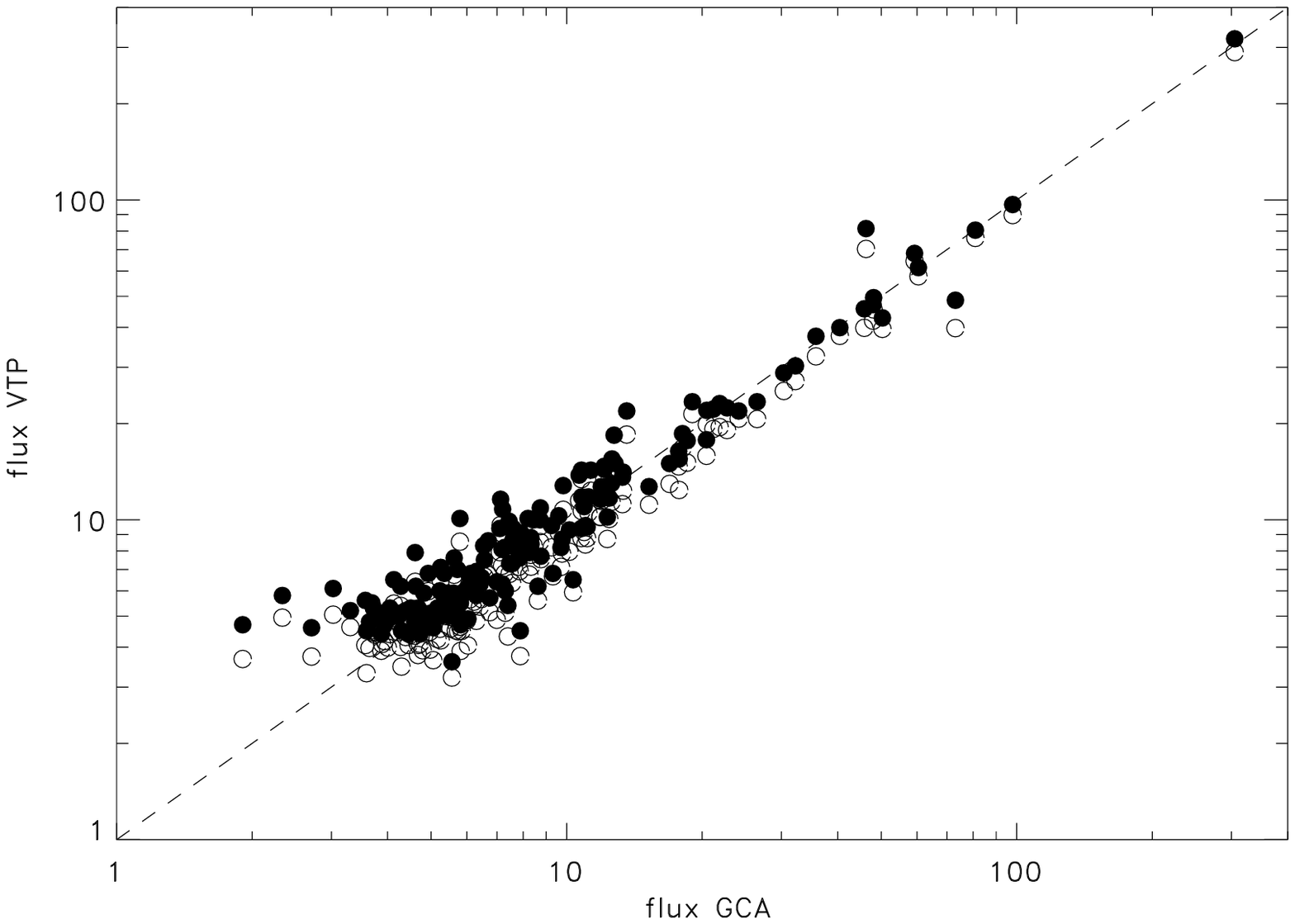}{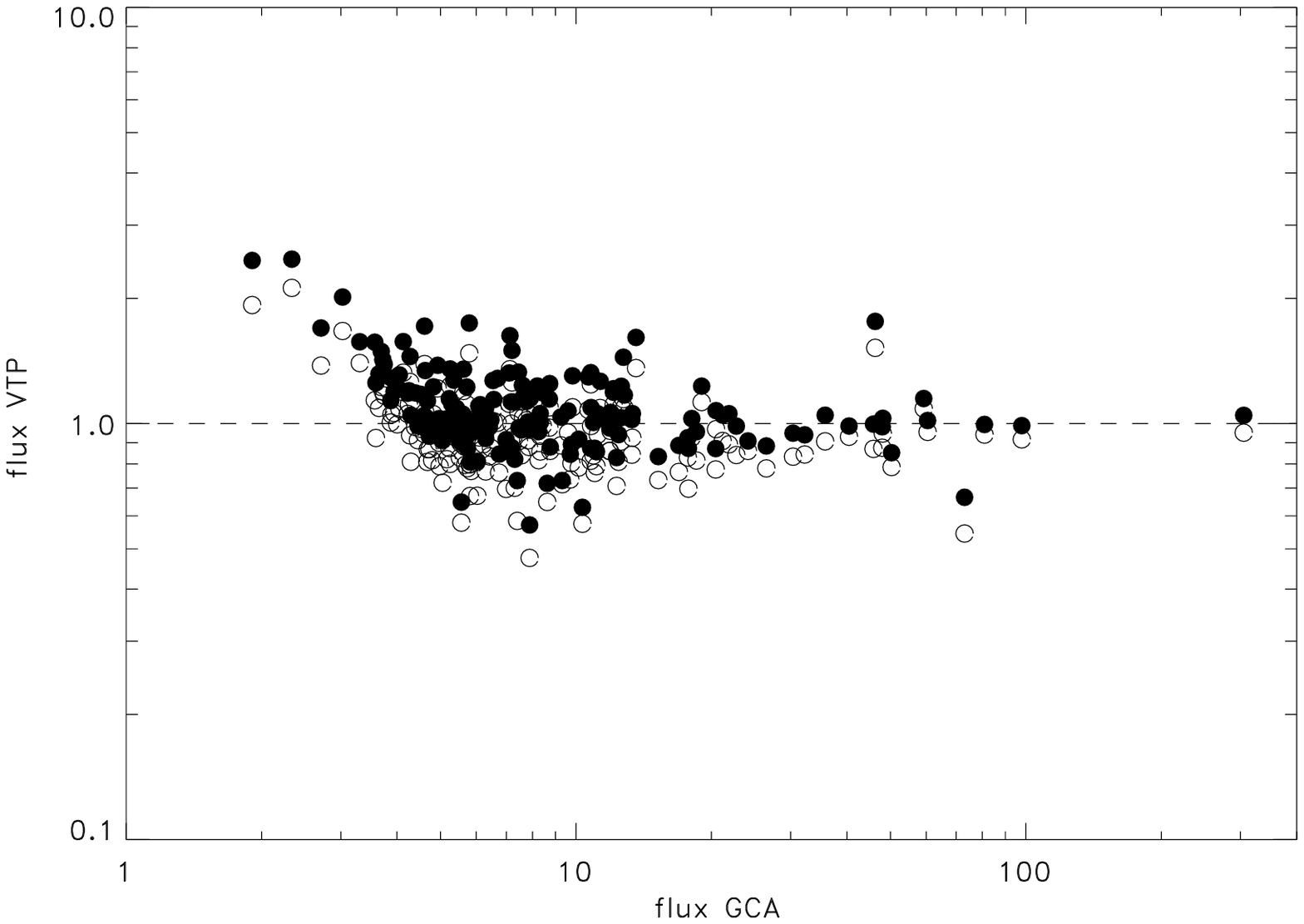}
\caption{Comparison of the X-ray fluxes derived in the present analysis 
with the results of Ebeling et al. (1998) for the 164 sources which are
common to both samples.}
\end{figure}

Two sources in the cluster list by Ebeling et al. (1998),
A2318 and Z2701, are classified as AGN in our work. The first
source appears as RXJ1905.7+7804 in Table 3. We have optically identified
a Seyfert 1 galaxy at the source position. 
The X-ray source is pointlike and has
a spectral hardness ratio that is too soft for
a cluster with a 3.8$\sigma$ deviation.
The X-ray source is also significantly offset from the optical cluster
center. Thus we identify the X-ray source with the AGN. 
The second source was found in our search for additional extended
sources in the 9$^h$ - 14$^h$ region and is not catalogued here.
It was dismissed because it is identified by as AGN by Bade et al. 1998.
The spectral hardness ratio shows again a 4.5$\sigma$ deviation
to the soft side. The KS-test yields a small extent of the source
and an inspection of the X-ray image indicates a smaller emission
contribution from the region of the optical cluster center. But
the main contribution comes obviously from the AGN. 

We can also make a comparison of the completeness of 
both surveys in the 9$^h$ - 14$^h$ region. While we recover
all the clusters in Ebeling et al. (except for Z2701 which is 
classified as AGN), two additional clusters are found in our search
with a flux larger than $L_X \ge 4.7 \times 10^{-12}$ erg s$^{-1}$
cm$^{-2}$ (in addition the two Virgo galaxies M60 and M49
which are summed into the Virgo cluster in Ebeling et al.).
Over the whole right ascension range we find 12 additional
objects above the flux limit of Ebeling et al. These sources
are RXCJ0005.3+1612 (A2703), RXCJ0209.5+1946 (A311), 
RXCJ0736.4+3925 (contains a BL Lac), RXCJ1121.7+0249 (SHK352), 
RXCJ1242.8+0241 (NGC4636), RXCJ1447.4+0827, 
RXCJ1501.2+0141, RXCJ1506.4+0136, RXCJ1617.5+3458 (NGC6107),
RXCJ1718.1+7801 (A2271), RXCJ1742.8+3900, RXCJ1900.4+6958 (A2315).
All the sources except for RXCJ1447.4+082 have a significant
extent. The Abell clusters have fluxes within about 15\% of the
flux limit of Ebeling et al. (1998) and most of the remaining sources
are nearby groups of galaxies being less rich than Abell clusters.
This number of extra sources is still compatible with the claimed
completeness of the Ebeling et al. (1998) sample. The above
comparison with the results of the southern RASS analysis in the 
REFLEX sample indicates, however, that both samples are still
less complete than the southern survey.
                                                                       
\section{Discussion and Summary}                                                
         
Using the early results of the RASS we compiled an X-ray
sample of galaxy clusters by selecting the RASS sources
which featured a significant extent in the first standard processing
of the RASS. The complete spectroscopic identification of these
sources (as far as no identification was already available)
leads to a compilation of a catalogue of 378 X-ray cluster sources.

A reanalysis of the X-ray properties of these sources shows that
the X-ray source properties can successfully be used in the source
identification process. In particular we found that in many cases the
hardness ratio can be used to flag sources which are severely
contaminated by an AGN. In this way we could also exclude 
previously identified X-ray clusters -- as for example the two cases
discussed in the previous section -- from identification as cluster
sources. 

One of the major points of concern of such catalogs of X-ray clusters
is the contamination of the observed X-ray luminosity by AGN.
As commented in Section 4.1 there are all combinations of clusters
and optically identified AGN or radio galaxies: AGN which do
not provide a significant contribution to the cluster X-ray flux,
AGN that contribute partially, and AGN which completely outshine the
clusters. These cases can only definitely be distinguished if
high resolution  X-ray images or X-ray spectra of high quality are
available. E.g. ROSAT HRI observations provide the means for this
distinction, but they are only available for a small fraction of 
the sources for survey sizes of the present survey. Lack of additional
X-ray data creates a twofold danger. Since X-ray sources are often
identified with the most plausible nearby optical counterpart,
there is a significant risk that the cluster ID will be discarded
when an AGN is found near or within the error circle. Conversely,
an AGN which is the primary X-ray emitter may not be recognized
and the emission falsely associated to the cluster. 

Also in the present case this problem of AGN cluster associations
cannot be solved in each case and we still expect some hidden 
misclassifications which will be hard to find until the whole sample 
is probed more deeply in X-rays. However, we believe
that the remaining uncertainties are small and not harmful for
the application of our sample for cosmological statistical studies     
and we will substantiate this believe in the following.
First of all we note that $74.4 \%$ of the sources in the catalog
in Table 1 are sources recognized as significantly extended.
This requires a major contribution to the X-ray emission from
diffuse cluster sources. Thus for this source
population we have a very high reliability that these sources are 
true X-ray clusters. We can now - as shown in Fig. 22 - compare
the spectral properties of these extended cluster sources to
those cluster sources where no significant extent could be
established in terms of the parameter $P_{ext}$. We note that
the two distributions are hardly different. Also both distributions
are almost symmetric around unity, as would be ideally expected.
For some reason the larger sample of extended sources shows 
a broader distribution of the $\Delta HR$ parameter. This seems 
to be caused by clusters with a larger number of photons for
which a small systematic error in the $HR$ estimate leads to
a significant additional scatter in $\Delta HR$ in  units
of $\sigma$ and also the low temperature groups will add to
this scatter.
A comparison to the
non-cluster sources shows a very large difference in the spectral
parameter distribution, however. While the non-cluster sources
have a median deviation of the parameter $\Delta HR$ of $-4.6\sigma$
the X-ray pointlike sources identified as clusters have
a median $\Delta HR \sim -0.02\sigma$. 

We can explore the contamination effect in somewhat more detail
by considering the effect on a typical cluster source. The median
number of source photons for the clusters listed in Table 1 is
about 94 source photons. For such a source the typical uncertainty
in the measured hardness ratio is about $\delta HR \sim 0.12$.
An AGN with median properties easily shows a deviation of 
$3 - 4 \sigma$ in the $\Delta HR$ parameter. Even if the contamination
fraction by AGN is only 50\% or 25\% the typical shift in the parameter
of  $\Delta HR \sim 1.7\sigma$ and  $\sim 0.9\sigma$ still constitutes
a recognizable distortion of the $\Delta HR$-distribution.
Another test is shown in Fig. 22b. Here we compare the distribution
of the $\Delta HR$ values for the clusters with the sample of point
like clusters artificially contaminated by removing 20\% of the cluster
sources and replacing them by a statistical sample of non-cluster sources.
There is a clearly visible change in the distribution of the spectral
parameters. A KS test shows that the two distributions are still
only distinguishable at the $\sim 90\%$ level. A test with a similar
contamination of 40\%, however, leads to a clear difference with a
KS probability for the two samples being statistically the same of
$\sim 10^{-4}$. Since a 20\% contamination of the point like clusters 
corresponds to a contamination of about 5\% of the total sample
we expect that the real misclassification fraction is not larger
than this percentage.

\begin{figure}
\plottwo{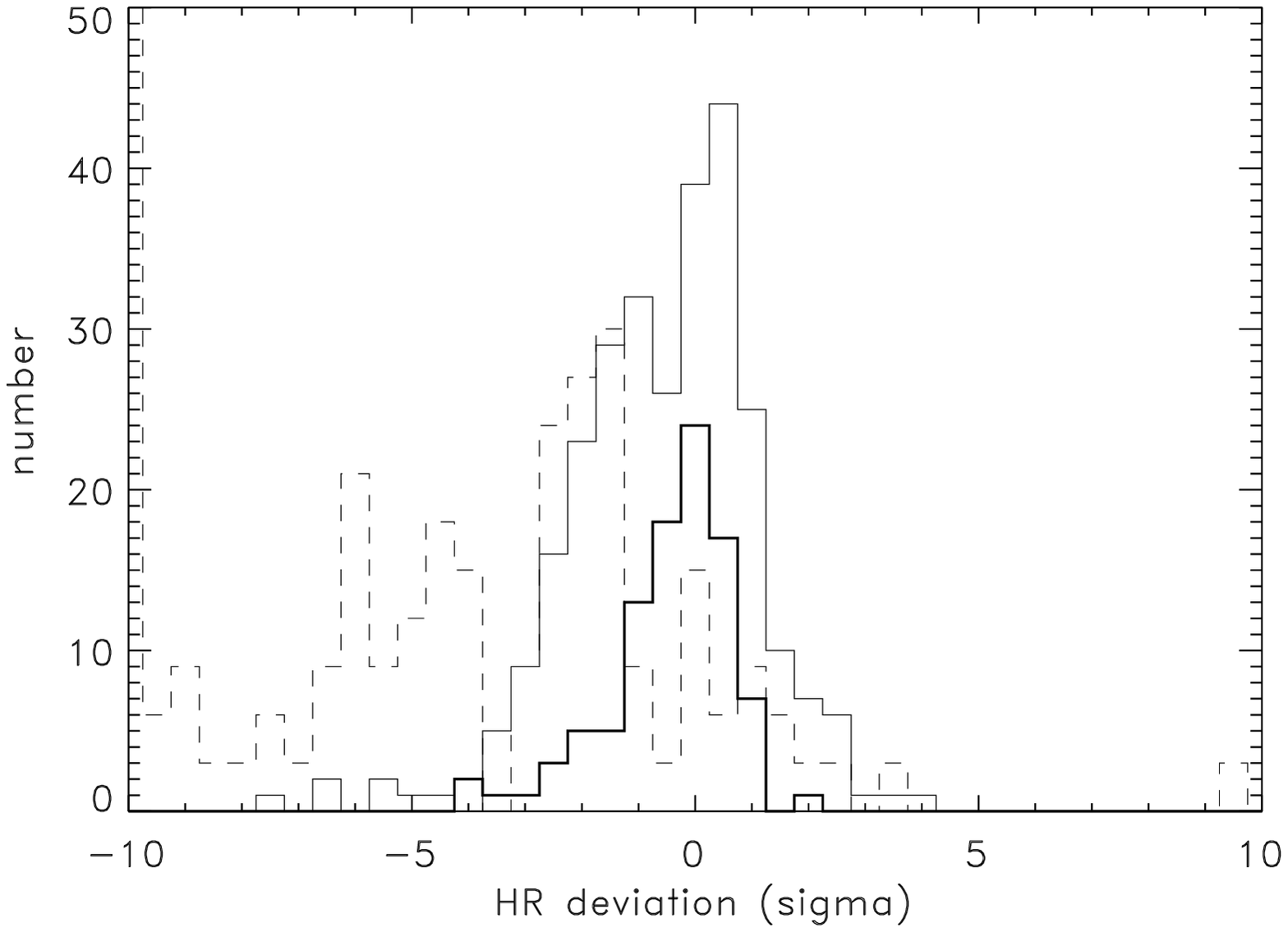}{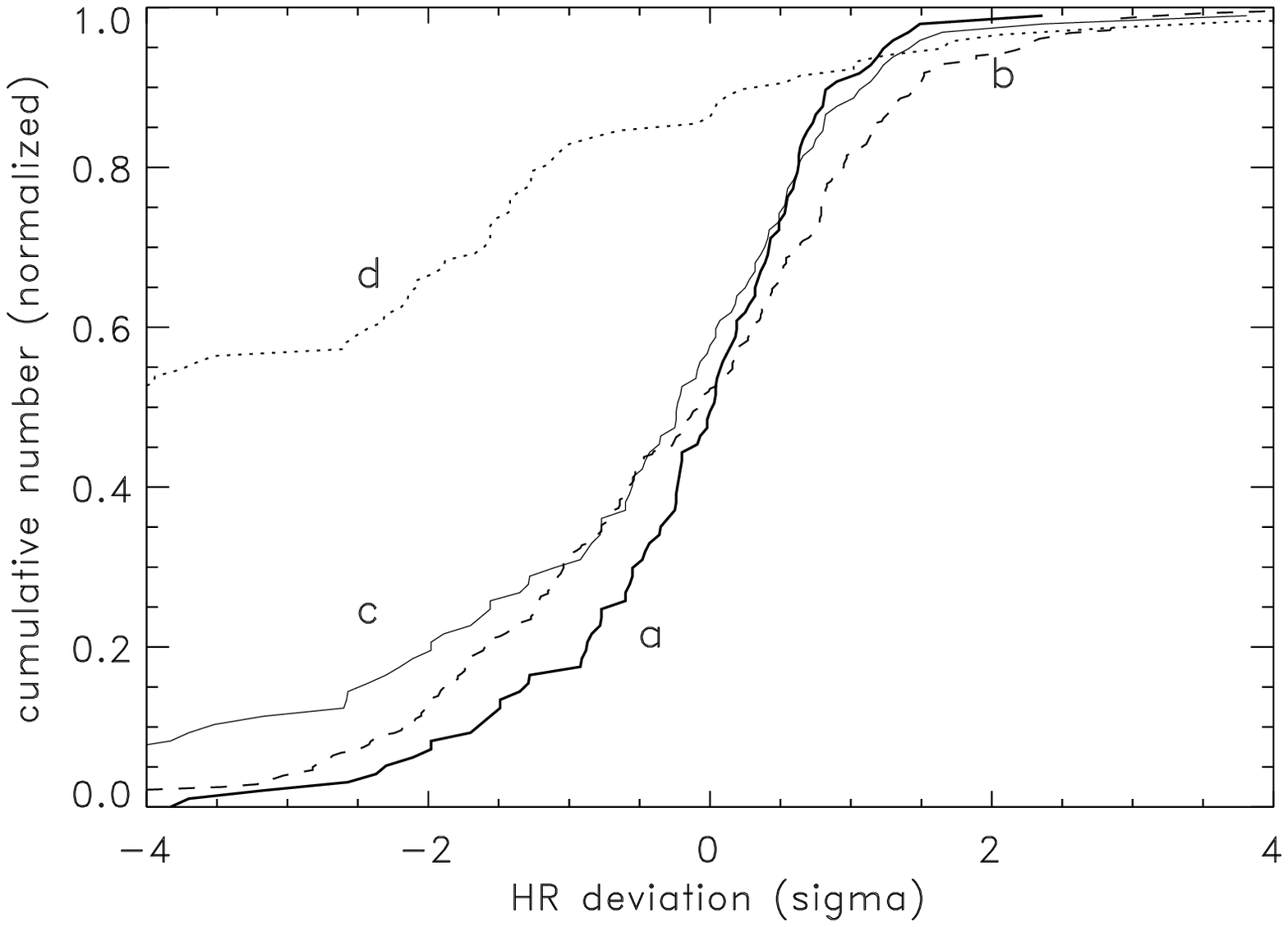}
\caption{Distribution of the spectral parameter, $\Delta HR$ in units of
sigma, for different subsamples of the present study. The parameter $\Delta HR$
is the difference of the measured from the estimated hardness ratio of the
source for given interstellar absorption scaled by the 1$sigma$-uncertainty
of the hardness ratio determination. The right pannel shows histograms of
the parameter $\Delta HR$ for pointlike cluster sources (thick line), extended
cluster X-ray sources (thin line), and non-cluster sources (broken line). The
right pannel shows normalized, cumulative histograms for (a) the pointlike cluster 
sources, (b) the extended cluster sources, (c) the sample of pointlike cluster
sources artificially contaminated by 20\% of non-cluster sources, and
(d) the non-cluster sources. The samples (b) and (c) have an about 10\% KS probability
of being sampled from the same parent distribution as (a).}
\end{figure}

The present X-ray cluster catalog provides a wealth of new data.
98 new cluster sources are listed in Table 1 and 
in addition new X-ray luminous groups associated with known 
giant ellipticals are reported in Table 1 and 8.

Test searches for additional X-ray clusters in the 9$^h$ - 14$^h$
region have shown that both, the selection of extended sources
by the RASS standard analysis and the finding of X-ray clusters by selecting
extended sources is quite incomplete in terms of the compilation
of flux-limited X-ray cluster catalogues - except for very high 
flux limits. We have shown that probably most of the missing clusters 
can be recovered by using a better search algorithm for extended
sources and by screening the optically known clusters. 
This may not be sufficient to produce a high quality, highly 
complete catalogue. Further screening of RASS sources, including CCD
imaging of sources not visibly extended in X-rays, is required.
The completion of the NORAS sample thus requires additional imaging
and spectroscopic observations.

\acknowledgments                                                                
We like to thank first of all the ROSAT Team who helped in providing
the RASS data fields, in discussions about the RASS source properties,
and by providing the EXSAS software which was essential for the
studies. We like to thank in particular J. Engelhauser, A. Vogler,
C. Rosso, and C. Izzo for help with the data and programs.
We would also like to thank the remote observers at FLWO, Perry
Berlind and Jim Peters, for observations made on the 1.5-m telescope,
and Susan Tokraz for help in the data reduction. 
This work has made use of the SIMBAD data base operated at CDS,
Strasbourg. In addition this research also made use of the 
NASA/IPAC Extragalactic Database (NED), which is operated by
the Jet Propulsion Laboratory, California Institute of Technology,
under contract with NASA. H.B. and P.S. acknowledge the support by 
the Verbundforschung under grant No. 50 OR 93065 and 50 OR 970835,
respectively. JPH and JM were supported by the Smithonian Institution.

\end{document}